\documentclass[journal]{IEEEtran}
\usepackage[utf8]{inputenc}
\usepackage{echodefs}
\usepackage{booktabs}
\usepackage[caption=false]{subfig}
\usepackage{import}
\usepackage{nccmath}

\usepackage{algorithm}

\algnewcommand\algorithmicforeach{\textbf{for each}}
\algdef{S}[FOR]{ForEach}[1]{\algorithmicforeach\ #1\ \algorithmicdo}

\newcommand{\rev}[1]{{\color{black}#1}}

\title{Total Least Squares Phase Retrieval}

\author{
Sidharth Gupta$^{\,\dagger}$ and Ivan Dokmani\'c$^{\,\ddagger,\, \dagger}$ \\
\normalsize{$^\dagger\,$University of Illinois at Urbana-Champaign, $^\ddagger\,$University of Basel} 
\\
\normalsize{\texttt{gupta67@illinois.edu, ivan.dokmanic@unibas.ch}}
\thanks{This work was supported by the European Research Council Starting Grant 852821---SWING.}
}

\begin{document}

\maketitle

\begin{abstract}

We address the phase retrieval problem with errors in the sensing vectors. A number of recent methods for phase retrieval are based on least squares (LS) formulations which assume errors in the quadratic measurements. We extend this approach to handle errors in the sensing vectors by adopting the total least squares (TLS) framework that is used in linear inverse problems with operator errors. We show how
gradient descent and 
the specific geometry of the phase retrieval problem can be used to obtain a simple and efficient TLS solution. Additionally, we derive the gradients of the TLS and LS solutions with respect to the sensing vectors and measurements which enables us to calculate the solution errors. By analyzing these error expressions we determine conditions under which each method should outperform the other. We run simulations to demonstrate that our method can lead to more accurate solutions. We further demonstrate the effectiveness of our approach by performing phase retrieval experiments on real optical hardware which naturally contains both sensing vector and measurement errors.

\end{abstract}

\begin{IEEEkeywords}
Phase retrieval, total least squares, operator error, sensing vector error, quadratic equations.
\end{IEEEkeywords}


\section{Introduction}

\IEEEPARstart{I}{n} the phase retrieval problem we seek to recover the signal $\vx \in \C^N$ from complex quadratic measurements
\begin{align}
    y_m \approx \abs{\inprod{\va_m, \vx}}^2, \qquad m = 1, \ldots, M \label{eq:quadratic_inverse_problem}
\end{align}
where $y_m \in \R$ are observed measurements and $\va_m \in \C^N$ are sensing vectors. 
This problem appears in a plethora of applied science applications such as x-ray diffraction crystallography or astronomy where the sensing vectors are Fourier basis vectors \cite{bendory2017fourier} and imaging through scattering media where the sensing vectors may be complex random Gaussian \cite{gupta2020fast}. 

In a prototypical phase retrieval problem, an object, $\vx$, is illuminated and the resulting optical field is measured with a detector. This optical field is complex-valued but common camera sensors only measure intensity, $\{\abs{\inprod{\va_m, \vx}}^2\}_{m=1}^M$, and thus the measurement phase information is lost. The left and right hand sides in \eqref{eq:quadratic_inverse_problem} are only approximately equal because in practical settings there can be errors in the measurements and sensing vectors. In this work we focus on gradient-based optimization strategies to solve \eqref{eq:quadratic_inverse_problem} where $M > N$. Gradient-based methods have proven successful when imaging through random scattering media \cite{gupta2020fast} or with coded diffraction Fourier patterns \cite{candes2015phase}.

Many recent approaches for solving the phase retrieval problem solve variants of the following nonlinear and nonconvex least squares (LS) problem,
\begin{align}
    \min_{\vx} \sum_{m=1}^M \left(y_m - \abs{\inprod{\va_m, \vx}}^2 \right)^2 , \tag{LS-PR} \label{eq:quadratic_LS}
\end{align}
which we can alternatively rewrite as
\begin{align}
    \min_{\substack{\vx, \\ r_1, \ldots, r_M}} \quad & \sum_{m=1}^M r_m^2 \label{eq:quadratic_OLS} \\
    \text{s.t.} \quad & y_m + r_m = \abs{\inprod{\va_m, \vx}}^2, \quad m = 1, \ldots, M \notag 
\end{align}
with $r_m \in \R$. Thus, LS seeks the smallest correction to the measurements so that $(y_m + r_m)$ can be obtained from quadratic measurements $\abs{\inprod{\va_m, \vx}}^2$ for each $m$. This is analogous to LS for linear inverse problems where corrections that bring the measurements into the range space of the linear operator are required instead.

In many practical settings, the sensing vectors, $\{\va_m\}_{m=1}^M$, are only approximately known via calibration.
In this work we show that properly accounting for errors in the sensing vectors may lead to a more accurate estimate of $\vx$. Inspired by the total least squares (TLS) framework for linear \cite{golub1980analysis, markovsky2007overview} and nonlinear \cite{boggs1987stable} inverse problems, we extend the LS formulation \eqref{eq:quadratic_OLS} to find corrections for both the measurements and the sensing vectors. In TLS phase retrieval, we optimize the objective
\begin{align}
    \min_{\substack{\vx,\\ r_1, \ldots, r_M,\\ \ve_1, \ldots, \ve_M}} \quad & \sum_{m=1}^M \lambda_{y} r_m^2 + \lambda_{a} \norm{\ve_m}_2^2 \label{eq:quadratic_TLS} \\
    \text{s.t.} \quad & y_m + r_m = \abs{\inprod{\va_m + \ve_m, \vx}}^2, \quad m = 1, \ldots, M \notag 
\end{align}
with corrections $\ve_m \in \C^{N}$ for $ 1\leq m \leq M$. Scalars $\lambda_{y} \in \R$ and $\lambda_{a} \in \R$ are nonnegative regularization weights. Now for each $m$ we want to find minimum weighted norm corrections so that $(y_m + r_m)$ can be obtained from quadratic measurements $\abs{\inprod{\va_m + \ve_m, \vx}}^2$. Efficiently obtaining the sensing vector corrections $\{\ve_m\}_{m=1}^M$ is a major challenge when moving from the LS problem \eqref{eq:quadratic_OLS} to the TLS problem \eqref{eq:quadratic_TLS}.


\subsection{Related work} \label{sec:related_work}

Algorithms by Gerchberg and Saxton \cite{gerchberg1972practical} and Fienup \cite{fienup1978reconstruction} are the most well-known approaches for solving the phase retrieval problem when the sensing vectors are the rows of the Fourier matrix, as in many practical imaging scenarios \cite{blahut2004theory}. These methods iteratively reduce the error between the observed measurements and the measurements generated from the solution at the current iterate. Another class of algorithms based on message passing have also been developed \cite{rajaei2017robust, sharma2019inverse}. Despite the nonconvexity of the problem, these error reduction and message passing algorithms work well in practice. They do not directly use gradient descent to obtain a solution.


Recently a series of works have shown that for suitable measurement models, the nonconvex LS objective \eqref{eq:quadratic_LS} can be globally optimized via gradient descent updates. The Wirtinger flow algorithm is one of the most well-known methods and proposes the framework comprising a spectral initialization followed by gradient descent updates \cite{candes2015phase}. Spectral initialization ensures that the iterates start in a convex basin near a global optimum when there are enough measurements in an error-free setting. This initialization was first proposed as part of the AltMinPhase algorithm \cite{netrapalli2013phase}.
Multiple works have extended this approach by modifying the initialization, gradient updates and objective for phase retrieval \cite{wang2017solving, chen2015solving}, and other quadratic problems with sensing matrices rather than sensing vectors \cite{huang2020solving} like the unassigned distance geometry problem \cite{huang2018reconstructing}.
There are also extensions that incorporate signal priors such as sparsity \cite{cai2016optimal, wang2017sparse}. 
None of these gradient descent approaches, however, account for sensing vector or \rev{sensing} matrix errors.

Another group of works have developed convex optimization approaches, which are closely related to low-rank matrix recovery techniques, for solving the phase retrieval problem \cite{davenport2016overview}. These methods use the fact that the measurements in \eqref{eq:quadratic_inverse_problem} can be expressed using the Frobenius matrix inner product, $y_m \approx \abs{\inprod{\va_m, \vx}}^2 = \vx^* \va_m \va_m^* \vx = \inprod{\va_m \va_m^*, \vx \vx^*}$. With this formulation, phase retrieval amounts to recovering a rank-1 positive semidefinite matrix, $\mX = \vx\vx^*$, from linear matrix inner product measurements, $\{\inprod{\va_m \va_m^*, \vx \vx^*}\}_{m=1}^M$ \cite{candes2015phasematrixcompletion, candes2013phaselift}. 
In practice, lifting the problem from recovering vectors in $\C^N$ to matrices in $\C^{N \times N}$ poses significant computational and memory challenges for even moderately sized problems.
Matrix sketching algorithms \cite{yurtsever2017sketchy} and convex methods which do not require lifting \cite{goldstein2018phasemax} have since been developed to address these challenges.

For linear inverse problems, the TLS method is an established approach for handling errors in both the measurements and the operator \cite{golub1980analysis, markovsky2007overview}. For linear problems, TLS can be efficiently solved using the singular value decomposition (SVD). For the quadratic case considered in this paper, such an approach is not apparent because of the magnitude-squared nonlinearity in \eqref{eq:quadratic_inverse_problem}. We therefore also cannot use the SVD to analyze the solution error as is done in the linear case \cite{van1989accuracy}. Linear TLS has been extended to settings with structured operator errors \cite{markovsky2005application, malioutov2014convex}, sparse signals \cite{zhu2011sparsity}, and signals with norm constraints \cite{sima2004regularized}. We note that Yagle and Bell use linear TLS to solve a particular subproblem in a phase retrieval algorithm which only addresses errors in the measurements \cite{yagle1999one}.

There also exist algorithms for nonlinear TLS which aim to solve a general optimization problem for inverse problems with arbitrary nonlinearities \cite{boggs1987stable, schwetlick1985numerical, powell1972rapidly}. The general optimization problem is similar to \eqref{eq:quadratic_TLS} except for the constraint which requires nonlinear rather than quadratic consistency. 
However, by using the specific structure of the phase retrieval problem \eqref{eq:quadratic_inverse_problem} we are able to obtain efficient algorithms and perform error analysis for TLS phase retrieval.

Our gradient descent strategy uses alternating updates to solve the TLS phase retrieval problem. While alternating updates have been successfully utilized to  solve the linear TLS problem \cite{zhu2011sparsity}, it is not straightforward to extend this approach to phase retrieval because of its quadratic nature. We show how to use the geometry of the optimization problem \eqref{eq:quadratic_TLS} to perform alternating updates for TLS phase retrieval.


\subsection{Contributions and paper organization}

We propose a TLS framework for solving the phase retrieval problem when there are errors in the sensing vectors. In Section \ref{sec:tls_pr} we explain our gradient descent strategy to solve the TLS phase retrieval problem which motivates an alternating updates procedure to solve the problem. With this approach there are additional computational challenges which we show can be made efficient by incorporating the geometry of the phase retrieval problem. In Section \ref{sec:linearization} we derive expressions for the reconstruction errors for the TLS and LS solutions. This gives us insight into when each method should perform well. This derivation requires the usage of theorems about differentiation of argmins and different matrix inversion lemmas. Through simulations in Section \ref{sec:simulations} we show that the TLS approach can lead to solutions of greater accuracy when there are sensing vector errors. We further verify the applicability of our framework through experiments on real optical hardware in Section \ref{sec:opu}. We see that TLS outperforms LS when aiming to recover random signals and real images. We conclude and motivate future work in Section \ref{sec:conclusion}.



\section{TLS for phase retrieval} \label{sec:tls_pr}

In this section we show how to solve the TLS phase retrieval problem. Recall \eqref{eq:quadratic_TLS},
\begin{align}
    \min_{\substack{\vx,\\ r_1, \ldots, r_M,\\ \ve_1, \ldots, \ve_M}} \quad & 
    \mfrac{1}{2M} \sum_{m=1}^M \lambda_{y} r_m^2 + \lambda_{a} \norm{\ve_m}_2^2  ,
    \\
    \text{s.t.} \quad & y_m + r_m = \abs{\inprod{\va_m + \ve_m, \vx}}^2, \quad m = 1, \ldots, M, \notag 
\end{align}
which has been normalized by the number of measurements by the scaling $\frac{1}{M}$. We can rearrange the constraint and substitute $r_m = \abs{\inprod{\va_m + \ve_m, \vx}}^2 - y_m$ for $1 \leq m \leq M$ to obtain,
\begin{align}
    \min_{\vx} \mfrac{1}{2M} \sum_{m = 1}^M \min_{\ve_m} \lambda_{a} \norm{\ve_m}_2^2 + \lambda_{y} \left(y_m -  \abs{\inprod{\va_m + \ve_m,\vx}}^2 \right)^2 .
\end{align}


Further we denote the $m$th corrected sensing vector as $\wh{\va}_m := (\va_m + \ve_m)$ to obtain the equivalent formulations
\begin{align}
    \min_{\vx} \mfrac{1}{2M} \sum_{m = 1}^M 
    \underbrace{
    \min_{\wh{\va}_m} \lambda_{a} \norm{\va_m - \wh{\va}_m}_2^2 + \lambda_{y} \left(y_m -  \abs{\inprod{\wh{\va}_m, \vx}}^2 \right)^2}_{\calI_m(\vx)} , \tag{TLS-PR1} \label{eq:quadratic_TLS_geometric}
\end{align}
and
\begin{align}
    \min_{\substack{\vx,\\ \wh{\va}_1, \ldots, \wh{\va}_M}} 
    \underbrace{
    \mfrac{1}{2M} \sum_{m = 1}^M  \lambda_{a} \norm{\va_m - \wh{\va}_m}_2^2 + \lambda_{y} \left(y_m -  \abs{\inprod{\wh{\va}_m, \vx}}^2 \right)^2}_{\calJ(\vx, \wh{\va}_1, \ldots, \wh{\va}_M)} . \tag{TLS-PR2} \label{eq:quadratic_TLS_geometric_2} 
\end{align}

As each data consistency term, $\left(y_m -  \abs{\inprod{\wh{\va}_m, \vx}}^2 \right)^2$, is proportional to $\norm{\vx}_2^4$ in an error-free setting, we set $\lambda_{y} = \frac{\lambda_{y}^\dag}{\norm{\vx^{(0)}}_2^4}$ in order to make the scaling of the objective invariant with respect to the norm of $\vx$. The vector $\vx^{(0)}$ is an initial guess for $\vx$ and $\lambda_{y}^\dag$ is a regularization parameter.
Furthermore, to account for the fact that the sensing vector corrections, $(\va_m - \wh{\va}_m)$, are $N$-dimensional and the data consistency terms are scalar we set $\lambda_{a} = \frac{\lambda_{a}^\dag}{N}$ where $\lambda_{a}^\dag$ is a regularization parameter.

In line with recent methods such as the Wirtinger flow algorithm \cite{candes2015phase}, our high level strategy is to obtain $\vx$ by solving
\begin{align}
    \argmin_{\vx} \mfrac{1}{2M} \sum_{m=1}^M \calI_m(\vx) ,
\end{align}
using gradient descent. To perform gradient descent with respect to $\vx$ we can use Wirtinger gradient updates \cite{candes2015phase},
\begin{align}
    \vx^{(\tau + 1)} &= \vx^{(\tau)} - \mfrac{\mu}{\norm{\vx^{(0)}}_2^2} \cdot
    \mfrac{1}{2M} \sum_{m=1}^M \nabla_{\vx} \calI_m\left(\vx^{(\tau)}\right) , \label{eq:gradient_update}
\end{align}
where $\mu$ is the step size and $\norm{\vx^{(0)}}_2$ is a guess for $\norm{\vx}_2$. The gradient is given by
\begin{align}
    \nabla_{\vx} \calI_m(\vx) = 2 \left( \abs{\inprod{\wh{\va}_m^\dag, \vx}}^2 - y_m \right) \wh{\va}_m^\dag \wh{\va}_m^{\dag *} \vx ,
\end{align}
where $\wh{\va}_m^\dag$ is the solution to the following nonconvex optimization problem
\begin{align}
    \wh{\va}_m^\dag = \argmin_{\va} \lambda_{a} \norm{\va_m - \va}_2^2 + \lambda_{y} \left(y_m -  \abs{\inprod{\va, \vx}}^2 \right)^2 .
    \label{eq:a_argmin}
\end{align}

This motivates the following alternating updates procedure to solve the TLS problem:
\begin{enumerate}
    \item Obtain an initial guess, $\vx^{(0)} \in \C^N$, for $\vx$. \label{strategy:initialization}
    \item Repeat steps \ref{strategy:a_update} and \ref{strategy:x_update} until convergence:
    \begin{enumerate}
        \item With $\vx$ fixed, obtain corrected sensing vectors, $\{\wh{\va}_m^\dag\}_{m=1}^M$, by solving \eqref{eq:a_argmin} for $1 \leq m \leq M$. \label{strategy:a_update}
        \item With $\{\wh{\va}_m^\dag\}_{m=1}^M$ fixed, take one gradient descent step to update $\vx$ \rev{using} \eqref{eq:gradient_update}. \label{strategy:x_update}
    \end{enumerate}
\end{enumerate}

The main challenge in our approach is obtaining corrected sensing vectors $\{\wh{\va}_m^\dag\}_{m=1}^M$ by solving \eqref{eq:a_argmin} so that we can perform gradient descent updates for $\vx$ \rev{using} \eqref{eq:gradient_update}. As \eqref{eq:quadratic_TLS_geometric_2} is nonconvex, a good initial guess\rev{, $\vx^{(0)}$,} can place us near a global minimum. There are multiple initialization options such as the spectral initialization for certain measurement models \cite{netrapalli2013phase}.

In the remainder of this section we will examine the geometry of the optimization problem in \eqref{eq:quadratic_TLS_geometric} and show how it can be leveraged to efficiently solve \eqref{eq:a_argmin} and obtain corrected sensing vectors. This is summarized by Proposition \ref{proposition:cubic} below. We will then present the complete TLS phase retrieval algorithm. Lastly, we also interpret the regularization parameters, $\lambda_{a}$ and $\lambda_{y}$, by showing that the TLS solution is the maximum likelihood estimator for a quadratic complex-valued error-in-variables (EIV) model.



\subsection{Optimization geometry}
\label{sec:geometry}

Moving from the LS formulation to the TLS formulation introduces significant computational issues. In addition to optimizing over \rev{vector} $\vx$, we must additionally optimize over $M$ sensing vectors in \eqref{eq:quadratic_TLS_geometric}, with typically $M > N$. We now study the optimization geometry of \eqref{eq:quadratic_TLS_geometric} and show that the $M$ inner minimizations over the $N$-dimensional vectors, $\{\wh{\va}_m\}_{m=1}^M$, can be simplified to minimizing over $M$ scalars which improves efficiency. For ease of visualization in this subsection, we consider the real-valued problem (all quantities in \eqref{eq:quadratic_inverse_problem}, \eqref{eq:quadratic_LS} and \eqref{eq:quadratic_TLS_geometric} are real) and we set $\lambda_{a} = \lambda_{y} = 1$. 

\begin{figure*}[t]
    \centering
    
    \subfloat{\includegraphics[width=0.3\linewidth]
        {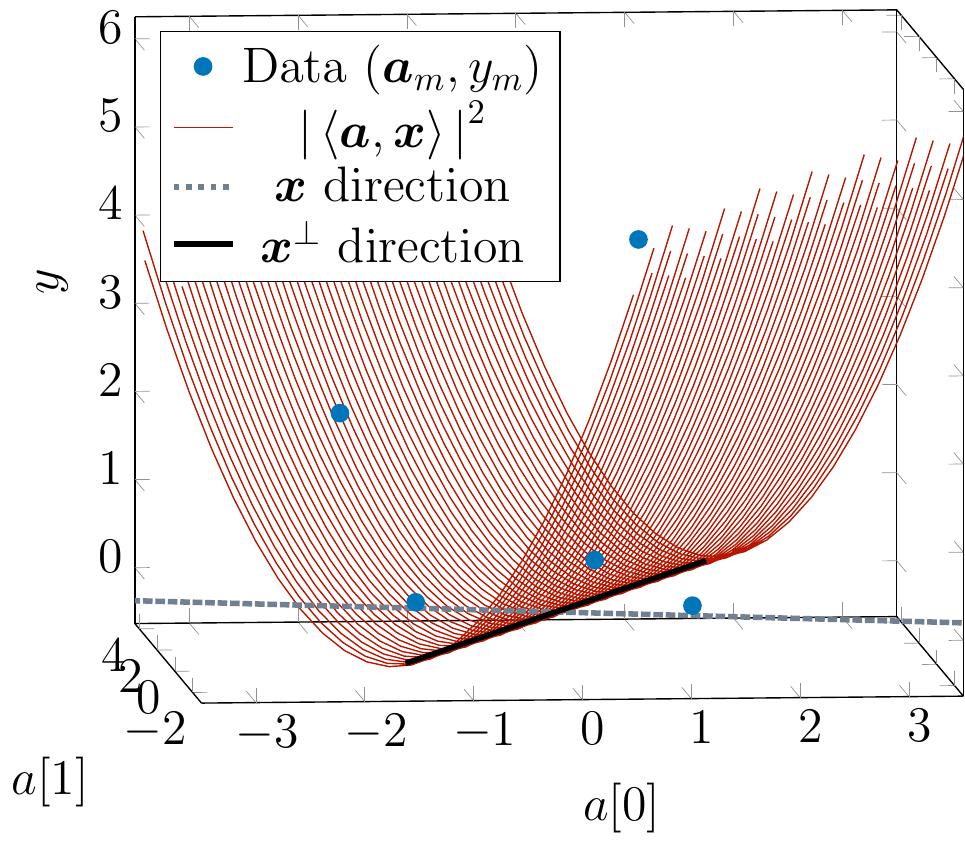}}
    \hfill
    \subfloat{\includegraphics[width=0.29\linewidth]
        {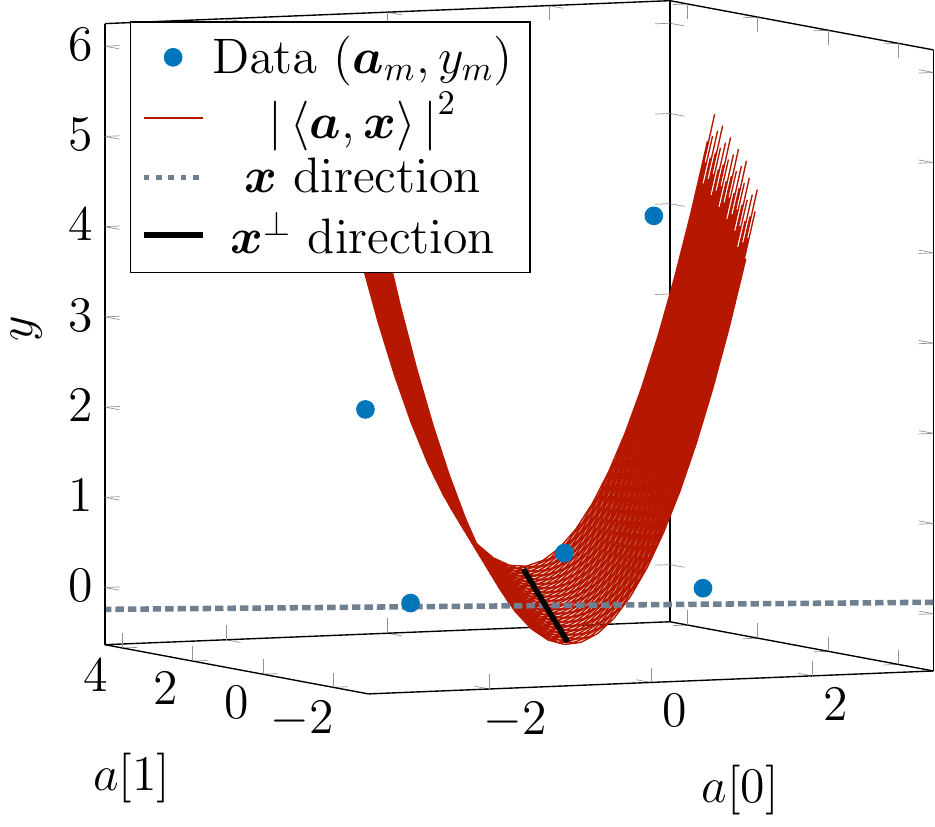}}
    \hfill
    \subfloat{\includegraphics[width=0.23\linewidth]
        {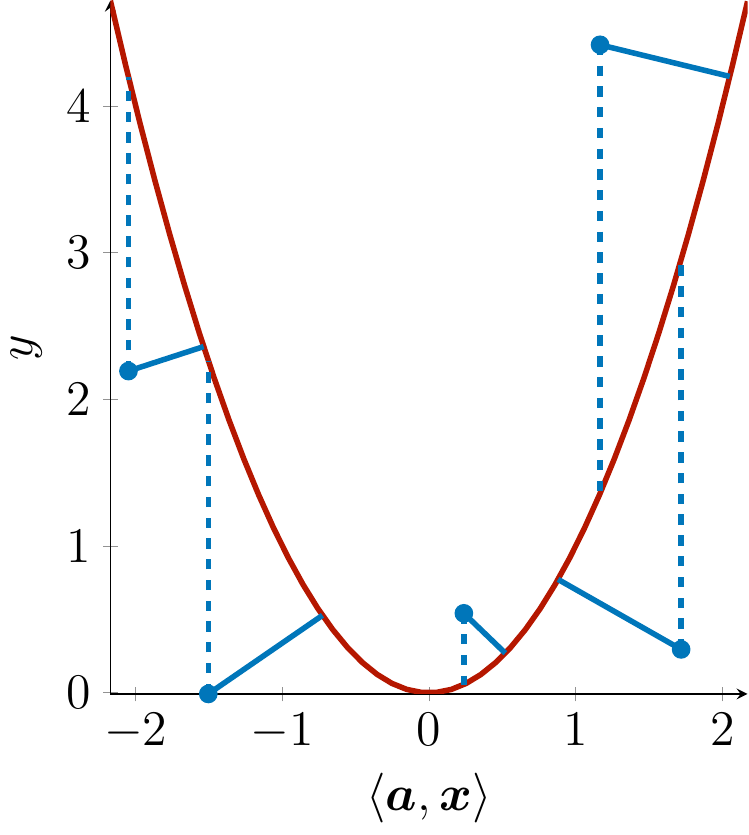}}
    
    \caption{Visualization of the phase retrieval problem when $\norm{\vx}_2 = 1$. The left column shows $M = 5$ data points, $\{(\va_m, y_m)\}_{m=1}^M$, when $N=2$. A paraboloid is fitted to the data points. The middle column shows the same paraboloid and data points from a different viewing angle. The right column shows the problem from a viewpoint that looks into the $\vx^\perp$ subspace. The dashed lines \rev{show} the distances minimized by the LS objective, \eqref{eq:quadratic_LS}. The solid lines show the distances minimized \rev{by the} TLS objective, \eqref{eq:quadratic_TLS_geometric}.}
    \label{fig:tls_geometry}
\end{figure*}

For a given vector $\vx$ we compare the values of the LS and TLS objectives, \eqref{eq:quadratic_LS} and \eqref{eq:quadratic_TLS_geometric}. The left column of Fig. \ref{fig:tls_geometry} visualizes the phase retrieval problem with $M = 5$ data points, $\{(\va_m, y_m)\}_{m=1}^M$, when $N=2$ and $\norm{\vx}_2=1$. The middle column shows the same data points from a different viewing angle. In phase retrieval we fit a paraboloid, $y(\va) = \abs{\inprod{\va, \vx}}^2$ that is parameterized by $\vx$ to the data points, $\{(\va_m, y_m)\}_{m=1}^M$. If there is no sensing vector or measurement error, the data points lie on the paraboloid (\eqref{eq:quadratic_inverse_problem} holds with equality). The left and middle figure show that the surface $y(\va) = \abs{\inprod{\va, \vx}}^2$ does not change in the subspace perpendicular to $\vx$, denoted as $\vx^\perp$. This can also be verified by considering the values of $\va$ that would result in the inner product $\inprod{\va, \vx}$ being zero. Crucially this means that the shortest paths between the data points and the paraboloid have no component in the $\vx^\perp$ subspace. As a result, we can view the problem in 2D from a viewpoint that looks into the $\vx^\perp$ subspace as shown in the right column of Fig. \ref{fig:tls_geometry}. This 2D plot shows two options for measuring closeness between the surface and the data points. The LS objective \eqref{eq:quadratic_LS}, is the sum of the squared vertical distance between the 2D parabola and each data point as indicated by the dashed lines. On the other hand, due to the minima over all $\wh{\va}_m$, the TLS objective \eqref{eq:quadratic_TLS_geometric}, is the sum of the squared Euclidean or orthogonal distance between the 2D parabola and each data point as shown by the solid lines. A similar geometrical interpretation is seen with linear TLS \cite{golub1980analysis, markovsky2007overview}.

Considering this geometry, to solve the inner minimizations in \eqref{eq:quadratic_TLS_geometric}, we find the closest point on the paraboloid to each data point. As the shortest path has no component in the $\vx^\perp$ subspace, our task of finding the closest point on a $(N+1)$-dimensional paraboloid to a $(N+1)$-dimensional data point reduces to a 2D geometry problem of finding the closest point on a parabola to a 2D data point. Rather than finding the minimizing $N$-dimensional $\wh{\va}_m^\dag$ for each data point, we instead only need to find the component of $\wh{\va}_m^\dag$ in the $\vx$ direction that is closest. This component is a scalar and is given by the inner product, $\nu_m = \inprod{\wh{\va}_m^\dag, \vx}$. We can then construct $\wh{\va}_m^\dag$ by adding the unchanged component in the $\vx^\perp$ subspace,
\begin{align}
    \wh{\va}_m^\dag := \wh{\va}_m^\dag(\nu_m) = \frac{\nu_m}{\norm{\vx}_2}\wh{\vx} + (\va_m - \inprod{\va_m, \wh{\vx}} \wh{\vx}), \label{eq:reconstruct_a}
\end{align}
where $\wh{\vx}$ is $\vx$ normalized.

If $\lambda_{a}$ and $\lambda_{y}$ are not one, 
a perpendicular distance is not minimized. As $\frac{\lambda_{a}}{\lambda_{y}}$ gets larger, the solid lines in the right column of Fig. \ref{fig:tls_geometry} become more vertical because there is a relatively larger penalty for correcting the sensing vectors and the problem moves towards a LS approach. Conversely, the lines become more horizontal as $\frac{\lambda_{a}}{\lambda_{y}}$ gets smaller. Irrespective of the values of $\lambda_{a}$ and $\lambda_{y}$, the shortest paths between the paraboloid and the data points still have no component in the $\vx^\perp$ subspace and \eqref{eq:reconstruct_a} can be used to obtain each $\wh{\va}_m^\dag$. We further note that this geometry also holds for the complex-valued phase retrieval problem \eqref{eq:quadratic_inverse_problem}.

\subsection{Correcting complex-valued sensing vectors} \label{sec:measurement_vector_update}


Our strategy is to set up each inner minimization over $\wh{\va}_m$ in \eqref{eq:quadratic_TLS_geometric} as the minimization of a fourth degree equation with respect to scalar $\nu_m = \inprod{\wh{\va}_m, \vx}$ rather than vector $\wh{\va}_m$.
We then directly obtain the minimizer of this equation.

The $M$ inner minimization problems in \eqref{eq:quadratic_TLS_geometric} are independent of each other and we can independently solve each summand for a fixed vector $\vx$. Consider the objective function of optimization problem $\calI_m(\vx)$,
\begin{align}
    f_m(\wh{\va}_m) 
    =& \lambda_{a}\norm{\va_m - \wh{\va}_m}_2^2 + \lambda_{y}\left(y_m -  \abs{\vx^*\wh{\va}_m}^2 \right)^2 .
\end{align}
Proposition \ref{proposition:cubic} states that $\argmin_{\wh{\va}_m} f_m(\wh{\va}_m)$ can be obtained by solving two scalar variable cubic equations and using \eqref{eq:reconstruct_a}.

\begin{proposition} \label{proposition:cubic}

Let sets $R_+$ and $R_-$ be the positive real solutions of
\begin{align}
    \alpha r^3 + \beta r \pm \abs{\gamma} = 0 
\end{align}
where  $\alpha = 2 \lambda_{y} \norm{\vx}_2^2 \in \R$, $\beta = \lambda_{a} - 2 \lambda_{y} y_m \norm{\vx}_2^2 \in \R$ and $\gamma = - \lambda_{a} \vx^*\va_m \in \C$. Further, with $\kappa$ denoting the phase of $\gamma$, let
\begin{align}
    S_+ = \{ \e^{j\kappa} r \,|\, r \in R_+\} 
    \,\,\,
    \text{and}
    \,\,\,
    S_- = \{ - \e^{j\kappa} r \,|\, r \in R_-\}. 
\end{align}
Then
$f_m(\wh{\va}_m)$ is minimized by $\wh{\va}_m^\dag \left(s^\dag \right)$ where 
\begin{align}
    s^\dag
    =
    \argmin_{s \in S_+ \cup \, S_-} f_m \left( \wh{\va}_m^\dag (s) \right)
\end{align}
and $\wh{\va}_m^\dag(\cdot)$ is defined in \eqref{eq:reconstruct_a}.
\end{proposition}

\begin{IEEEproof}
Expanding $f_m(\wh{\va}_m)$ gives
\begin{align}
    f_m(\wh{\va}_m) 
    =& \lambda_{a}(\norm{\va_m}_2^2 -\va_m^* \wh{\va}_m - \wh{\va}_m^* \va_m + \wh{\va}_m^* \wh{\va}_m) \notag \\
    &+ \lambda_{y}(y_m^2 -2y_m \wh{\va}_m^*\vx\vx^*\wh{\va}_m + (\wh{\va}_m^*\vx\vx^*\wh{\va}_m)^2) .
\end{align}

We can use Wirtinger derivatives to calculate the derivative of \rev{the} real-valued $f_m(\wh{\va}_m)$ with respect to \rev{the} complex vector $\wh{\va}_m$ \cite{candes2015phase},
\begin{align}
    \nabla_{\wh{\va}_m}f_m =& \left(\vphantom{\abs{\vx^* \wh{\va}_m}^2} \lambda_{a} (-\va_m^* + \wh{\va}_m^*) \right. \notag \\
    &\left. + \lambda_{y}( - 2y_m \wh{\va}_m^* \vx \vx^* + 2 \abs{\vx^* \wh{\va}_m}^2 \wh{\va}_m^* \vx \vx^* )\right)^* \notag \\
    =& \lambda_{a} (\wh{\va}_m - \va_m) \notag \\
    & + \lambda_{y}(- 2y_m \vx \vx^* \wh{\va}_m + 2 \abs{\vx^* \wh{\va}_m}^2 \vx \vx^* \wh{\va}_m) .
\end{align}

Setting the derivative to zero, and then left-multiplying by nonzero $\vx^*$ gives
\begin{align}
    2 \lambda_{y} \norm{\vx}_2^2 \abs{\vx^* \wh{\va}_m}^2 (\vx^* \wh{\va}_m) &- 2 \lambda_{y} y_m \norm{\vx}_2^2 (\vx^* \wh{\va}_m) \notag \\
    &+ \lambda_{a}(\vx^*\wh{\va}_m) - \lambda_{a}\vx^*\va_m = 0 \notag \\
    2 \lambda_{y} \norm{\vx}_2^2 \abs{\vx^* \wh{\va}_m}^2 (\vx^* \wh{\va}_m) &+ (\lambda_{a} - 2 \lambda_{y} y_m \norm{\vx}_2^2) (\vx^* \wh{\va}_m) \notag \\
    &- \lambda_{a}\vx^*\va_m = 0.
\end{align}
The left hand side is now scalar-valued and is a function of scalar $\nu_m = \inprod{\wh{\va}_m, \vx} = \vx^*\wh{\va}_m \in \C$ instead of a vector. Recalling our analysis of the optimization geometry in Section \ref{sec:geometry}, we can solve for $\nu_m$ and then obtain $\wh{\va}_m^\dag = \wh{\va}_m^\dag(\nu_m)$ using \eqref{eq:reconstruct_a}. If we substitute $\alpha = 2 \lambda_{y} \norm{\vx}_2^2 \in \R$, $\beta = \lambda_{a} - 2 \lambda_{y} y_m \norm{\vx}_2^2 \in \R$ and $\gamma = - \lambda_{a} \vx^*\va_m \in \C$ we wish to solve the following for $\nu_m$,
\begin{align}
    \alpha \abs{\nu_m}^2 \nu_m + \beta \nu_m + \gamma = 0.
\end{align}

Because the sensing vectors and ground truth signal are complex, this cubic equation is a function of $\nu_m \in \C$ and its conjugate $\bar{\nu}_m$ ($\abs{\nu_m}^2 = \nu_m \bar{\nu}_m$). We therefore cannot use standard cubic root finding formulae. Further note that the coefficients $\alpha$ and $\beta$ are always real and $\gamma$ may be complex. To solve, first multiply by $\bar{\nu}_m$,
\begin{align}
    \alpha \abs{\nu_m}^4 + \beta \abs{\nu_m}^2 + \gamma \bar{\nu}_m = 0.
\end{align}

Next, with complex-exponential representation, $\nu_m = r\e^{j\phi}$ and $\gamma = \abs{\gamma}\e^{j\kappa}$ (recall $\gamma$ is known), the equation becomes
\begin{align}
    \alpha r^3 + \beta r + \abs{\gamma} \e^{j(\kappa - \phi)} &= 0 .
\end{align}

The real and imaginary parts of the left hand side should both equate to zero. Using Euler's identity, $\e^{j\theta} = \cos(\theta) + j\sin(\theta)$, we arrive at the following simultaneous equations,
\begin{equation}
\begin{cases}
    \sin(\kappa - \phi) = 0 \\
    \alpha r^3 + \beta r + \abs{\gamma} \cos(\kappa - \phi) = 0 .
\end{cases}
\end{equation}
For the first equation to hold, $\cos(\kappa - \phi) = \pm 1$ and so the phase of $\nu_m$ has two possible values; $\phi = \kappa$ or $\phi = (\kappa - \pi)$. To obtain the magnitude of $\nu_m$ we can solve the following two cubic equations for $r$ to get six values, three from each,
\begin{align}
    \alpha r^3 + \beta r + \abs{\gamma} &= 0 \quad \text{and} \quad \phi = \kappa 
    \label{eq:phi_equal_kappa_proof} \\
    \alpha r^3 + \beta r - \abs{\gamma} &= 0 \quad \text{and} \quad \phi = \kappa - \pi. 
    \label{eq:phi_equal_kappa_minus_pi_proof}
\end{align}

As the solutions of these two cubic equations are magnitudes of complex numbers, we let sets $R_+$ and $R_-$ be the positive real solutions of \eqref{eq:phi_equal_kappa_proof} and \eqref{eq:phi_equal_kappa_minus_pi_proof} respectively. To obtain values for $\nu_m$ we combine $R_+$ and $R_-$ with their phases to get $S_+$ and $S_-$---multiply the elements of $R_+$ by $\e^{j\kappa}$ and multiply the elements of $R_-$ by $\e^{j(\kappa - \pi)} = - \e^{j \kappa}$. We then construct candidate minimizers of $f_m(\cdot)$ by using the possible values for $\nu_m$, the set, $S_+ \cup \, S_-$,  as the argument for \eqref{eq:reconstruct_a}. Finally, the global minimizer is the candidate minimizer that gives the minimum value as the argument of $f_m(\cdot)$.
\end{IEEEproof}

To solve \eqref{eq:phi_equal_kappa_proof} and \eqref{eq:phi_equal_kappa_minus_pi_proof} for $r$, Cardano's formula for cubic equations or a general cubic root formula derived from Cardano's formula can be used (see Appendix \ref{sec:cubic_roots}). Furthermore we note that the procedure to update the sensing vectors is independent of the sensing vector measurement model.


\subsection{TLS phase retrieval algorithm}

Now that we have a method for solving the inner minimizations in \eqref{eq:quadratic_TLS_geometric}, we present the complete TLS phase retrieval algorithm in Algorithm \ref{algo:tls_pr_algorithm}. We say that the algorithm has converged if the value of $\calJ(\vx, \wh{\va}_1^\dag, \ldots, \wh{\va}_M^\dag)$ in \eqref{eq:quadratic_TLS_geometric_2} between consecutive iterates is less than some threshold. In practice all sensing vectors can be updated (lines \ref{algo:start_vector_update}-\ref{algo:end_vector_update}) in parallel for a given $\vx$ because all sensing vectors are independent of each other.

\begin{algorithm}[t]
\caption{TLS phase retrieval.}
\begin{algorithmic}[1]
\Require Erroneous sensing vectors $\{\va_m\}_{m=1}^M$; Erroneous observations $\{y_m\}_{m=1}^M$; Convergence threshold $T$; Step size $\eta$; Regularization parameters $\lambda_{y}$ and $\lambda_{a}$.
\Ensure Recovered signal $\vx \in \C^N$.

\State $\vx \gets$ \texttt{Initialization}$(y_1, \ldots, y_M, \va_1, \ldots, \va_M)$ \label{algo:initialization}

\State \texttt{loss\_previous} $\gets -\infty$
\State \texttt{loss\_current} $\gets \infty$
\While {$|$\texttt{loss\_current} - \texttt{loss\_previous}$|> T$}

// Update each sensing vector for a given $\vx$
\ForEach{$m \in \{1, \ldots, M \}$} \label{algo:start_vector_update}
\State $\alpha \gets 2 \lambda_{y} \norm{\vx}_2^2$
\State $\beta \gets \lambda_{a} - 2 \lambda_{y} y_m \norm{\vx}_2^2$
\State $\gamma \gets - \lambda_{a} \vx^*\va_m$
\State $\kappa \gets \texttt{Angle}(\gamma)$

\State $R_+ \gets$ \texttt{PositiveRealRoots}$\left(\alpha r^3 + \beta r + \abs{\gamma} \right)$
\State $R_- \gets$ \texttt{PositiveRealRoots}$\left(\alpha r^3 + \beta r - \abs{\gamma} \right)$

\State $S_+ \gets \e^{j\kappa} \cdot R_+$
\State $S_- \gets - \, \e^{j\kappa} \cdot R_-$





\State $s^\dag = \argmin_{s \in S_+ \cup \, S_-} f_m \left( \wh{\va}_m^\dag (s) \right)$

\State $\wh{\va}_m^\dag \gets \wh{\va}_m^\dag \left(s^\dag \right)$





    

\EndFor \label{algo:end_vector_update}

// Update $\vx$ with sensing vectors fixed
\State $\vx  \gets$ \texttt{x\_gradient\_step}$(\vx, \wh{\va}_1^\dag, \ldots, \wh{\va}_M^\dag)$ \label{algo:gradient_step}
\State \texttt{loss\_previous} $\gets$ \texttt{loss\_current}
\State \texttt{loss\_current} $\gets  \calJ(\vx, \wh{\va}_1^\dag, \ldots, \wh{\va}_M^\dag)$
\EndWhile
\end{algorithmic}
\label{algo:tls_pr_algorithm}
\end{algorithm}

\subsection{ML estimator for EIV models}

Proposition \ref{proposition:ml_estimator} below provides an interpretation of the regularization parameters in \eqref{eq:quadratic_TLS_geometric} by connecting them to the error level. It states that under certain assumptions the solution to \eqref{eq:quadratic_TLS_geometric} is the maximum likelihood (ML) estimator for the complex-valued EIV model given by
\begin{align}
    y_m = \abs{\inprod{\wt{\va}_m, \wt{\vx}}}^2 + (-\eta_m), \quad \va_m = \wt{\va}_m + (-\vdelta_m) \label{eq:quadratic_EIV}
\end{align}
for $1 \leq m \leq M$. With this EIV model we aim to recover $\wt{\vx}$ and $\{\wt{\va}_m\}_{m=1}^M$ from $\{y_m\}_{m=1}^M$ and $\{\va_m\}_{m=1}^M$ which are known. The quantities $\{\eta_m\}_{m=1}^M$ and $\{\vdelta_m\}_{m=1}^M$ are random error perturbations. This result is an extension of the relationship between linear TLS and the linear error-in-variables model \cite{markovsky2005application, wiesel2006maximum}. Similarly, this result is a specific instance of what is seen for nonlinear TLS \cite{boggs1987stable}.

\begin{proposition} \label{proposition:ml_estimator}
Assume in \eqref{eq:quadratic_EIV} that $\{\eta_m\}_{m=1}^M$ are iid zero-mean Gaussian with covariance $\sigma_{\eta}^2 \mI$, $\{\vdelta_m\}_{m=1}^M$ are independent of each other and each is an iid zero-mean complex Gaussian vector with covariance $2\sigma_{\vdelta}^2$, i.e. $\mathrm{vec}([\mathrm{Re}(\vdelta_m) \, | \, \mathrm{Im}(\vdelta_m)]) \sim \calN(\vzero, \sigma_{\vdelta}^2 \mI)$. Further assume that $\{\eta_m\}_{m=1}^M$ and $\{\vdelta_m\}_{m=1}^M$ are independent of each other and that $\{\wt{\va}_m\}_{m=1}^M$ and $\wt{\vx}$ are deterministic. Under these assumptions, the solution to optimization problem \eqref{eq:quadratic_TLS_geometric}, when $\lambda_{a} = \frac{1}{\sigma_{\vdelta}^2}$ and $\lambda_{y} = \frac{1}{\sigma_{\eta}^2}$, is the maximum likelihood estimator for \eqref{eq:quadratic_EIV}.
\end{proposition}

\begin{IEEEproof}
The proof follows a standard procedure and is provided in Appendix \ref{sec:ml_estimator_proof_appendix}.
\end{IEEEproof}

\section{TLS and LS solution reconstruction errors} \label{sec:linearization}

In this section we evaluate the reconstruction error for the TLS and LS phase retrieval solutions by deriving their Taylor expansions. Through these expressions we are able to gain insight into the behavior of the TLS solution relative to the LS solution and understand when each method performs well.
\rev{We also use these expressions to understand how the reconstruction errors rely on the level of the measurement and the sensing vector errors when all the errors are Gaussian.}
Since this analysis is cumbersome, in this section we will consider the real-valued phase retrieval problem where \rev{the ground truth signal}, the sensing vectors and the sensing vector errors in \eqref{eq:quadratic_inverse_problem} are real. Simulations in Section \ref{sec:simulations} show that the reasoning carries through to the complex problem. In our derivations we will use theorems about differentiation of argmins and various matrix inversion lemmas.

We denote the ground truth signal as $\vx^\#$ and the TLS and LS solutions as $\vx^\dag_{\mathrm{TLS}}$ and $\vx^\dag_{\mathrm{LS}}$. If there are no errors in the sensing vectors or measurements, $\vx^\#$ and $-\vx^\#$ are both optimum LS and TLS solutions for \eqref{eq:quadratic_LS} and \eqref{eq:quadratic_TLS_geometric_2} (with the $m$th corrected sensing vector being $\va_m$). Due to this inherent sign ambiguity it is standard to define the reconstruction errors as
\begin{align}
    \min_{\sigma} \norm{\vx^\# - \sigma \cdot \vx^\dag_{\mathrm{TLS}}}_2
    \,\,\, \mathrm{and} \,\,\,
    \min_{\sigma} \norm{\vx^\# - \sigma \cdot \vx^\dag_{\mathrm{LS}}}_2 \label{eq:errors}
\end{align}
where $\sigma \in \{1 , -1 \}$. Our results are unchanged if the analysis is done with optimum solution $\vx^\#$ ($\sigma=1$) or with optimum solution $-\vx^\#$ ($\sigma=-1$). Consequently, we choose optimum solution $\vx^\#$ with $\sigma = 1$ in the following analysis.




\subsection{Reconstruction error analysis}

The erroneous sensing vectors and measurements in \eqref{eq:quadratic_inverse_problem} can be expressed as perturbed versions of error-free sensing vectors and measurements, $\{\wt{\va}_m\}_{m=1}^M$ and $\{\wt{y}_m\}_{m=1}^M$. We denote the sensing vector and measurement error perturbations as $\{\vdelta_m\}_{m=1}^M$ and $\{\eta_m\}_{m=1}^M$. Stacking these into vectors we define,
\begin{align}
    \wt{\vt} &= \left[\wt{\va}_1^\T, \ldots , \wt{\va}_M^\T, \wt{y}_1, \ldots, \wt{y}_M \right]^\T \in \R^{(MN + M)}, \\
    \vgamma &= \left[\vdelta_1^\T, \ldots , \vdelta_M^\T, \eta_1, \ldots, \eta_M \right]^\T \in \R^{(MN + M)}, \label{eq:gamma} \\
    \vt &= \wt{\vt} + \vgamma \notag \\
    &= \left[\va_1^\T, \ldots , \va_M^\T, y_1, \ldots, y_M \right]^\T \in \R^{(MN + M)}.
\end{align}

In order to calculate the reconstruction errors we need access to expressions for $\vx^\dag_{\mathrm{TLS}}$ and $\vx^\dag_{\mathrm{LS}}$. We begin by noting that the solutions are functions of the sensing vectors and measurements, $\vx^\dag_{\mathrm{TLS}}(\vt)$ and $\vx^\dag_{\mathrm{LS}}(\vt)$. If there are no errors in the sensing vectors or measurements, an optimum LS solution for \eqref{eq:quadratic_LS} is $\vx^\dag_{\mathrm{LS}}(\wt{\vt}) = \vx^\#$. Similarly an optimum TLS solution in \eqref{eq:quadratic_TLS_geometric_2} for $\vx^\dag_{\mathrm{TLS}}(\wt{\vt}) = \vx^\#$ with the $m$th corrected sensing vector being $\va_m$ (no correction \rev{needed}). Now, if we instead have sensing vector and measurement errors, our solutions are $\vx^\dag_{\mathrm{TLS}}(\wt{\vt} + \vgamma)$ and $\vx^\dag_{\mathrm{LS}}(\wt{\vt} + \vgamma)$ which we can interpret as perturbed versions of $\vx^\dag_{\mathrm{LS}}(\wt{\vt}) = \vx^\dag_{\mathrm{TLS}}(\wt{\vt}) = \vx^\#$. Assuming $\norm{\vgamma}$ is small, we can study the first-order terms in the Taylor series expansions of $\vx^\dag_{\mathrm{TLS}}(\vt)$ and $\vx^\dag_{\mathrm{LS}}(\vt)$ to measure the perturbation from $\vx^\#$.

The Taylor series expansion of $\vx^\dag_{\mathrm{TLS}}(\vt) = \vx^\dag_{\mathrm{TLS}}(\wt{\vt} + \vgamma)$ at the no error point, $\wt{\vt}$, is
\begin{align}
    \vx^\dag_{\mathrm{TLS}}(\wt{\vt} + \vgamma) 
    &= 
    \vx^\dag_{\mathrm{TLS}}(\wt{\vt}) + \nabla_{\vt}\vx^\dag_{\mathrm{TLS}}(\vt)\big|_{\vt=\wt{\vt}}\, \vgamma + \mO(\norm{\vgamma}_2^2) \notag \\
    &= 
    \vx^\# + \nabla_{\vt}\vx^\dag_{\mathrm{TLS}}(\vt)\big|_{\vt=\wt{\vt}}\, \vgamma + \mO(\norm{\vgamma}_2^2)
    , \label{eq:general_taylor}
\end{align}
where $\mO(\norm{\vgamma}_2^2)$ represents terms with norm of order $\norm{\vgamma}_2^2$. The Taylor series expansion for $\vx^\dag_{\mathrm{LS}}(\vt)$ can be written similarly. Using these expansions, to the first-order when $\norm{\vgamma}$ is small, the reconstruction errors for the TLS and LS problems are
\begin{align}
    e_{\mathrm{TLS}} &:= \norm{\nabla_{\vt}\vx^\dag_{\mathrm{TLS}}(\vt)\big|_{\vt=\wt{\vt}}\, \vgamma}_2 \\
    e_{\mathrm{LS}} &:= \norm{\nabla_{\vt}\vx^\dag_{\mathrm{LS}}(\vt)\big|_{\vt=\wt{\vt}}\, \vgamma}_2
\end{align}

To evaluate $e_{\mathrm{TLS}}$ and $e_{\mathrm{LS}}$ we must calculate the derivatives $\nabla_{\vt}\vx^\dag_{\mathrm{TLS}}(\vt) \in \R^{N \times (MN + M)}$ and $\nabla_{\vt}\vx^\dag_{\mathrm{LS}}(\vt) \in \R^{N \times (MN + M)}$ which are the derivatives of the argmins of \eqref{eq:quadratic_TLS_geometric_2} and \eqref{eq:quadratic_LS}. We use the method by Gould et al. to take derivatives of argmin problems \cite{gould2016differentiating}.

With the substitution $\ve_m = \wh{\va}_m - \va_m$ and multiplicative constants absorbed into $\lambda_{a}$ and $\lambda_{y}$, the TLS optimization problem \eqref{eq:quadratic_TLS_geometric_2} can be rewritten as
\begin{align}
    \vq^\dag =& \argmin_{\vq}
    \underbrace{\sum_{m = 1}^M \lambda_{a}\norm{\ve_m}_2^2 + \lambda_{y}\left(y_m -  \abs{\inprod{\va_m + \ve_m, \vx}}^2 \right)^2}_{f(\vq, \vt)} \notag \\
    & \quad \text{s.t.} \quad \vq = \begin{bmatrix}\ve_1^\T
    & \cdots & \ve_M^\T & \vx^\T
    \end{bmatrix}^\T \in \R^{MN + N} . \label{eq:error_analysis_objective}
\end{align}

The solution, $g(\vt) := \vq^\dagger$, is a function of $\vt$ and $\vx^\dag_{\mathrm{TLS}}(\vt)$ is the last $N$ entries of $g(\vt)$, denoted as $g(\vt)_{-N}$,
\begin{align}
    & g(\vt) := \vq^\dag = \argmin_{\vq} f(\vq, \vt) \in \R^{MN +N}, \\
    & \vx^\dag_{\mathrm{TLS}}(\vt) = g(\vt)_{-N} \in \R^N. \label{eq:last_rows_for_x}
\end{align}

The derivatives of $g(\vt)$ with respect to the $k$th sensing vector and measurement can be computed after specific second derivatives of $f(\vq, \vt)$ are computed \cite{gould2016differentiating},
\begin{align}
    \nabla_{\va_k} g(\vt) &= - (\nabla^2_{\vq\vq} f(\vq, \vt))^{-1} (\nabla^2_{\va_k \vq} f(\vq, \vt)) \in \R^{(MN+N) \times N}. \label{eq:argmin_wrt_a} \\
    \frac{d}{d y_k}g(\vt) &= - (\nabla^2_{\vq\vq} f(\vq, \vt))^{-1} \left(\frac{d}{d y_k} \nabla_{\vq} f(\vq, \vt)\right) \in \R^{MN+N} .
    \label{eq:argmin_wrt_y}
\end{align}

We can then obtain $\nabla_{\vt}\vx^\dag_{\mathrm{TLS}}(\vt)$ by vertically stacking the derivatives \eqref{eq:argmin_wrt_a} and \eqref{eq:argmin_wrt_y} for $1 \leq k \leq M$ to form $\nabla_{\vt}g(\vt) \in \R^{(MN +N) \times (MN + M)}$,
\begin{align}
    \nabla_{\vt}g(\vt) = 
    \begin{bmatrix}
    \nabla_{\va_1} g(\vt), \cdots, \nabla_{\va_M} g(\vt),
    \frac{d}{d y_1}g(\vt), \cdots, \frac{d}{d y_M}g(\vt)
    \end{bmatrix} , \label{eq:stacked_derivatives}
\end{align}
and taking the last $N$ rows. Appendix \ref{sec:tls_gradients} contains the derivations for the last $N$ rows of \eqref{eq:argmin_wrt_a} and \eqref{eq:argmin_wrt_y}.

The same approach can be used for the LS problem by considering its optimization problem,
\begin{align}
    \vx^\dag_{\mathrm{LS}} &= \argmin_{\vx} \sum_{m = 1}^M \left(y_m -  \abs{\inprod{\va_m, \vx}}^2 \right)^2 . \label{eq:error_analysis_ls_objective}
\end{align}
The corresponding derivative derivations are in Appendix \ref{sec:ls_gradients}.

Proposition \ref{proposition:linearized_error} below states the expressions for $e_{\mathrm{TLS}}$ and $e_{\mathrm{LS}}$. We denote
\begin{align}
    \wt{\mY} &= \diag(\wt{y}_1, \ldots, \wt{y}_M) \in \R^{M \times M} \\
    \wt{\mA} &= \begin{bmatrix} \text{--- } \wt{\va}_1^\T \text{ ---} \\
    \vdots \\
    \text{--- } \wt{\va}_M^\T \text{ ---} \end{bmatrix} \in \R^{M \times N} \\
    \mE_{\mY} &= \diag(\eta_1, \ldots, \eta_M) \in \R^{M \times M} \\
    \mE_{\mA} &= \begin{bmatrix} \text{--- } \vdelta_1^\T \text{ ---} \\
    \vdots \\
    \text{--- } \vdelta_M^\T \text{ ---} \end{bmatrix} \in \R^{M \times N}
\end{align}
and use these quantities to define diagonal matrix, $\mD$, and vector, $\vw$,
\begin{align}
    \mD &= \left( \mI_M + 4 \frac{\lambda_{y}}{\lambda_{a}} \norm{\vx^\#}_2^2 \wt{\mY} \right)^{-1} \in \R^{M \times M} \\
    \vw &= \left( 
    (2 \wt{\mY})^{-1} \mE_{\mY} \wt{\mA} - \mE_{\mA}
    \right) \vx^\# \in \R^M . \label{eq:w_definition}
\end{align}

\begin{proposition} \label{proposition:linearized_error}
To the first-order, the reconstruction errors for the solution $\vx_{\mathrm{TLS}}^\dag$ to the TLS optimization problem \eqref{eq:error_analysis_objective}, and, the solution $\vx_{\mathrm{LS}}^\dag$ to the LS optimization problem \eqref{eq:error_analysis_ls_objective} are
\begin{align}
    e_{\mathrm{TLS}} &= \norm{\left( \wt{\mA}^\T \wt{\mY} \mD \wt{\mA} \right)^{-1} \wt{\mA}^\T \wt{\mY} \mD \vw}_2 \label{eq:tls_linearized_error} \\
    e_{\mathrm{LS}} &= \norm{\left( \wt{\mA}^\T \wt{\mY} \wt{\mA} \right)^{-1} \wt{\mA}^\T \wt{\mY} \vw}_2 . \label{eq:ls_linearized_error}
\end{align}
\end{proposition}
\begin{IEEEproof}
Lemma \ref{lemma:taylor_expansions} in Appendix \ref{sec:taylor_series} states the Taylor series expansions around the no error point, $\wt{\vt}$, for the TLS and LS solutions. The result in this proposition follows by considering only the zeroth and first-order terms.
\end{IEEEproof}

As expected, when $\vgamma \to 0$, the errors $\mE_{\mA}$ and $\mE_{\mY}$ tend to zero which makes \rev{the vector} $\vw$ zero and the reconstruction errors are zero. The difference \rev{between} the TLS and LS reconstruction errors in Proposition \ref{proposition:linearized_error} is due to the diagonal matrix $\mD$. As $\frac{\lambda_{y}}{\lambda_{a}} \to 0$, $\mD \to \mI_M$ and $e_{\mathrm{TLS}} \to e_{\mathrm{LS}}$. This is because the relative weighting of the sensing error consistency terms in \eqref{eq:quadratic_TLS_geometric_2}, $\norm{\va_m - \wh{\va}_m}_2^2$ for all $m$, increases which makes modifying the sensing vectors increasingly costly and the TLS problem moves closer to the LS problem.
\rev{Additionally, there are also error models under which the reconstruction errors are equal. For example, if $\mE_{\mY} = r_y \wt{\mY}$ and $\mE_{\mA} = r_A \wt{\mA}$ where $r_y,\, r_A \in \R$.}

\rev{
Furthermore, if $M = N$ and $\wt{\mA}$ is invertible, we can again have $e_{\mathrm{TLS}} = e_{\mathrm{LS}}$. However, having $M = N$ is not a practical setting for the real-valued phase retrieval problem because the map from $\vx^\#$ to $\left[\inprod{\wt{\va}_1, \vx^\#}^2, \ldots, \inprod{\wt{\va}_m, \vx^\#}^2 \right]^\T$ is not injective, even after accounting for the sign ambiguity \cite{balan2006signal}. The same holds for the complex-valued phase retrieval problem, even after accounting for the global phase shift \cite{conca2015algebraic}. Therefore, we can expect to require more measurements to obtain a unique solution to the phase retrieval problem with the TLS framework.

}

\rev{
The reconstruction errors in Proposition \ref{proposition:linearized_error} can be further interpreted by assuming a distribution for the measurement and sensing vectors errors; in Proposition \ref{proposition:expected_squared_error} we assume that the nonzero entries of $\mE_{\mY}$ and $\mE_{\mA}$ are iid zero-mean Gaussian (with different variances for $\mE_{\mY}$ and $\mE_{\mA}$).

\begin{proposition} \label{proposition:expected_squared_error}

With the setting of Proposition \ref{proposition:linearized_error}, assume that the diagonal elements of the diagonal matrix $\mE_{\mY}$ are iid zero-mean Gaussian with variance $\sigma^2_\eta$ and that the rows of $\mE_{\mA}$ are independent zero-mean Gaussian random vectors with covariance $\sigma^2_{\vdelta} \mI$. If $\mE_{\mY}$ and $\mE_{\mA}$ are independent of each other, the expected squared first-order reconstruction errors are

\begin{align}
    \mathbb{E}\left[e_{\mathrm{TLS}}^2 \right] =& \
    \sigma^2_{\vdelta} \cdot \norm{\vx^\#}_2^2
    \norm{\left( \wt{\mA}^\T \wt{\mY} \mD \wt{\mA} \right)^{-1} \wt{\mA}^\T \wt{\mY} \mD}^2_F \notag \\
    &+
    \frac{\sigma^2_\eta}{4} \cdot
    \norm{\left( \wt{\mA}^\T \wt{\mY} \mD \wt{\mA} \right)^{-1} \wt{\mA}^\T \wt{\mY}^{\frac{1}{2}} \mD}^2_F
    \label{eq:tls_linearized_error_expectation} \\
    \mathbb{E}\left[e_{\mathrm{LS}}^2 \right] =& \
    \sigma^2_{\vdelta} \cdot \norm{\vx^\#}_2^2
    \norm{\left( \wt{\mA}^\T \wt{\mY} \wt{\mA} \right)^{-1} \wt{\mA}^\T \wt{\mY}}^2_F \notag \\
    &+
    \frac{\sigma^2_\eta}{4} \cdot
    \norm{\left( \wt{\mA}^\T \wt{\mY} \wt{\mA} \right)^{-1} \wt{\mA}^\T \wt{\mY}^{\frac{1}{2}}}^2_F. \label{eq:ls_linearized_error_expectation}
\end{align}

\end{proposition}

\begin{IEEEproof}
The expectations are computed in Appendix \ref{sec:expected_squared_error_derivation}.
\end{IEEEproof}

Just like in Proposition \ref{proposition:linearized_error}, the difference between the TLS and LS expressions in Proposition \ref{proposition:expected_squared_error} are due to the diagonal matrix $\mD$. Each expression is a sum of two terms---the first term shows how the expectations depend on $\sigma^2_{\vdelta}$ and the second term shows how they depend on $\sigma^2_{\eta}$.


}

\subsection{Reconstruction error numerical experiments}

The expressions in \rev{Proposition \ref{proposition:linearized_error}} provide a means to understand when each approach should perform well. \rev{Furthermore, their squared-expectations in Proposition \ref{proposition:expected_squared_error} allow us to verify the optimal maximum likelihood parameters stated in Proposition \ref{proposition:ml_estimator}.}

\paragraph{Impact of varying error strength and number of measurements}

We compare TLS and LS by numerically evaluating \eqref{eq:tls_linearized_error} and \eqref{eq:ls_linearized_error} with different measurement and sensing vector error levels while varying the number of measurements.

These experiments only consider the first-order error. \rev{The} actual error is computed in a variety of experiments in Section \ref{sec:simulations}. We will use SNR to quantify the measurement and sensing vector error level. The measurement SNR is $-20 \log_{10} (\norm{\mE_{\mY}}_F / \|\wt{\mY}\|_F )$ and similarly the sensing vector SNR is $-20 \log_{10} (\norm{\mE_{\mA}}_F / \|\wt{\mA}\|_F )$. Furthermore we define the relative reconstruction errors, $\mathrm{rel.}e_{\mathrm{TLS}} = \frac{e_{\mathrm{TLS}}}{\norm{\vx^\#}}$ and $\mathrm{rel.}e_{\mathrm{LS}} = \frac{e_{\mathrm{LS}}}{\norm{\vx^\#}}$.

\begin{figure}[!t]
    \centering
    \subfloat[Measurement SNR is 65 dB. \label{fig:linearization_varying_m_higher_ysnr}]
    {\includegraphics[width=0.75\linewidth]
        {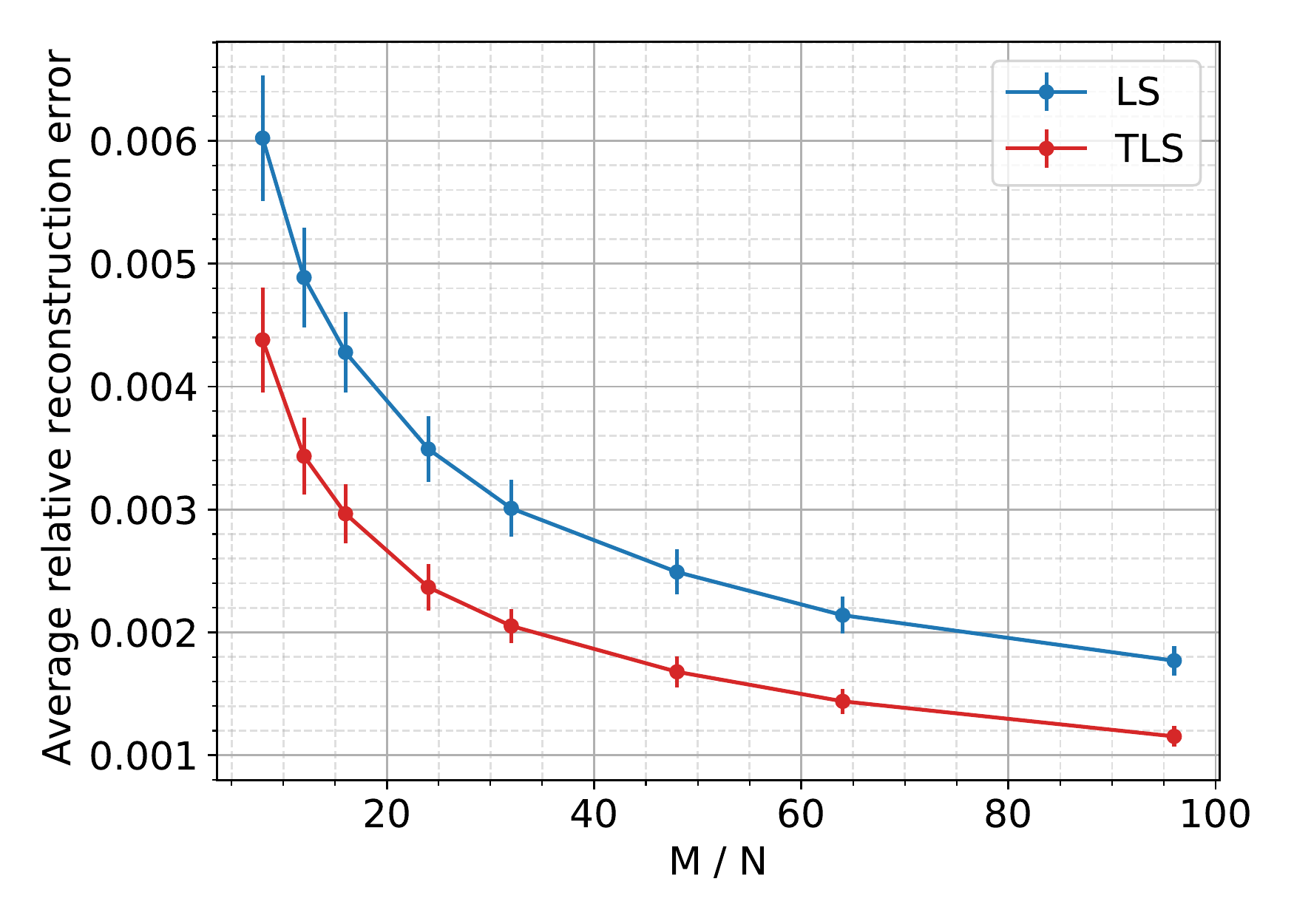}}
    
    \subfloat[Measurement SNR is 40 dB. \label{fig:linearization_varying_m}]
    {\includegraphics[width=0.75\linewidth]
        {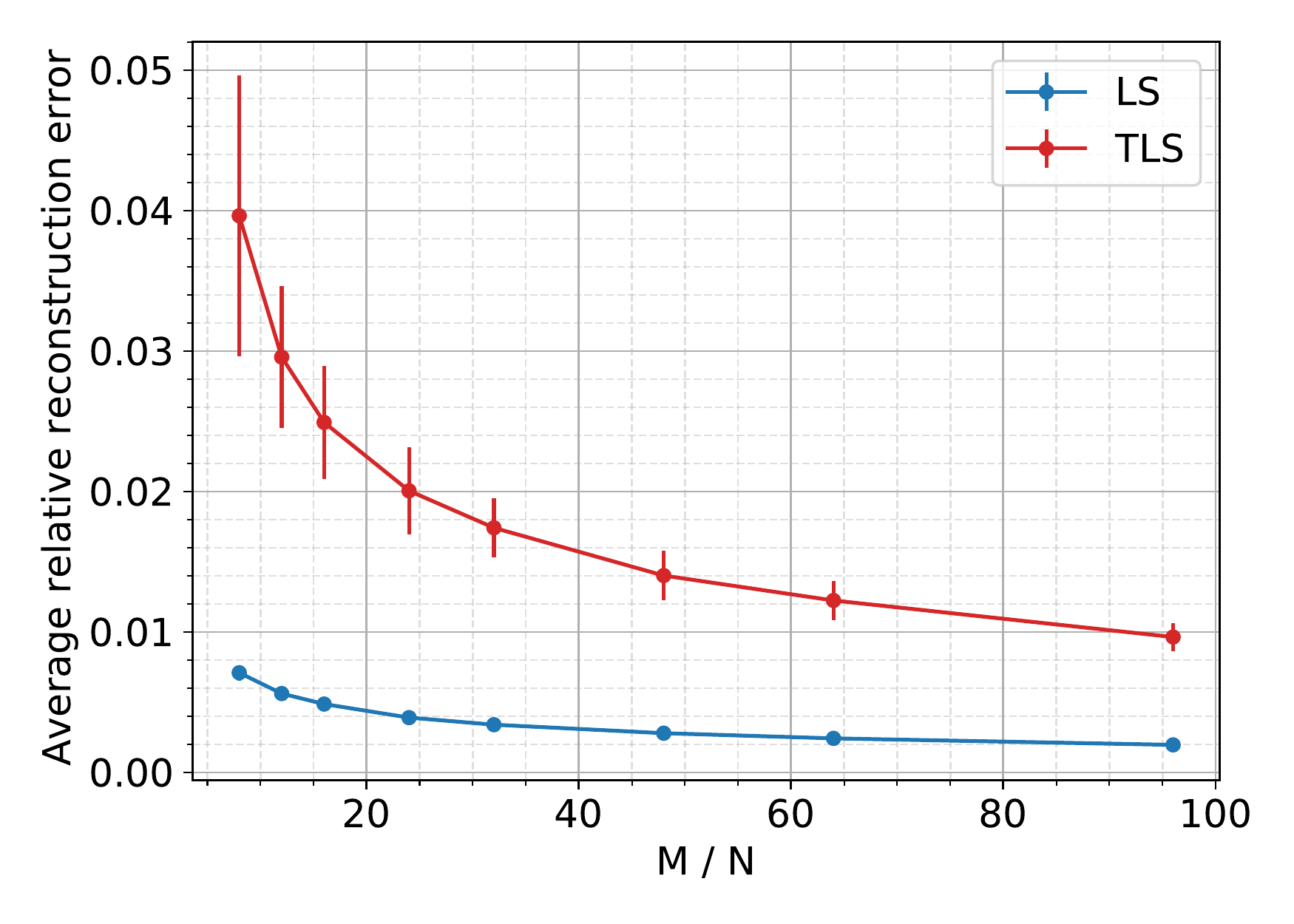}}
    
    
    \caption{Relative reconstruction errors, \eqref{eq:tls_linearized_error} and \eqref{eq:ls_linearized_error} for different values of $\frac{M}{N}$ when sensing vector SNR is 40 dB \rev{and measurement SNR is varied. All errors are Gaussian.}}
\end{figure}

We plot the relative reconstruction errors as the oversampling ratio $\frac{M}{N}$ is varied with $N = 100$. Regularization parameters $\lambda_{y}$ and $\lambda_{a}$ are set to one. For each value of $\frac{M}{N}$ we do 100 trials and each trial uses new sensing vectors, ground truth signals and errors. \rev{The standard deviation of the trials is indicated by error bars in the upcoming plots.} The sensing vectors and ground truth signal are iid standard real Gaussian. Furthermore, the measurement and sensing vector errors are iid zero-mean real Gaussian with variance such that the sensing vector SNR is 40 dB. In Fig. \ref{fig:linearization_varying_m_higher_ysnr} the measurement SNR is 65 dB and TLS has lower reconstruction error than LS. When the measurement SNR decreases to 40 dB in Fig. \ref{fig:linearization_varying_m}, LS outperforms TLS. Although these experiments use the first-order error, \rev{they are consistent with our intuition.} The relative performance of TLS is better when most of the error is due to sensing vector error. \rev{We also see that the performance of both methods improves as the number of measurements increases. Lastly,} from Fig. \ref{fig:linearization_varying_m}, TLS may improve relatively faster than LS as the number of measurements increase.

\rev{

\begin{figure*}[t]
    \centering
    \includegraphics[width=0.875\linewidth]{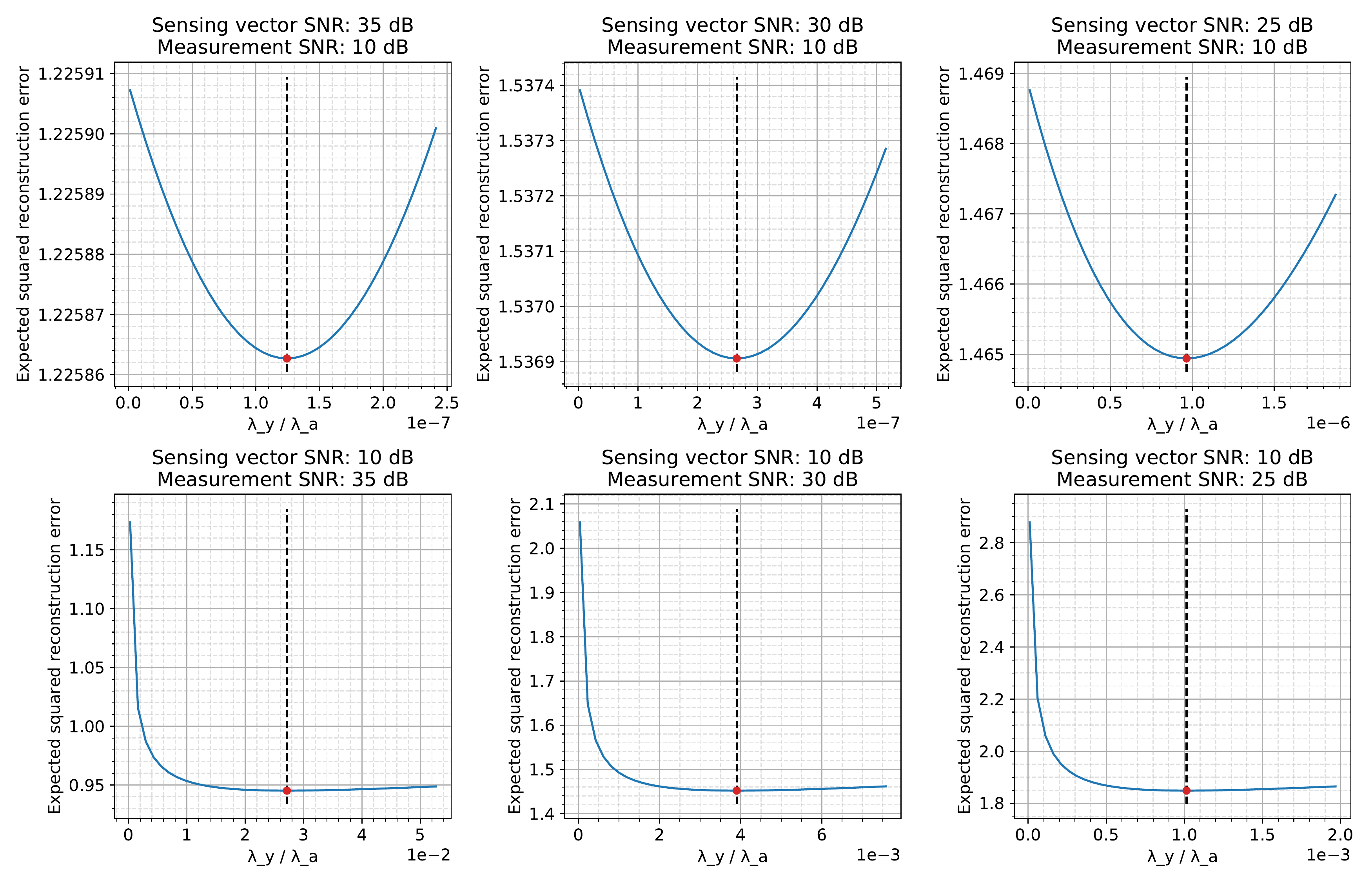}
    \caption{\rev{The TLS expected squared reconstruction error \eqref{eq:tls_linearized_error_expectation} is plotted for different ratios, $\frac{\lambda_y}{\lambda_a}$, to verify the optimal maximum likelihood parameters. Each subplot shows a different combination of sensing vector and measurement SNR. The minima are marked in red and the theoretically optimal ratio is indicated by the dashed black lines.}}
    \label{fig:ml_parameter_verification}
\end{figure*}

\paragraph{Verification of optimal ML parameters}

The expression for TLS \eqref{eq:tls_linearized_error_expectation} in Proposition \ref{proposition:expected_squared_error} enables us to verify the optimal maximum likelihood parameters for $\lambda_y$ and $\lambda_a$ from Proposition \ref{proposition:ml_estimator}. Although Proposition \ref{proposition:ml_estimator} is stated for the complex-valued phase retrieval problem, the same procedure shows that the optimal parameters are the same for real-valued phase retrieval we considered here. The theoretically optimal parameter ratio is $\dfrac{\lambda_y}{\lambda_a} = \dfrac{\sigma^2_{\vdelta}}{\sigma^2_{\eta}}$.

To verify numerically whether this agrees with Proposition \ref{proposition:expected_squared_error}, we vary $\frac{\lambda_y}{\lambda_a}$ (which is contained in $\mD$) around the optimal ratio and plot the TLS expression \eqref{eq:tls_linearized_error_expectation}. We do this multiple times and in each run use a different iid standard real Gaussian ground truth signal and a different set of iid standard real Gaussian sensing vectors. As in Proposition \ref{proposition:expected_squared_error}, the errors in each run are iid zero-mean Gaussian and their variances are set to obtain different SNRs. Fig. \ref{fig:ml_parameter_verification} shows the different runs with the minima marked in red and the theoretically optimal ratio indicated by the dashed black lines. We can see that all the minima are at the optimal ratio which verifies Proposition \ref{proposition:ml_estimator}. In the top row, most of the error is due to measurement error and the optimal ratio is low. This further highlights that TLS sensing vector corrections are less important when most of the error is due to measurement error.

We further note that the consistency between Propositions \ref{proposition:ml_estimator} and \ref{proposition:expected_squared_error} demonstrates that the first-order expressions can be used to explain the performance of our TLS framework. Section \ref{sec:simulations} shows that the real reconstruction errors follow the same trends as the numerical simulations in this section.
}

\section{TLS phase retrieval simulations}
\label{sec:simulations}

We compare the performance of TLS phase retrieval against LS phase retrieval through simulations.\footnote{Code available at \url{https://github.com/swing-research/tls_phase}.} To obtain a LS solution we use the Wirtinger flow method \cite{candes2015phase}.

In this section we set the regularization parameters of \eqref{eq:quadratic_TLS_geometric_2} to $\lambda_{a} = \frac{1}{N}$ and $\lambda_{y} = \frac{1}{\norm{\vx^{(0)}}_2^4}$ in all experiments with $\vx^{(0)}$ being an initial guess for $\vx^\#$. These regularization parameters are tuned later in Section \ref{sec:opu}. We fix the ground truth signal to be iid complex Gaussian with $N=100$. Furthermore, the TLS and LS iterations are stopped when their objective function values \rev{change by less than $10^{-6}$ between successive iterates}. The ground truth signal, TLS solution and LS solution are denoted as $\vx^\#$, $\vx^\dag_{\mathrm{TLS}}$ and $\vx^\dag_{\mathrm{LS}}$. In all experiments we generate $M$ quadratic measurements using $M$ clean sensing vectors. The TLS and LS methods must then recover the signal $\vx^\#$ from erroneous measurements and sensing vectors. We use SNR, as defined in Section \ref{sec:linearization}, to quantify measurement and sensing vector error.
\rev{Also as in Section \ref{sec:linearization}, the plots in this section indicate the standard deviation of the trials using error bars.}


\subsection{Measurement models}

In our experiments we will consider the complex-valued Gaussian and coded diffraction pattern measurement models. However, Algorithm \ref{algo:tls_pr_algorithm} is not restricted to these measurement models. Recently in optical computing applications, random Gaussian scattering media have been used to do rapid high-dimensional randomized linear algebra, kernel classification and dimensionality reduction using laser light \cite{gupta2019don, saade2016random}. The coded diffraction pattern model modulates the signal with different patterns before taking the Fourier transform. It is inspired by the fact that in coherent x-ray imaging the field at the detector is the Fourier transform of the signal \cite{blahut2004theory}. 

When using the Gaussian measurement model, the $n$th entry of sensing vector $m$, $a_{mn}$, is distributed by the complex normal distribution for the complex-valued problem, $a_{mn} \sim \calN(0, 1) + j\calN(0,1)$. For the real-valued problem it is the standard normal distribution, $a_{mn} \sim \calN(0, 1)$. The Gaussian measurement model sensing vector entries are independent of each other and the sensing vectors are also independent of each other. A description of the coded diffraction pattern measurement model is in Appendix \ref{sec:cdp_appendix}.

In this section of the main paper\rev{,} the complex Gaussian measurement model is used. In Appendix \ref{sec:cdp_experiments_appendix} these experiments are repeated for the coded diffraction pattern measurement model and the same behavior is seen.


\subsection{Algorithm initialization}

In our experiments we opt to do the initialization of the signal being recovered (line \ref{algo:initialization} of Algorithm \ref{algo:tls_pr_algorithm}) via a spectral initialization. This method comprising a spectral initialization followed by gradient descent updates has been proven to lead to globally optimal solutions for the LS phase retrieval problem \eqref{eq:quadratic_LS} in an error-free setting under the Gaussian and coded diffraction pattern models \cite{netrapalli2013phase, candes2015phase}.

The spectral initialization is the leading eigenvector of the matrix $\sum_m y_m \va_m \va_m^* \in \C^{N \times N}$ which we efficiently compute using 50 power method iterations. This eigenvector is scaled appropriately by estimating the norm of the signal of interest as $\left(\frac{1}{2M}\sum_m y_m \right)^{1/2}$.

\rev{
\subsection{Signal recovery}

To evaluate performance we compute the distance between the ground truth signal and the recovered signal. As the value of the objective function \eqref{eq:quadratic_TLS_geometric} is the same for $\vx$ and phase shifted $\e^{j \varphi}\vx$, we cannot distinguish between $\vx$ and its phase-shifted variant. We therefore use a standard definition of distance that is invariant to phase shifts \rev{which is detailed in Definition \ref{def:rel_distance}}.

\begin{definition} \label{def:rel_distance}
Denote the ground truth as $\vx^\# \in \C^N$ and let $\vx^\dag \in \C^N$ be a solution to the phase retrieval problem. The distance between $\vx^\#$ and $\vx^\dag$, $\mathrm{dist}(\vx^\#, \vx^\dag)$, is defined as $\mathrm{dist}(\vx^\#, \vx^\dag) = \min_{\varphi \in [0,2\pi)} \norm{\vx^\# - \e^{j \varphi}\vx^\dag}_2$. Furthermore, the relative distance is defined as $\mathrm{rel.dist}(\vx^\#, \vx^\dag) = \frac{\mathrm{dist}(\vx^\#, \vx^\dag)}{\norm{\vx^\#}_2}$ and the reconstruction SNR in dB is defined as $-20\log_{10}(\mathrm{rel.dist}(\vx^\#, \vx^\dag))$. 

\end{definition}
}

\paragraph{Combinations of sensing vector and measurement error}

To understand how performance changes with different amounts of sensing vector and measurement error, we add different amounts of random iid complex Gaussian error to sensing vectors and random iid real Gaussian error to measurements. For each combination of sensing vector error and measurement error we perform 100 phase retrieval trials. In each trial we generate a new ground truth signal and $M$ new sensing vectors to produce $M$ new error-free measurements. In each trial we then add new random error perturbations to the sensing vectors and measurements. We evaluate performance by subtracting the relative distance of the TLS solution from that of the LS solution, $( \mathrm{rel.dist}(\vx^\#, \vx^\dag_{\mathrm{LS}}) - \mathrm{rel.dist}(\vx^\#, \vx^\dag_{\mathrm{TLS}}) )$, and average across all 100 trials. If this average is positive, TLS \rev{has} outperformed LS.

\begin{figure}[!t]
    \centering
    \subfloat[$\frac{M}{N}=8$]{\includegraphics[width=0.75\linewidth]
        {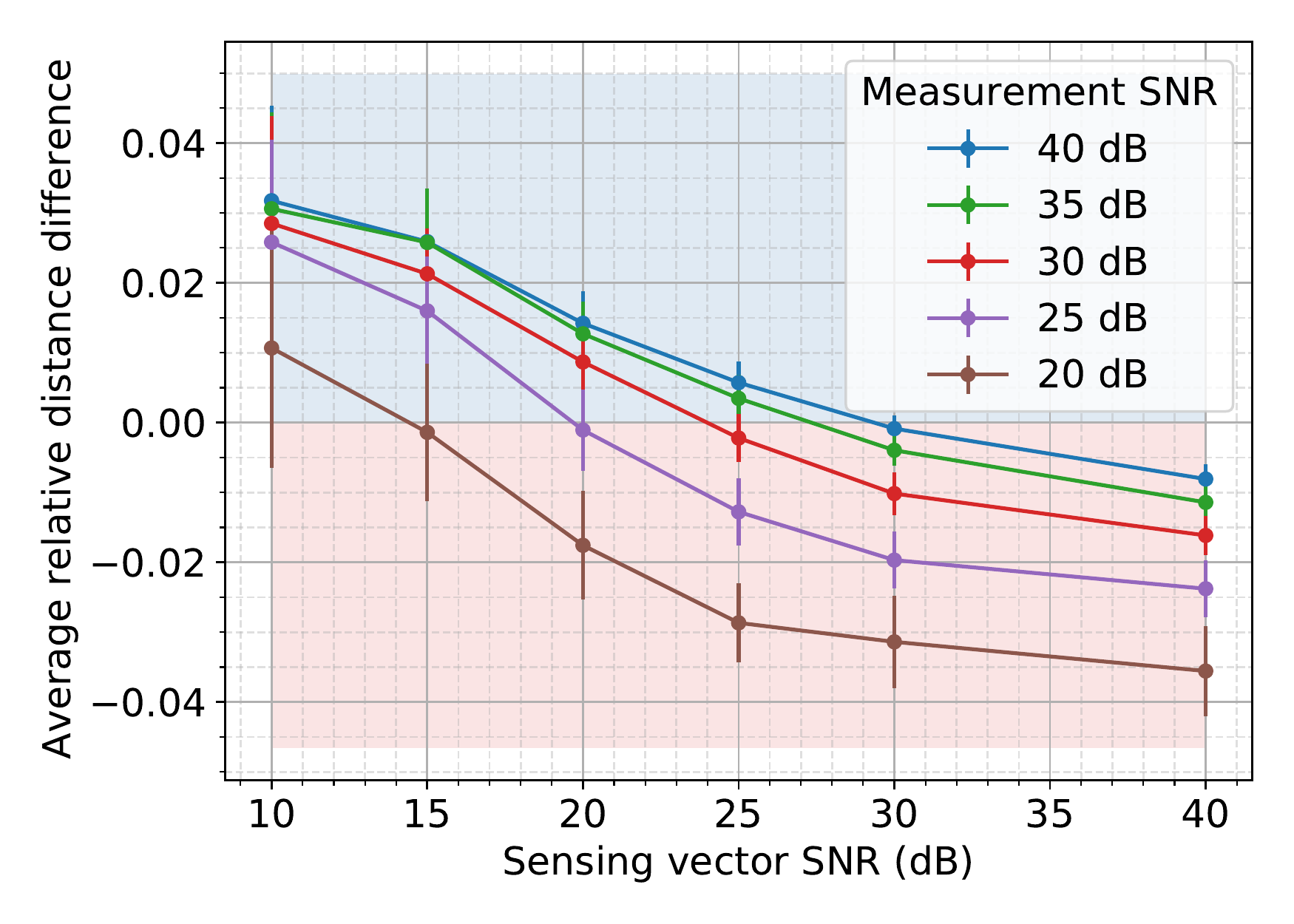}}
    
    \subfloat[$\frac{M}{N}=16$]{\includegraphics[width=0.75\linewidth]
        {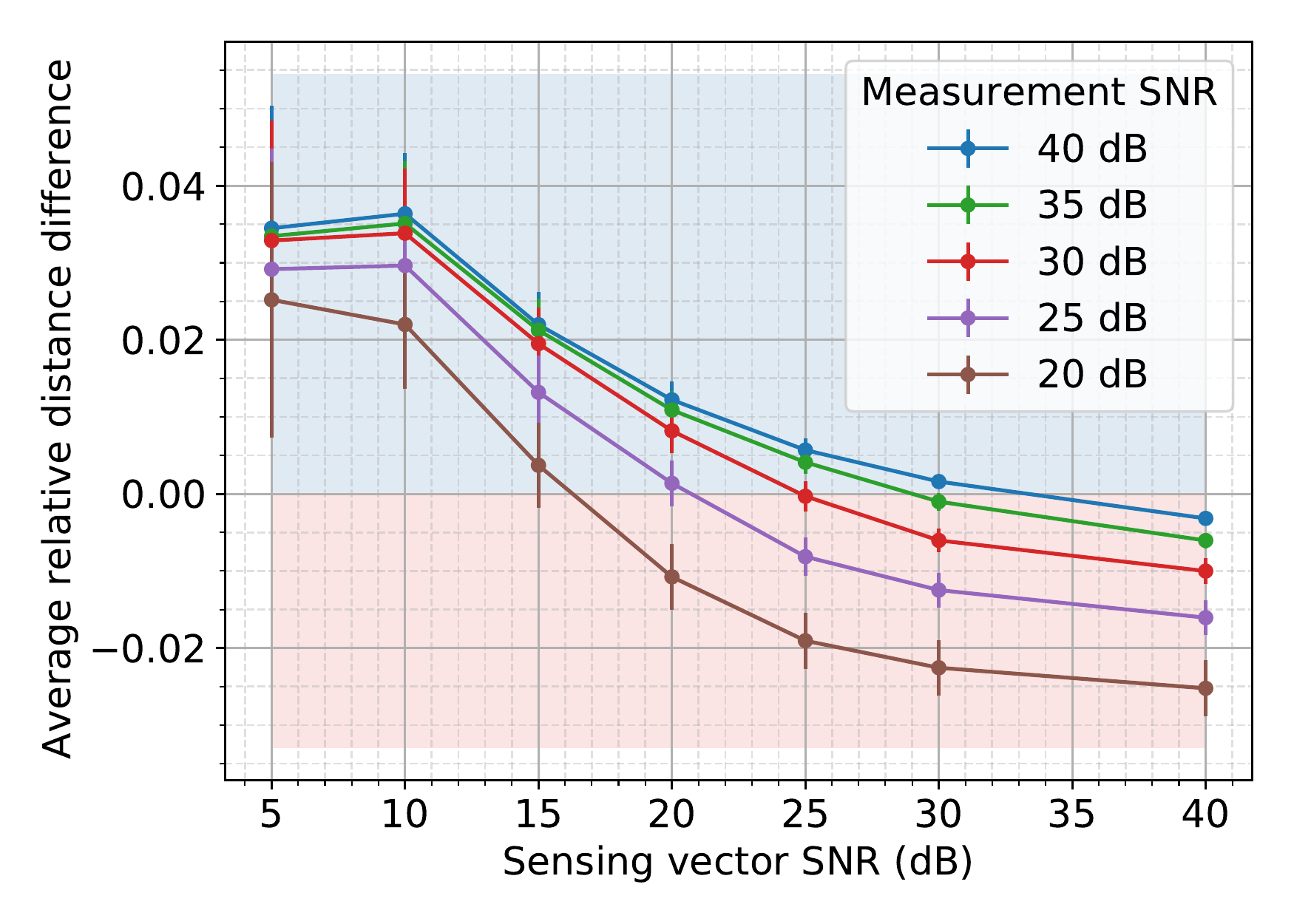}}
    
    \subfloat[$\frac{M}{N}=32$]{\includegraphics[width=0.75\linewidth]
        {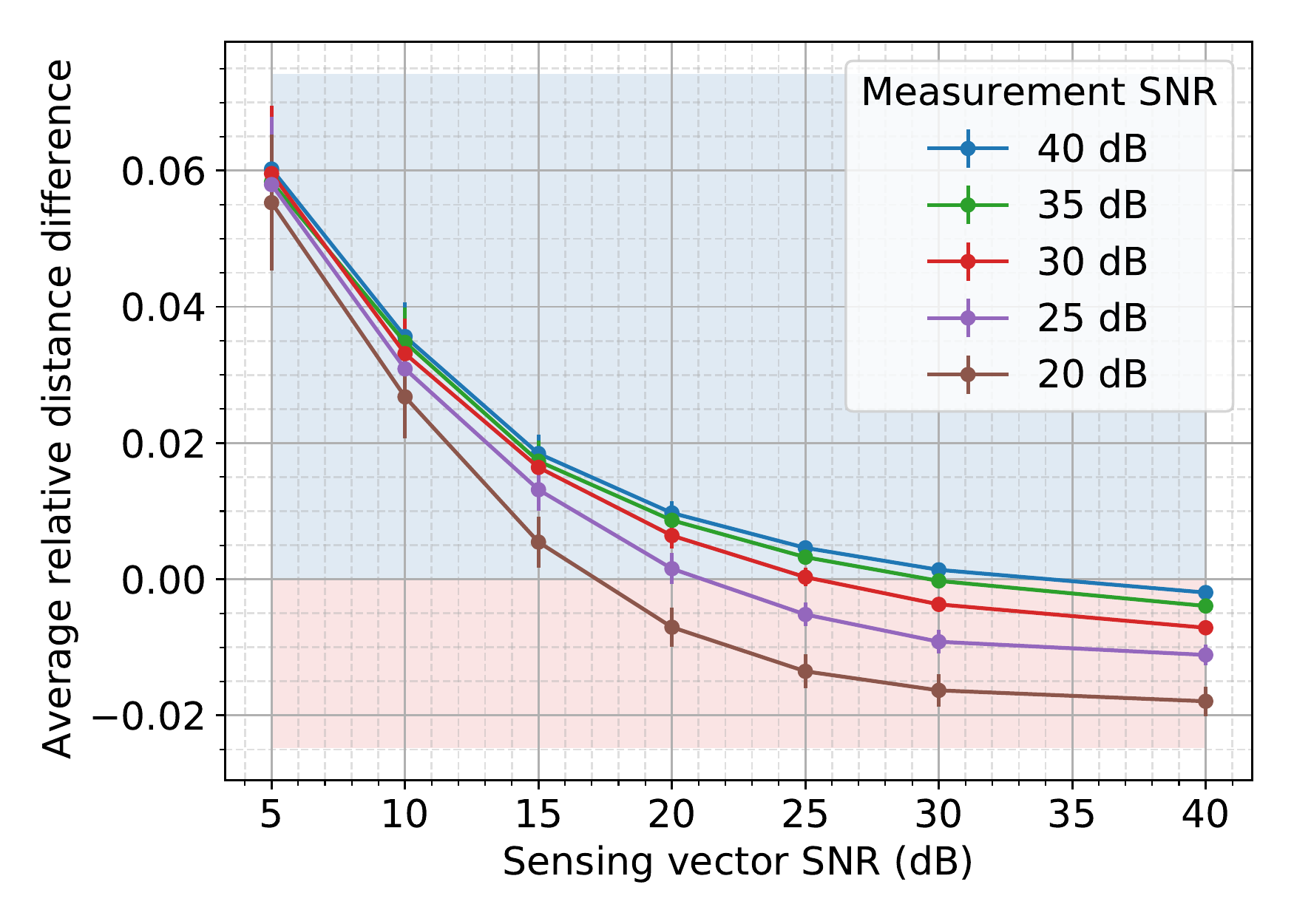}}
    
    \caption{Average difference in relative distance of TLS and LS solutions, $\mathrm{rel.dist}(\vx^\#, \vx^\dag_{\mathrm{LS}}) - \mathrm{rel.dist}(\vx^\#, \vx^\dag_{\mathrm{TLS}})$, for the Gaussian measurement model for different measurement and sensing vector SNR combinations when $\frac{M}{N} \in \{8, 16, 32\}$.}
    \label{fig:gaussian_model_random_perturbations}
\end{figure}

We use a step size of $\mu = \frac{0.5}{\lambda_{a}}$ for TLS and $\mu = 0.02$ for LS to perform the gradient update for $\vx$ in \eqref{eq:gradient_update}. The TLS step size is inversely proportional to $\lambda_{a}$ because the relative importance of the data consistency term is inversely proportional to the sensing vector consistency term in \eqref{eq:quadratic_TLS_geometric_2}. Fig. \ref{fig:gaussian_model_random_perturbations} shows the performance for $\frac{M}{N} \in \{8, 16, 32\}$. Note that the minimum sensing vector SNR is 10 dB when $\frac{M}{N}=8$ and 5 dB in the other cases. For a fixed sensing vector SNR, the performance of TLS decreases when the measurement SNR decreases. This is expected because more of the error is in the measurements which LS is designed for. In general TLS is better when the sensing vector SNR decreases for a fixed measurement SNR because TLS phase retrieval accounts for sensing vector error. 
However, this trend starts to break for very low sensing vector SNR as shown at 5 dB when $\frac{M}{N} = 16$. 
Increasing the number of measurements overcomes this issue and in general improves TLS performance as was indicated by the first-order reconstruction errors with Gaussian error in \rev{Figs. \ref{fig:linearization_varying_m_higher_ysnr} and \ref{fig:linearization_varying_m}}.

\paragraph{Impact of varying the number of measurements}

\begin{figure}[!t]
    \centering
    \subfloat[Sensing vector SNR is 10 dB. \label{fig:gaussian_varymn}]
    {\includegraphics[width=0.75\linewidth]
        {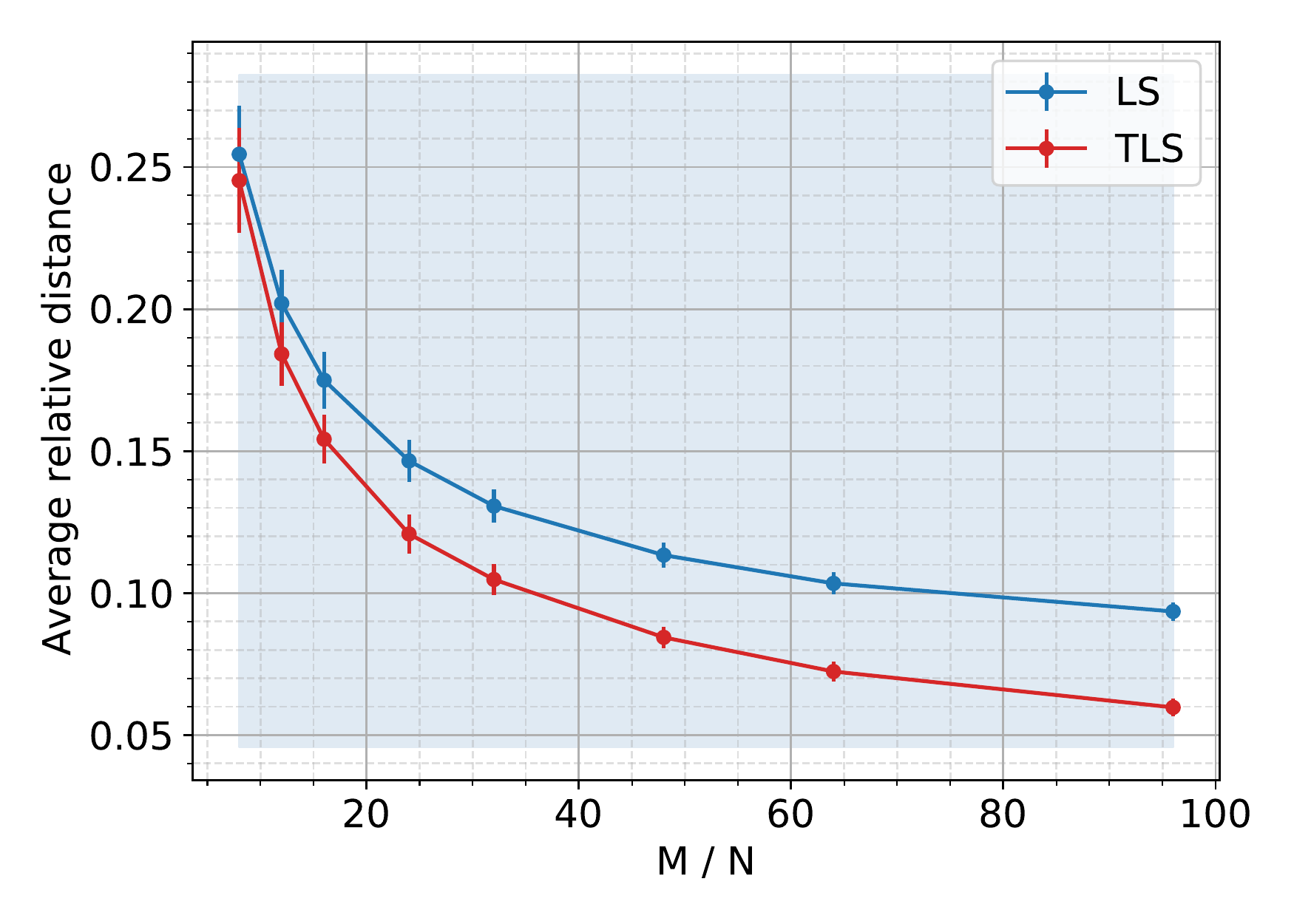}}
    
    \subfloat[Sensing vector SNR is 30 dB. \label{fig:gaussian_varymn_largerA_SNR}]
    {\includegraphics[width=0.75\linewidth]
        {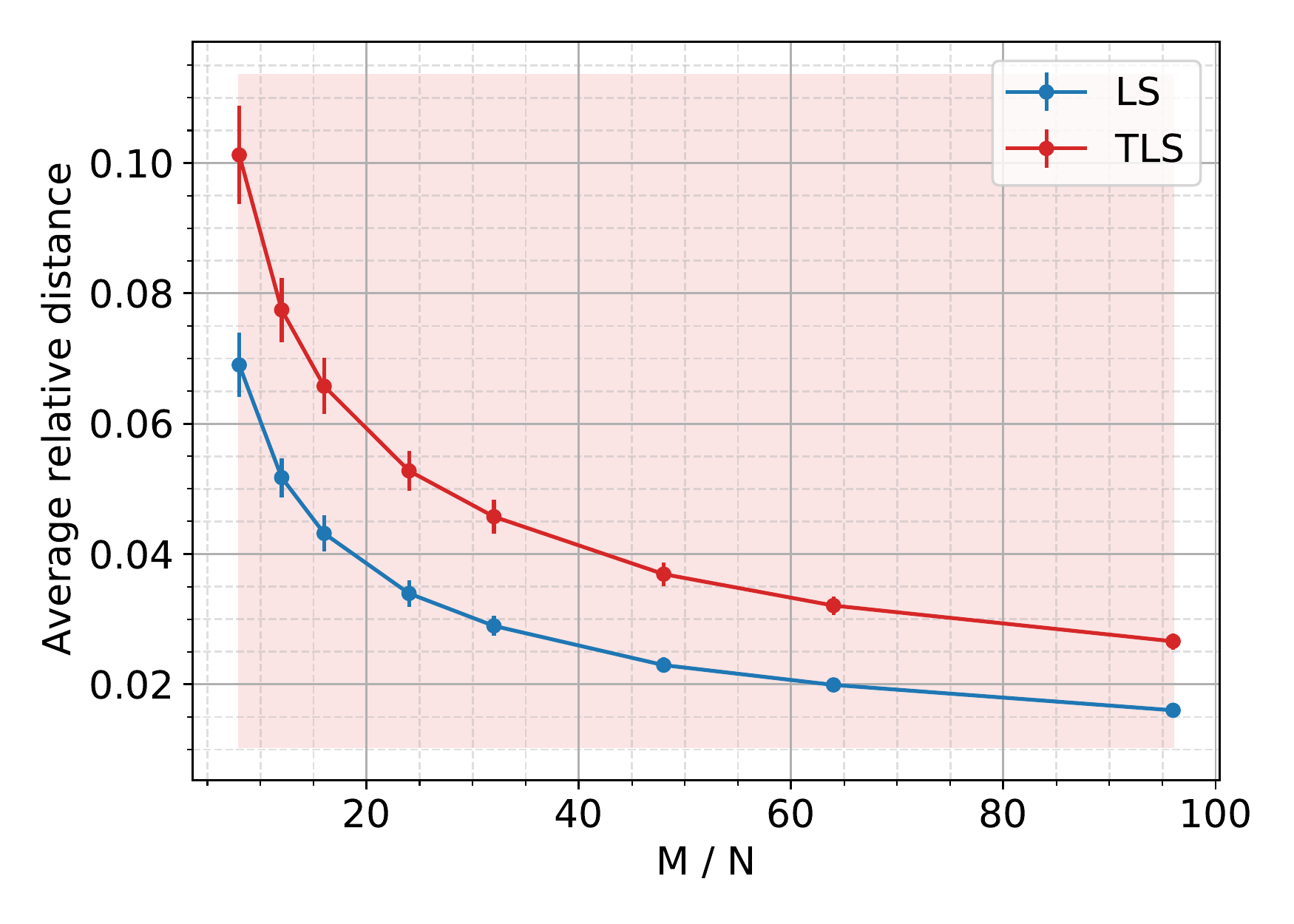}}
    
    
    \caption{Relative distance of reconstructions using TLS and LS for the Gaussian measurement model for different $\frac{M}{N}$ when measurement SNR is 20 dB \rev{and measurement SNR is varied. All errors are Gaussian.}}
\end{figure}

To clearly see the impact of varying the number of measurements we fix the measurement SNR to 20 dB and sensing vector SNR to 10 dB and plot the reconstruction relative distance for TLS and LS in Fig. \ref{fig:gaussian_varymn}. We do 100 trials for each value of $\frac{M}{N}$. The performance improvement of TLS over LS increases as the number of measurements are increased. In Fig. \ref{fig:gaussian_varymn_largerA_SNR} we increase the sensing vector SNR to 30 dB. When the balance of the Gaussian error shifts more towards the measurements, LS performs better.
\rev{This is identical to what was seen with the first-order reconstruction errors in \rev{Figs. \ref{fig:linearization_varying_m_higher_ysnr} and \ref{fig:linearization_varying_m}}.}

\rev{
\paragraph{Accuracy of first-order reconstruction errors}

Appendix \ref{sec:taylor_verification} contains the description of an experiment where we verify the accuracy of the reconstruction error expressions in Proposition \ref{proposition:linearized_error} against the real errors. As expected, it shows that the first-order expressions in Proposition \ref{proposition:linearized_error} increase in accuracy as the sensing vector and measurement error decreases---this corresponds to the norm of $\vgamma$ in \eqref{eq:gamma} decreasing, but that these expressions may only serve as a rough rule of thumb when errors are large.
}

\rev{
\paragraph{Sensing vectors and measurements error model}

The simulations in this section use iid random Gaussian errors. In Appendix \ref{sec:handcrafted_errors} we design errors that require access to the ground truth signal norm and to the error-free measurements. We show that the improvement of TLS over LS can be larger in this artificial scenario.
}

\rev{
\subsection{Corrected sensing vector verification}

To characterize the sensing vector corrections performed by our algorithm, we define a metric sensitive to the relative correction error of the sensing vectors. The metric only considers corrections in the direction of the recovered signal because our algorithm only corrects the component of the sensing vectors in the direction of this signal due to the optimization geometry (Section \ref{sec:geometry}).

\begin{definition} \label{def:rel_correction}
Denote the complex-valued ground truth signal and sensing vectors as $\vx^\#$ and $\va^\# := \{\va^\#_m\}_{m=1}^M$. Let their counterparts obtained by solving the TLS phase retrieval problem be $\vx^\dag$ and $\va^\dag := \{\va^\dag_m\}_{m=1}^M$. Further let $\varphi = \argmin_{\varphi \in [0,2\pi)} \norm{\vx^\# - \e^{j \varphi}\vx^\dag}_2$. 
Then, denoting $\mathsf{y}(\va, \vx) = \left[ \inprod{\va_1, \vx}, \ldots, \inprod{\va_M, \vx}\right]$, the relative sensing vector correction error between $\va^\#$ and $\va^\dag$ is defined as $\mathrm{rel.corr}(\{\va^\#,\, \vx^\#\}, \{\va^\dag, \,\vx^\dag\}) = \dfrac{\norm{\mathsf{y}(\va^\#,\, \vx^\#) - \mathsf{y}(\va^\dag,\, \e^{j \varphi}\vx^\dag)}_2}{\norm{\mathsf{y}(\va^\#,\, \vx^\#)}_2}$.

\end{definition}

To evaluate performance, we denote the ground truth and TLS corrected sensing vectors as $\{\wt{\va}_m\}_{m=1}^M$ and $\{\wh{\va}^\dag_m\}_{m=1}^M$. For the sensing vectors in the previous experiments of Fig. \ref{fig:gaussian_model_random_perturbations} we compute $\mathrm{rel.corr}(\{\wt{\va},\, \vx^\#\}, \{\wh{\va}^\dag, \,\vx^\dag_{\mathrm{TLS}}\})$ and average across the 100 trials. Fig. \ref{fig:gaussian_model_random_perturbations_A_correction} shows the relative correction error when $\frac{M}{N} \in \{16, 32\}$. We see that as the sensing vector SNR increases, the relative correction error decreases. Furthermore, as the measurement SNR decreases, the relative correction error increases. These relative correction error increases are more pronounced when the sensing vector SNR is high, a setting where sensing vector correction is needed less. This observation is consistent with Fig. \ref{fig:gaussian_model_random_perturbations}---when sensing vector SNR is high, the TLS sensing vector corrections hinder TLS performance and LS outperforms TLS.

\begin{figure}[!t]
    \centering

    \subfloat[$\frac{M}{N}=16$]{\includegraphics[width=0.75\linewidth]
        {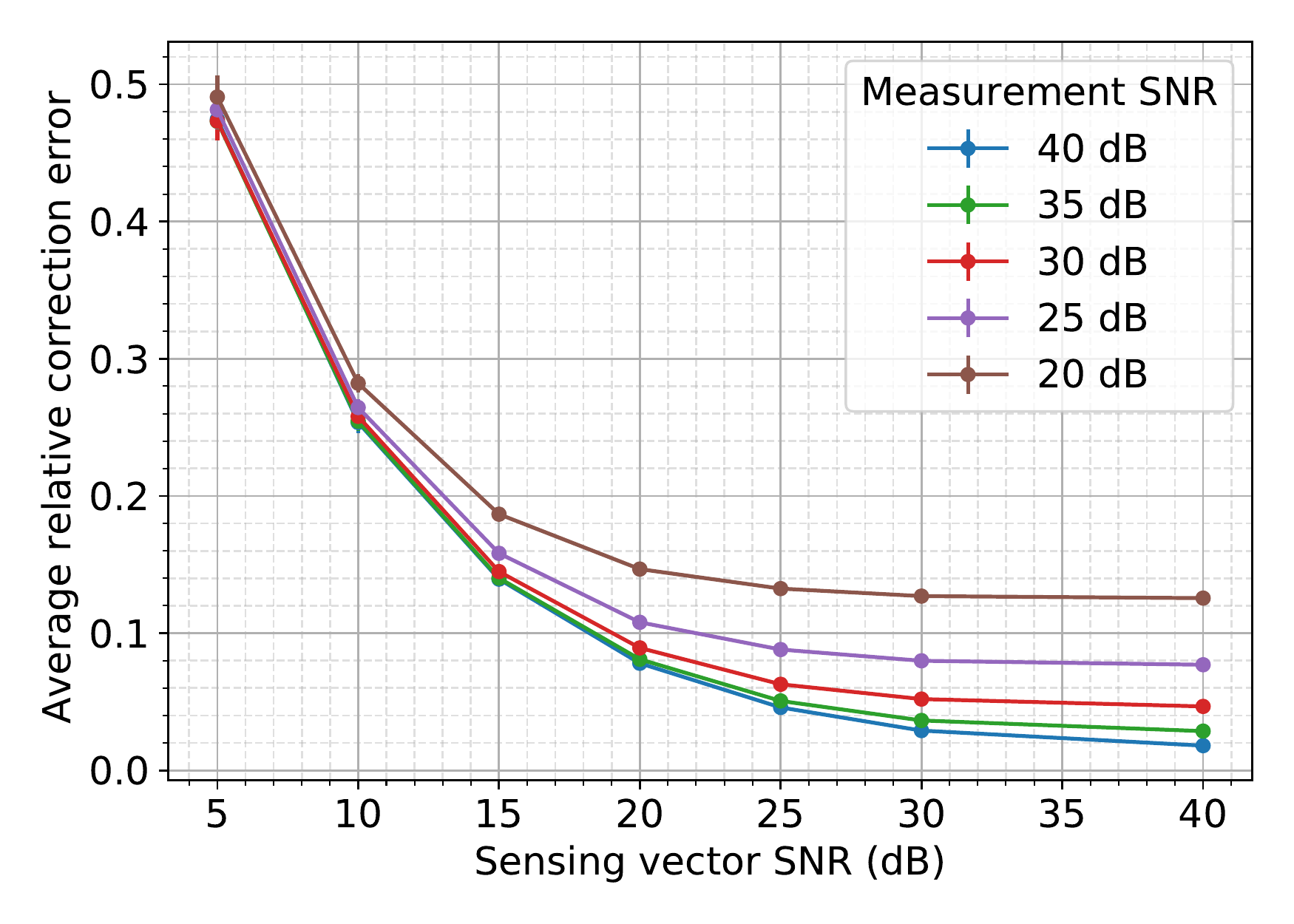}}
    
    \subfloat[$\frac{M}{N}=32$]{\includegraphics[width=0.75\linewidth]
        {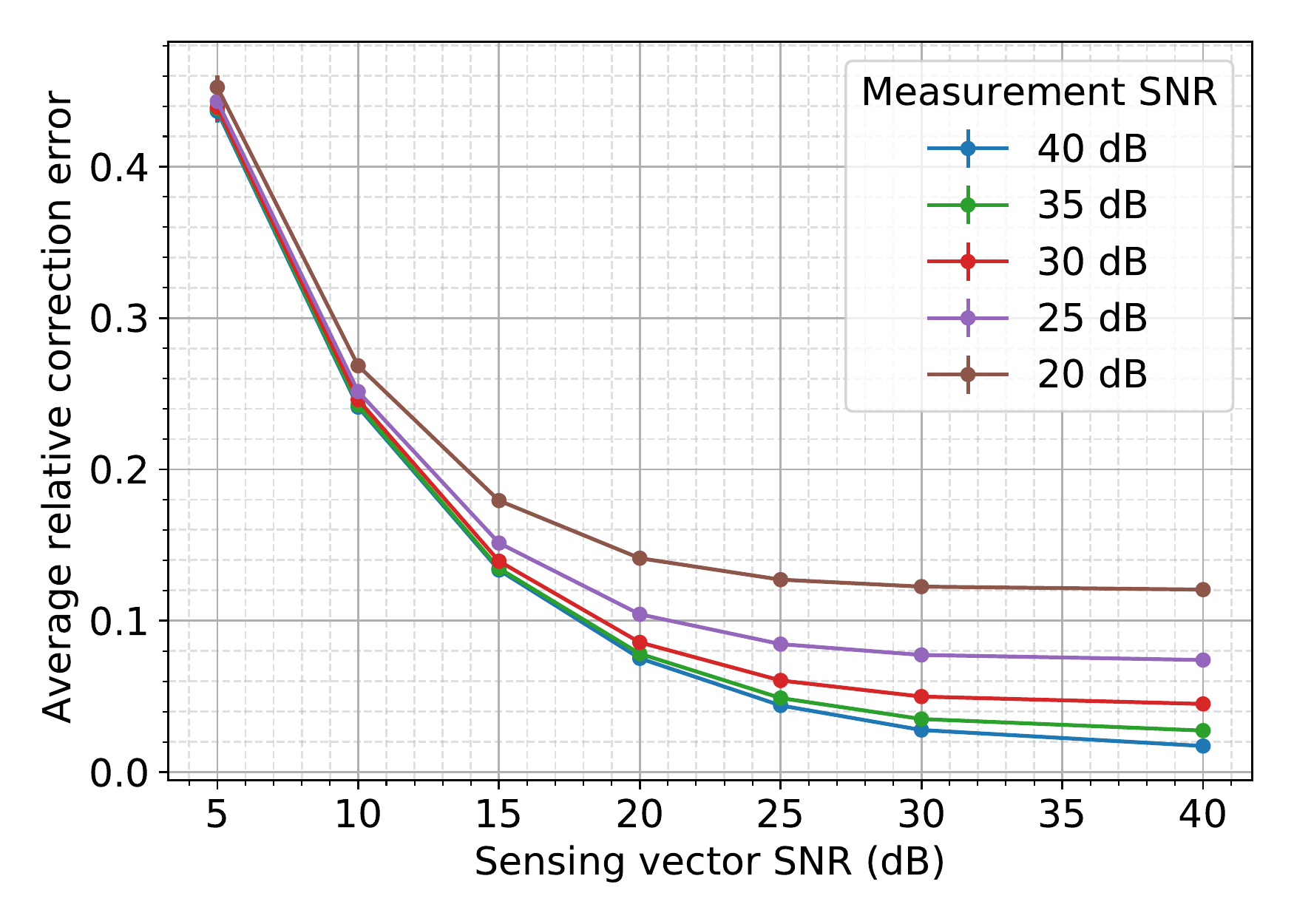}}
    
    \caption{\rev{Average relative sensing vector correction error when using TLS, $\mathrm{rel.corr}(\{\wt{\va},\, \vx^\#\}, \{\wh{\va}^\dag, \,\e^{j \varphi} \vx^\dag_{\mathrm{TLS}})$, for the Gaussian measurement model for different measurement and sensing vector SNR combinations when $\frac{M}{N} \in \{16, 32\}$.}}
    \label{fig:gaussian_model_random_perturbations_A_correction}
\end{figure}

Next we investigate how the number of measurements impacts the relative correction error. We do this with the sensing vectors from the previous experiments in Figs. 5a and 5b. Fig. \ref{fig:gaussian_varymn_A_error} shows the averages over the 100 trials. Here the measurement SNR was fixed to 20 dB and the sensing vector SNR was 10 dB or 30 dB. Consistent with what was seen previously, the performance of TLS improves with increasing number of measurements. Additionally, increasing the number of measurements provides greater gains when the sensing vector SNR is lower.

\begin{figure}[t]
    \centering
    \includegraphics[width=0.75\linewidth]{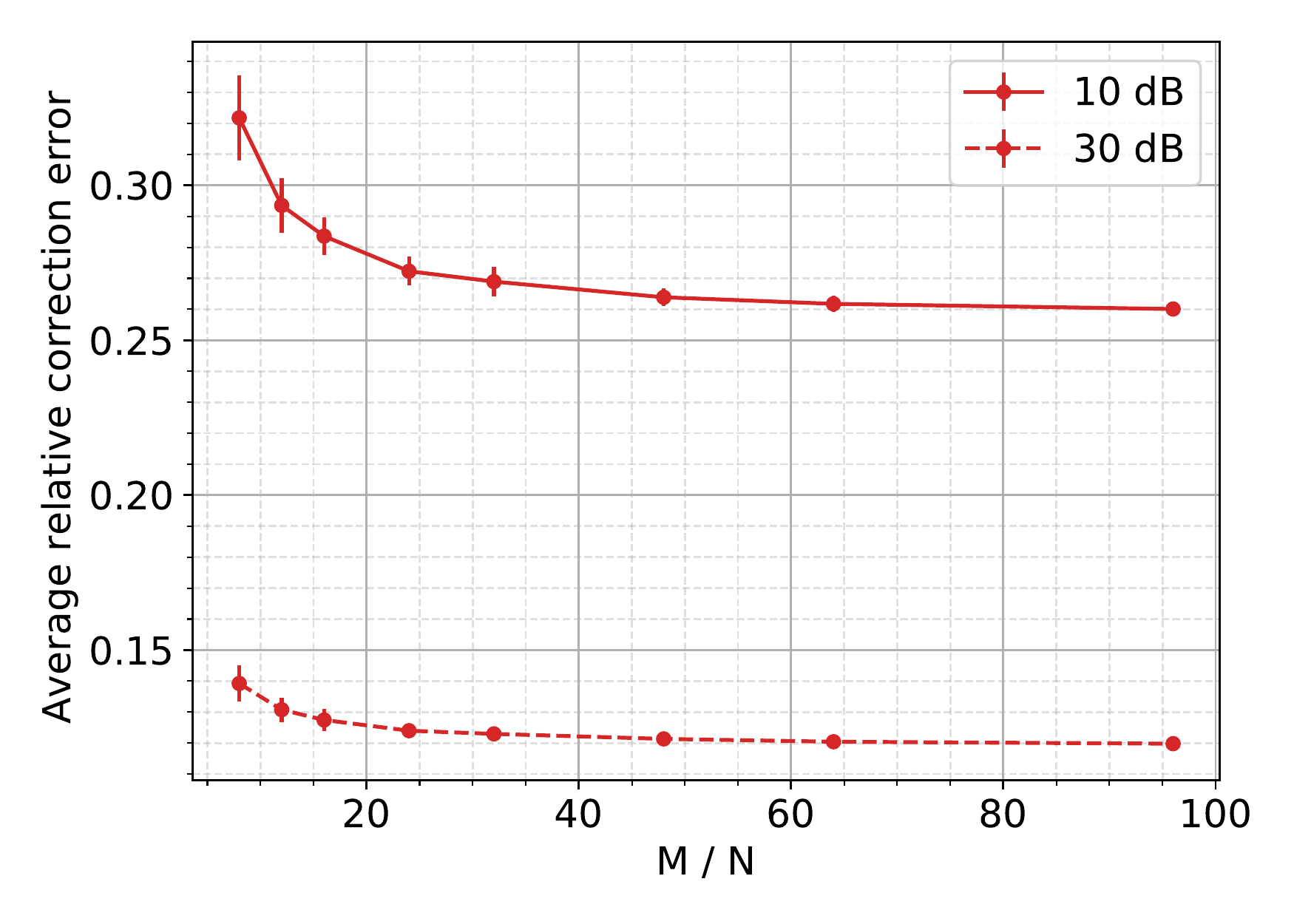}
    \caption{\rev{Relative sensing vector correction error when using TLS for the Gaussian measurement model for different $\frac{M}{N}$ when measurement SNR is 20 dB. The sensing vector SNR is 10 dB or 30 dB.}}
    \label{fig:gaussian_varymn_A_error}
\end{figure}

}

\section{Experiments on real optical hardware} \label{sec:opu}

\begin{figure}[t]
\centering
\def\svgwidth{0.8\linewidth}
\fontsize{8}{8}\selectfont
\begingroup%
  \makeatletter%
  \providecommand\color[2][]{%
    \errmessage{(Inkscape) Color is used for the text in Inkscape, but the package 'color.sty' is not loaded}%
    \renewcommand\color[2][]{}%
  }%
  \providecommand\transparent[1]{%
    \errmessage{(Inkscape) Transparency is used (non-zero) for the text in Inkscape, but the package 'transparent.sty' is not loaded}%
    \renewcommand\transparent[1]{}%
  }%
  \providecommand\rotatebox[2]{#2}%
  \newcommand*\fsize{\dimexpr\f@size pt\relax}%
  \newcommand*\lineheight[1]{\fontsize{\fsize}{#1\fsize}\selectfont}%
  \ifx\svgwidth\undefined%
    \setlength{\unitlength}{535.20256463bp}%
    \ifx\svgscale\undefined%
      \relax%
    \else%
      \setlength{\unitlength}{\unitlength * \real{\svgscale}}%
    \fi%
  \else%
    \setlength{\unitlength}{\svgwidth}%
  \fi%
  \global\let\svgwidth\undefined%
  \global\let\svgscale\undefined%
  \makeatother%
  \begin{picture}(1,0.51816919)%
    \lineheight{1}%
    \setlength\tabcolsep{0pt}%
    \put(0,0){\includegraphics[width=\unitlength,page=1]{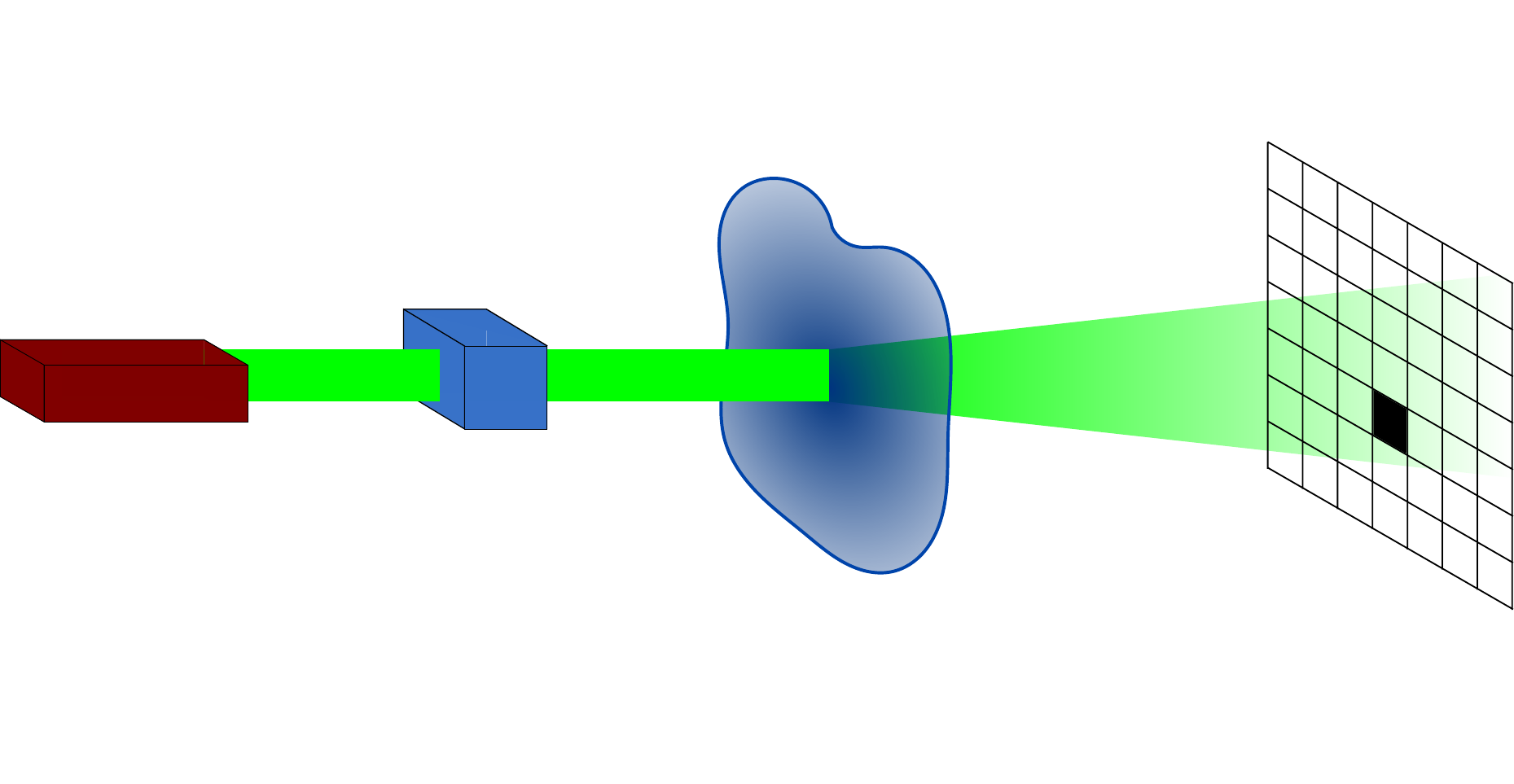}}%
    \put(0.19982436,0.19497716){\color[rgb]{0,0,0}\makebox(0,0)[lt]{\lineheight{0.25}\smash{\begin{tabular}[t]{l}DMD encoding\\of $\vx^\# \in \R^N$ \end{tabular}}}}%
    \put(0.48641946,0.11764824){\color[rgb]{0,0,0}\makebox(0,0)[lt]{\begin{minipage}{0.66158414\unitlength}\raggedright $\abs{\inprod{\va_m, \vx^\#}}^2 \approx y_m$\end{minipage}}}%
    \put(0.39210725,0.48608718){\color[rgb]{0,0,0}\makebox(0,0)[lt]{\lineheight{0.125}\smash{\begin{tabular}[t]{l}Random\\scattering\\medium\end{tabular}}}}%
    \put(0.75664195,0.50433827){\color[rgb]{0,0,0}\makebox(0,0)[lt]{\lineheight{0.25}\smash{\begin{tabular}[t]{l}Camera\\taking 8-bit\\measurements\end{tabular}}}}%
    \put(0,0){\includegraphics[width=\unitlength,page=2]{opu.pdf}}%
    \put(0.48641925,0.05345349){\color[rgb]{0,0,0}\makebox(0,0)[lt]{\begin{minipage}{0.54995776\unitlength}\raggedright $\va_m \overset{\text{iid}}{\sim} \mathcal{N}(0,\mI) + j\mathcal{N}(0,\mI)$\end{minipage}}}%
    \put(0.0309683,0.37045065){\color[rgb]{0,0,0}\makebox(0,0)[lt]{\lineheight{0.25}\smash{\begin{tabular}[t]{l}Laser\\light\\source\end{tabular}}}}%
  \end{picture}%
\endgroup%

\caption{The optical processing unit (OPU). A coherent laser beam spatially encodes a signal, $\vx$, via a digital micro-mirror device (DMD) which is then shined through a random medium. A camera measures the squared magnitude of the scattered light.}
\label{fig:opu}
\end{figure}

In this section we show that TLS phase retrieval outperforms LS phase retrieval when using real optical hardware. We use an Optical Processing Unit (OPU) which enables rapid random high-dimensional matrix-vector multiplication.\footnote{Visit \url{https://www.lighton.ai/lighton-cloud/} for a publicly available cloud OPU with a scikit-learn interface.} A known signal $\vx^\# \in \R^N$ is encoded onto coherent laser light using a digital micro-mirror device (DMD) which is then shined through a Gaussian multiple scattering medium as shown in Fig. \ref{fig:opu}. We denote the transmission matrix of the Gaussian medium as $\mA \in \C^{M \times N}$. The $M$ rows of the transmission matrix are sensing vectors, $\va_m \in \C^N$ for $1 \leq m \leq M$. The intensity of the scattered light in the sensor plane, $y_m \approx \abs{\inprod{\va_m, \vx^\#}}^2$ for all $m$, is then measured using a camera. We do phase retrieval using the optical measurements to reconstruct the input, $\vx^\#$. The input signals are limited to real-valued binary images due to the DMD.

The OPU measurements and sensing vectors both contain errors. Errors in the optical measurements are caused by 8-bit quantized camera measurements
\rev{and Poisson noise which scales with the square root of the mean intensity of the scattered light. Additionally, there are measurement errors due to thermal effects and other system properties that result in a noise floor. Thus the lowest intensity that can be measured is not zero, even if an all-zero signal, $\vx^\# = \vzero$, is encoded on the laser light and shined through the scattering medium.}
The sensing vectors are erroneous because the entries of $\mA$ are unknown and must be calibrated from erroneous optical measurements \cite{gupta2020fast}. There may also be other sources of experimental error. Unlike in the computer simulations of Section \ref{sec:simulations}, \rev{when using the OPU}, we do not know the exact error model and \rev{we also do not know} the levels of the errors in the measurements and sensing vectors. 

In the experiments, the TLS and LS step sizes are tuned to $\frac{0.4}{\lambda_{a}}$ and 0.005. The initialization method and termination criteria are the same as in Section \ref{sec:simulations}. Additionally, we use the fact that the images being reconstructed are real-valued and binary to regularize both the TLS and LS methods. After the initialization (Algorithm \ref{algo:tls_pr_algorithm}, Step \ref{algo:initialization} for TLS) and each $\vx$ update step (Algorithm \ref{algo:tls_pr_algorithm}, Step \ref{algo:gradient_step} for TLS) we take the elementwise absolute value of the signal to set the phase of all elements to zero. We then normalize the entries of $\vx^\dag_{\mathrm{TLS}}$ and $\vx^\dag_{\mathrm{LS}}$ with absolute value larger than one to one.

Appendix \ref{sec:opu_appendix} contains details of the sensing vector calibration method used and further OPU experimental details.

\paragraph{Random ground truth signals}

\begin{figure}[t]
    \centering
    \includegraphics[width=0.75\linewidth]{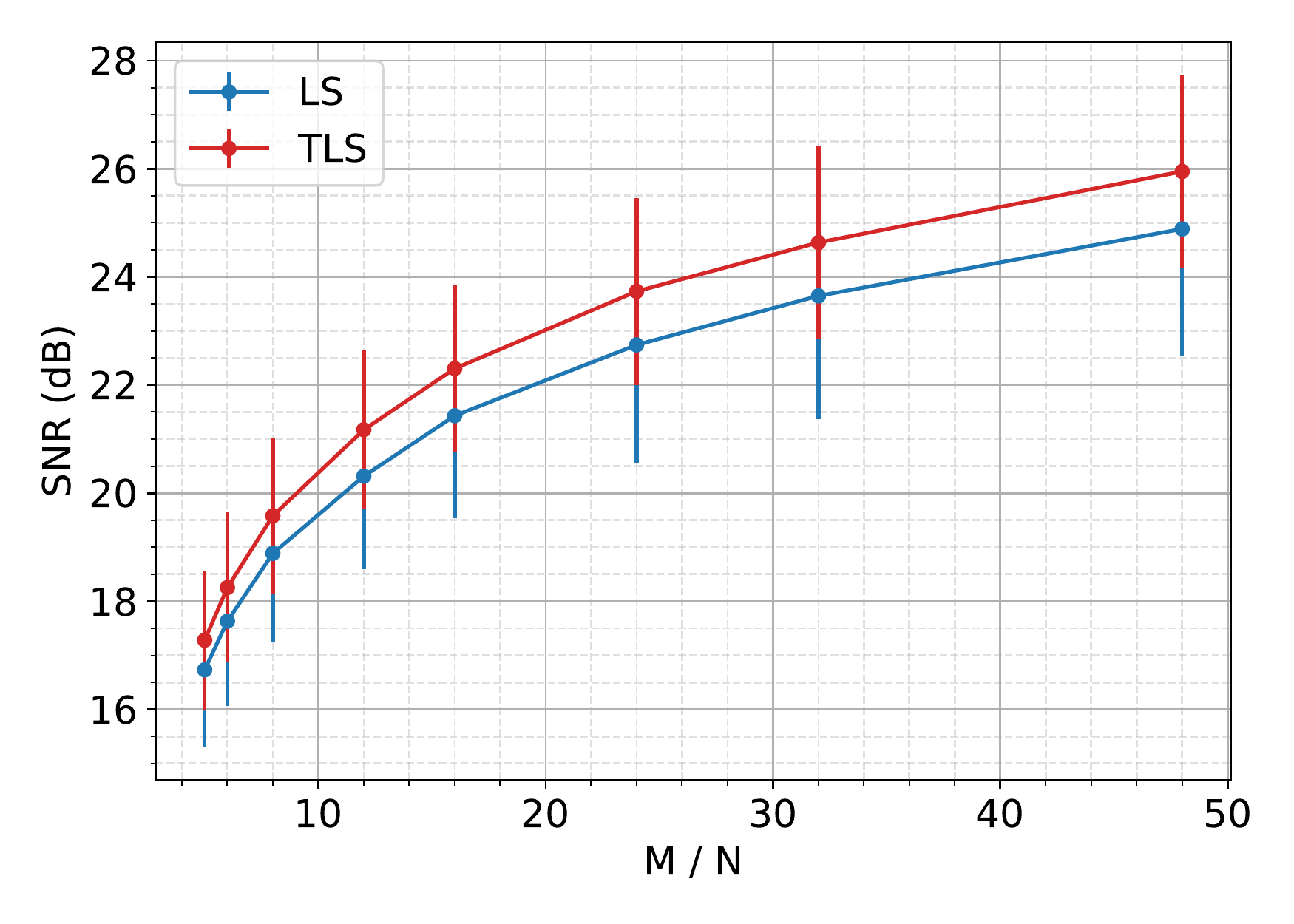}
    \caption{SNR of reconstructed random binary images when using TLS and LS for phase retrieval on the OPU. Values of $\frac{M}{N}$ between five and 48 are used.}
    \label{fig:opu_trials}
\end{figure}

\begin{figure}[t]
    \centering
    \includegraphics[width=0.75\linewidth]{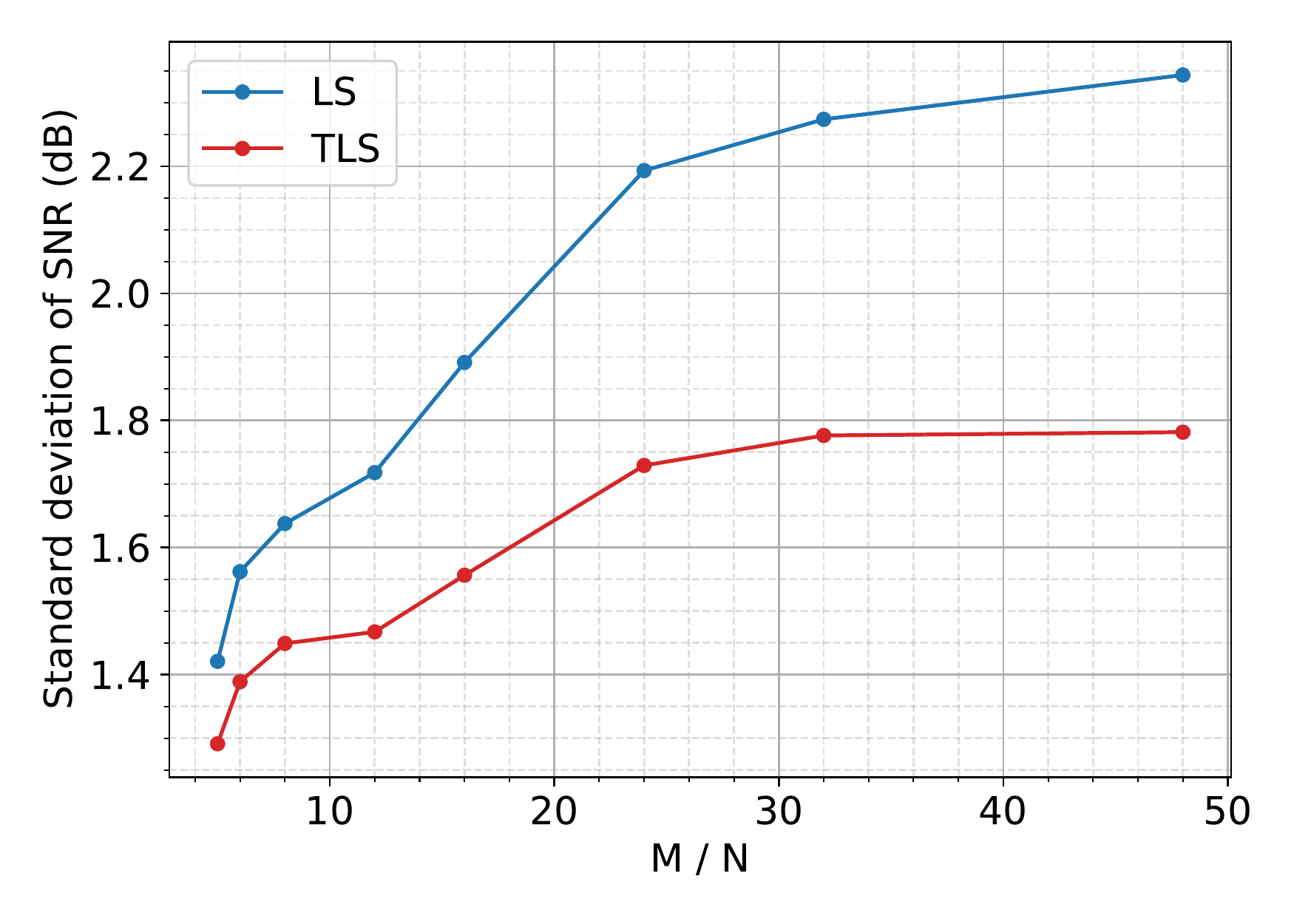}
    \caption{\rev{Standard deviation of the SNR in Fig. \ref{fig:opu_trials} when random binary images are reconstructed on the OPU.}}
    \label{fig:opu_trials_std}
\end{figure}

Our ground truth signals are real-valued random binary images of size $N = 16 \times 16 = 256$. We vary the oversampling ratio, $\frac{M}{N}$, and perform 100 trials for each ratio. In each trial a new ground truth image and set of calibrated sensing vectors is used. On a held out set of ten images and with $\frac{M}{N} = 8$ we tune $\lambda_{a} = 40$ and $\lambda_{y} = \norm{\vx^{(0)}}_2^{-4}$ in \eqref{eq:quadratic_TLS_geometric_2} where $\vx^{(0)}$ is the initialization. Fig. \ref{fig:opu_trials} shows that the SNR of the reconstructed images using TLS is higher than when using LS for all numbers of measurements. \rev{Additionally, in Fig. \ref{fig:opu_trials_std} we plot the standard deviation of the results in Fig. \ref{fig:opu_trials} and show that the TLS method has lower variability.}

\paragraph{Real image ground truth signals}

\begin{figure}[t]
    \centering
    \includegraphics[width=0.99\linewidth]{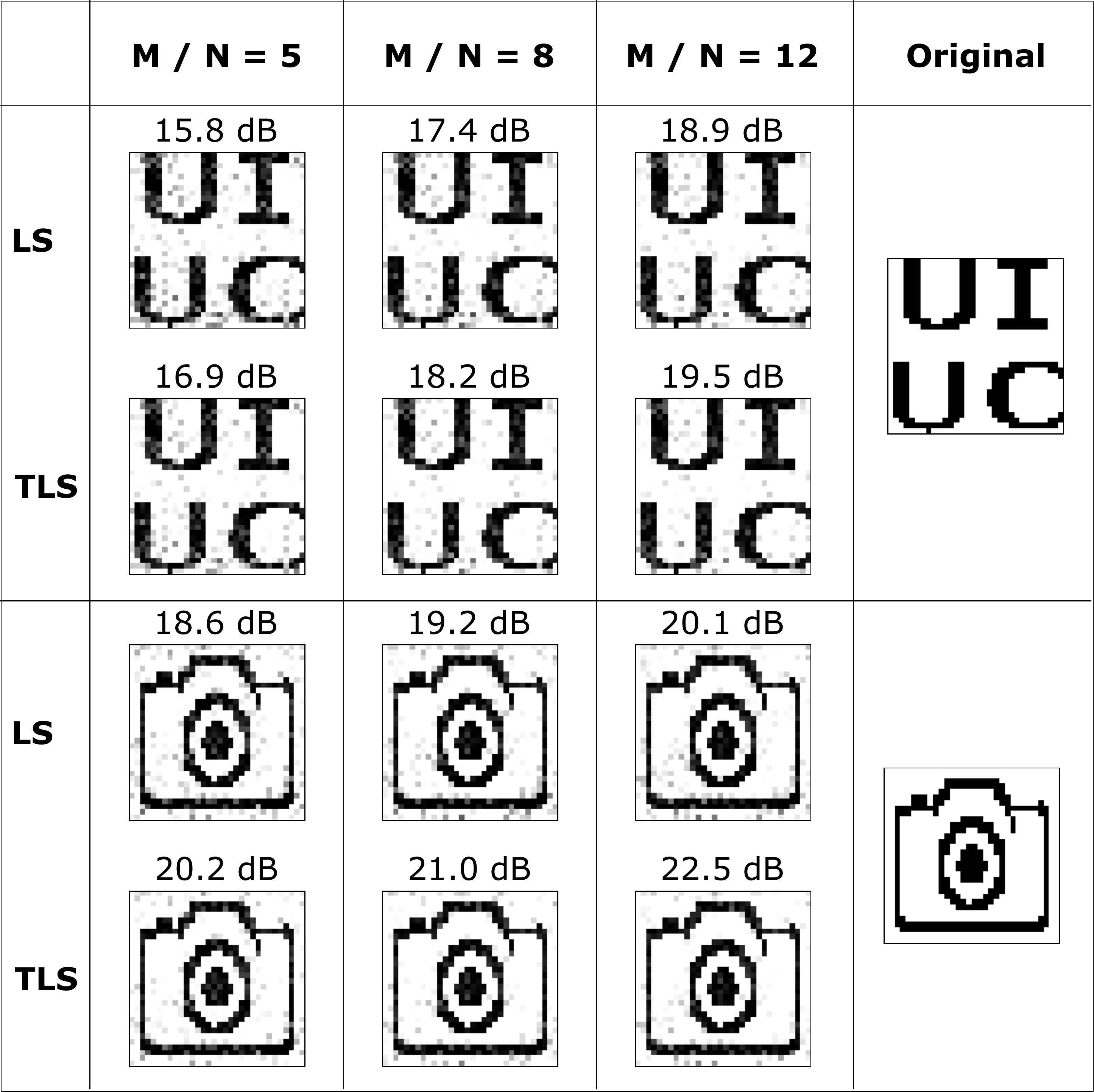}
    \caption{Reconstructions of  $32 \times 32$ images and reconstruction SNR when using TLS and LS on the OPU. The oversampling ratio is varied.}
    \label{fig:opu_images}
\end{figure}

We reconstruct binary images of size $N = 32 \times 32 = 1024$ for $\frac{M}{N} \in \{5, 8, 12 \}$. On a held out set of five images and with $\frac{M}{N} = 5$ we tune $\lambda_{a} = 20$ and again $\lambda_{y} = \norm{\vx^{(0)}}_2^{-4}$ in \eqref{eq:quadratic_TLS_geometric_2} where $\vx^{(0)}$ is the initialization. Fig. \ref{fig:opu_images} shows the original images, their reconstructions and their reconstruction SNR. For a given oversampling ratio, the TLS approach reports better SNR values and reconstructs images of better visual quality as compared to the LS approach.

\section{Conclusion}
\label{sec:conclusion}

We have developed a TLS framework for solving the phase retrieval problem that accounts for both sensing vector error and measurement error. One of the keys to solving the TLS problem via gradient descent was studying the geometry of the TLS optimization problem to realize that the sensing vectors can be efficiently updated by solving a scalar variable optimization problem instead of a vector variable optimization problem. By deriving the Taylor series expansions for the TLS and LS solutions we have also obtained approximate expressions for their reconstruction error. These expressions enabled us to anticipate the accuracy of the TLS solution relative to the LS solution and understand when which approach will lead to a better solution. We verify the TLS method through a range of computer simulations. Furthermore, in experiments with real optical hardware, TLS outperforms LS.

Presently we correct the sensing vectors based on the one signal that we wish to recover. An interesting future line of work lies in multi-signal TLS for sensing vector denoising so that \rev{subsequent} signals can be recovered without requiring sensing vector corrections if their measurements use the same set of sensing vectors. There are multiple areas where this is required. An upcoming application is in optical neural network backpropagation where unknown random sensing vectors are the rows of weight matrices of fully-connected layers. Denoised and more accurate rows will enable better machine learning performance.

\rev{There exist other applications of phase retrieval with uncertain sensing vectors. Ptychography, which can be modeled analogously to (1), is a prime example \cite{pfeiffer2018x}. Ptychography has recently been addressed by least squares, comprising spectral initialization followed by gradient descent \cite{valzania2021accelerating}. It would be interesting to see whether our TLS phase retrieval algorithm brings about improvements. Other ptychography methods use alternating updates to recover the object and the sensing vectors \cite{maiden2017further}. Our geometric intuition may help reduce the number of unknowns in sensing vector updates and thus improve the overall computational efficiency of these algorithms.}


\appendices

\section{Roots of cubic equations} \label{sec:cubic_roots}

Consider finding the roots of the following cubic equation
\begin{align}
    ax^3 + bx^2 + cx + d = 0 .
\end{align}

Denote
\begin{align}
    \psi_0 &= b^2 - 3ac \\
    \psi_1 &= 2b^3 - 9abc + 27a^2 d \\
    \psi_3 &= \sqrt[3]{\frac{\psi_1 + \sqrt{\psi_1^2 - 4\psi_0^3}}{2}} .
\end{align}

Then for $k \in \{0, 1, 2\}$ the three roots, $x_k$, are 
\begin{align}
    x_k = -\frac{1}{3a} \left(b + \theta^k \psi_3 + \frac{\psi_0}{\theta^k \psi_3} \right)
\end{align}
where $\theta$ is the cube root of unity, $\theta = \frac{-1 + \sqrt{-3}}{2}$.

Note that the cubic equations in this paper are with $b=0$ which simplifies the above expressions.

\section{Proof of Proposition \ref{proposition:ml_estimator}}
\label{sec:ml_estimator_proof_appendix}

The ML estimator estimates both $\wt{\vx}$ and $\{\wt{\va}_m\}_{m=1}^M$ from the data $\{y_m\}_{m=1}^M$ and $\{\va_m\}_{m=1}^M$ by minimizing the negative conditional log-likelihood
\begin{align*}
    & \argmin_{\vx, \ve_1, \ldots, \ve_M} -\ln\left(\prod_{m=1}^M \mathrm{Pr}\{y_m, \va_m | \wt{\vx} = \vx, \wt{\va}_m = \va_m + \ve_m \}\right)
\end{align*}

With $\wt{\vx}$ and $\{\wt{\va}_m\}_{m=1}^M$ given, the only randomness in each $y_m$ and $\va_m$ are due to $\eta_m$ and $\vdelta_m$. Furthermore as $\{\eta_m\}_{m=1}^M$ and $\{\vdelta_m\}_{m=1}^M$ are independent, the negative conditional log-likelihood is,
\begin{align}
    \sum_{m=1}^M - \ln(\mathrm{Pr}\{\eta_m =& \abs{\inprod{\va_m + \ve_m, \vx}}^2 - y_m \}) \notag \\
    & - \ln(\mathrm{Pr}\{\vdelta_m = \ve_m\}) . \label{eq:neg_cond_log_likelihood}
\end{align}

Using the assumptions on the error distributions,
\begin{align}
    \ln(\mathrm{Pr}\{\eta_m =& \abs{\inprod{\va_m + \ve_m, \vx}}^2 - y_m \}) \notag \\
    =& K_{\eta} -\frac{1}{2\sigma_{\eta}^2} \left( y_m - \abs{\inprod{\va_m + \ve_m, \vx}}^2 \right)^2
\end{align}
and
\begin{align}
    \ln\left(\mathrm{Pr}\left\{\vdelta_m = \ve_m\right\}\right) = K_{\vdelta} - \frac{1}{2 \sigma^2_{\vdelta}} \norm{\ve_m}_2^2 
\end{align}
where $K_{\eta}$ and $K_{\vdelta}$ are constants independent of $\vx$ and $\{\ve_m\}_{m=1}^M$. Substituting these into \eqref{eq:neg_cond_log_likelihood} gives
\begin{align*}
    \argmin_{\vx, \ve_1, \ldots, \ve_M} \sum_{m = 1}^M  \frac{1}{ \sigma^2_{\vdelta}} \norm{\ve_m}_2^2 + \frac{1}{ \sigma^2_{\eta}}\left(y_m -  \abs{\inprod{\va_m + \ve_m, \vx}}^2 \right)^2 .
\end{align*}

\clearpage
\bibliographystyle{IEEEtran}
\bibliography{refs}
\clearpage
\section*{Supplementary material}

\section{Taylor series expansions} \label{sec:taylor_series}

Lemma \ref{lemma:taylor_expansions} states the Taylor series expansions around the no error point $\wt{\vt}$ for the TLS and LS solutions. The notation defined in Section \ref{sec:linearization} is used.

\begin{lemma} \label{lemma:taylor_expansions}

The Taylor series expansions for the solution $\vx_{\mathrm{TLS}}^\dag$ to the TLS optimization problem \eqref{eq:error_analysis_objective}, and, the solution $\vx_{\mathrm{LS}}^\dag$ to the LS optimization problem \eqref{eq:error_analysis_ls_objective} at the no error point $\vt = \wt{\vt}$ with perturbation $\vgamma$ are
\begin{align}
    \vx_{\mathrm{TLS}}^\dag &= \vx^\# 
    +
    \left( \wt{\mA}^\T \wt{\mY} \mD \wt{\mA} \right)^{-1} \wt{\mA}^\T \wt{\mY} \mD \vw + \mO(\norm{\vgamma}_2^2)\\
    \vx_{\mathrm{LS}}^\dag &= \vx^\# 
    +
    \left( \wt{\mA}^\T \wt{\mY} \wt{\mA} \right)^{-1} \wt{\mA}^\T \wt{\mY} \vw + \mO(\norm{\vgamma}_2^2) .
\end{align}
\end{lemma}
\begin{IEEEproof}
To compute the Taylor series expansions we require $\vx^\dag_{\mathrm{TLS}}(\wt{\vt})$ and $\nabla_{\vt}\vx^\dag_{\mathrm{TLS}}(\vt)\big|_{\vt=\wt{\vt}}\, \vgamma$ for the TLS problem \eqref{eq:general_taylor} and the corresponding terms for the LS problem, $\vx^\dag_{\mathrm{LS}}(\wt{\vt})$ and $\nabla_{\vt}\vx^\dag_{\mathrm{LS}}(\vt)\big|_{\vt=\wt{\vt}}\, \vgamma$

In the no error setting, when $\vgamma=0$, $\vx^\dag_{\mathrm{TLS}}(\wt{\vt}) = \vx^\#$. This is because with no error the minimum objective function \eqref{eq:error_analysis_objective} value of zero is achievable with $\ve_m = 0$ for all $m$ and $\vx^\dag_{\mathrm{TLS}}(\vt) = \vx^\#$. Similarly for the LS problem, when $\vgamma=0$, the LS solution, $\vx^\dag_{\mathrm{LS}}(\wt{\vt})$, is also $\vx^\#$, as this achieves the minimum LS objective function \eqref{eq:error_analysis_ls_objective} value of zero.

The full derivations for the derivatives $\nabla_{\vt}\vx^\dag_{\mathrm{TLS}}(\vt)$ and $\nabla_{\vt}\vx^\dag_{\mathrm{LS}}(\vt)$ using \eqref{eq:stacked_derivatives} are contained in Appendices \ref{sec:tls_gradients} and \ref{sec:ls_gradients}. Appendices \ref{sec:tls_taylor_derivation} and \ref{sec:ls_taylor_derivation} then evaluate these derivatives at $\wt{\vt}$ and multiply them by $\vgamma$. We again use the fact that at $\wt{\vt}$ the solutions are $\vx^\#$ and that the TLS sensing vector corrections, $\ve_m$, are zero for all $m$.
\end{IEEEproof}

\subsection{Gradients for TLS problem} \label{sec:tls_gradients}

For convenience, we restate the optimization problem \eqref{eq:error_analysis_objective},
\begin{align}
    \vq^\dag = & \argmin_{\vq}
    \underbrace{\sum_{m = 1}^M \lambda_{a}\norm{\ve_m}_2^2 + \lambda_{y}\left(y_m -  \abs{\inprod{\va_m + \ve_m, \vx}}^2 \right)^2}_{f(\vq, \vt)} \notag \\
    & \quad \text{s.t.} \quad \vq = \begin{bmatrix}\ve_1^\T
    & \cdots & \ve_M^\T & \vx^\T
    \end{bmatrix}^\T \in \R^{MN + N} .
\end{align}

We denote the quantities
\begin{align}
    \wh{\va}_m &= \va_m + \ve_m \in \R^N \\
    d_m &= (\abs{\inprod{\wh{\va}_m, \vx}}^2 - y_m)\inprod{\wh{\va}_m, \vx} \in \R \\
    l_m &= \abs{\inprod{\wh{\va}_m, \vx}}^2 - y_m \in \R \\
    m_m &= 2 \abs{\inprod{\wh{\va}_m, \vx}}^2 \in \R \\
    h_m &= l_m + m_m = 3\abs{\inprod{\wh{\va}_m, \vx}}^2 - y_m \in \R \\
    p_m &= \frac{2\lambda_{y}}{\lambda_{a}} d_m \in \R \\
    \phi_m &= \frac{h_m}{1 + \frac{2\lambda_{y}}{\lambda_{a}} h_m \norm{\vx}_2^2} \in \R
\end{align}
which are used to denote
\begin{align}
    \mB &= \frac{\lambda_{a}}{2\lambda_{y}}\mI_{MN} + \begin{bmatrix} h_1 \vx \vx^\T && \\
    & \ddots & \\
    && h_M \vx \vx^\T\end{bmatrix} \in \R^{MN \times MN} \\
    \mC &= \begin{bmatrix} d_1\mI_N + h_1 \vx (\va_1 + \ve_1)^\T \\
    \vdots \\
    d_M\mI_N + h_M \vx (\va_M + \ve_M)^\T
    \end{bmatrix} \in \R^{MN \times N} \\
    \mT &= \sum_{m=1}^M h_m \wh{\va}_m \wh{\va}_m^\T \in \R^{N \times N} \\
    \wh{\mA} &= (\mA + \mE)^\T \in \R^{N \times M} \\
    \vp &= \begin{bmatrix} p_1 & \cdots p_M \end{bmatrix}^\T \in \R^M \\
    \mPhi &= \diag\left(\phi_1, \ldots, \phi_M \right) \in \R^{M \times M} .
\end{align}

The first derivative of the objective function with respect to $\vq$, $ \nabla_{\vq} f(\vq, \vt) \in \R^{MN +N}$, is
\begin{align}
    \nabla_{\vq} f(\vq, \vt) = 2 \lambda_{a}
    \begin{bmatrix} \ve_1 \\ \vdots \\ \ve_M \\ \vzero \end{bmatrix}
    + 4 \lambda_{y}
    \begin{bmatrix} d_1 \vx \\ \vdots \\ d_M \vx \\ \sum_{m=1}^M d_m \wh{\va}_m \end{bmatrix} .
\end{align}

The second derivative of the objective function with respect to $\vq$, $\nabla^2_{\vq\vq} f(\vq, \vt) \in \R^{(MN+N) \times (MN+N)}$, is
\begin{align}
    &\nabla^2_{\vq\vq} f(\vq, \vt) \notag \\
    =& 2 \lambda_{a} \begin{bmatrix} \mI_N & & & \vdots \\
    & \ddots & & \vzero_{MN\times N} \notag \\
    & &\mI_N & \vdots \\
    \cdots & \vzero_{N\times MN} & \cdots & \vzero_{N\times N}
    \end{bmatrix} \\
    &+
    4 \lambda_{y} \begin{bmatrix} \ddots & & & d_1 \mI_N \\
    & \vzero_{MN \times MN} & & \vdots \\
    & & \ddots &  d_M \mI_N \\
    d_1 \mI_N & \cdots & d_M \mI_N & \vzero_{N\times N}
    \end{bmatrix} \notag \\
    &+
    4 \lambda_{y} \begin{bmatrix} l_1 \vx \vx^\T& & & l_1 \vx \wh{\va}_1^\T \\
    & \ddots & & \vdots \\
    & & l_M \vx \vx^\T &  l_M \vx \wh{\va}_M^\T \\
    l_1 \wh{\va}_1\vx^\T & \cdots & l_M \wh{\va}_M\vx^\T & \sum_{i=1}^M l_i \wh{\va}_i \wh{\va}_i^\T
    \end{bmatrix} \notag \\
    &+
    4 \lambda_{y} \begin{bmatrix} m_1 \vx \vx^\T& & & m_1 \vx \wh{\va}_1^\T \\
    & \ddots & & \vdots \\
    & & m_M \vx \vx^\T &  m_M \vx \wh{\va}_M^\T \\
    m_1 \wh{\va}_1\vx^\T & \cdots & m_M \wh{\va}_M\vx^\T & \sum_{i=1}^M m_i \wh{\va}_i \wh{\va}_i^\T
    \end{bmatrix} \notag \\
    =&
    4 \lambda_{y} \begin{bmatrix} \mB & \mC \\ \mC^\T & \mT \end{bmatrix} . \label{eq:tls_hessian}
\end{align}

The second derivative with respect to $y_k$, $\frac{d}{d y_k} \nabla_{\vq} f(\vq, \vt) \in \R^{MN+N}$, is
\begin{align}
    \frac{d}{d y_k} \nabla_{\vq} f(\vq, \vt) 
    &= -4\lambda_{y} \begin{bmatrix} \vzero_{(k-1)N \times N} \\ 
    \vx \wh{\va}_k^\T \vx \\
    \vzero_{(M-k)N \times N} \\ 
    \wh{\va}_k\wh{\va}_k^\T \vx \end{bmatrix} . \label{eq:second_gradient_wrt_y}
\end{align}

The second derivative with respect to $\va_k$, $\nabla^2_{\va_k \vq} f(\vq, \vt) \in \R^{(MN+N) \times N}$, is
\begin{align}
    & \nabla^2_{\va_k\vq} f(\vq, \vt) \notag \\
    &= 4 \lambda_{y} l_k 
    \begin{bmatrix} \vzero_{(k-1)N \times N} \\
    \vx\vx^\T \\
    \vzero_{(M-k)N \times N} \\ 
    \wh{\va}_k\vx^\T 
    + \inprod{\wh{\va}_k, \vx} \mI_N
    \end{bmatrix}
    +
    4 \lambda_{y} m_k
    \begin{bmatrix} \vzero_{(k-1)N \times N} \\
    \vx\vx^\T \\
    \vzero_{(M-k)N \times N} \\ 
    \wh{\va}_k\vx^\T \end{bmatrix} \notag \\
    &= 4 \lambda_{y}
    \begin{bmatrix} \vzero_{(k-1)N \times N} \\
    h_k \vx\vx^\T \\
    \vzero_{(M-k)N \times N} \\ 
    h_k \wh{\va}_k\vx^\T 
    + d_k \mI_N 
    \end{bmatrix} . \label{eq:second_gradient_wrt_a}
\end{align}

We will require the inverse of the second derivative \eqref{eq:tls_hessian}, $(\nabla^2_{\vq\vq} f(\vq, \vt))^{-1} \in \R^{(MN+N) \times (MN+N)}$, in our calculations \eqref{eq:argmin_wrt_a} \eqref{eq:argmin_wrt_y}. We can use blockwise matrix inversion to invert the block matrix \eqref{eq:tls_hessian},
\begin{align}
    &(\nabla^2_{\vq\vq} f(\vq,\vt))^{-1} \notag \\
    &= \frac{1}{4 \lambda_{y}} \begin{bmatrix} \mB^{-1} + \mQ_{CB}^\T \mQ_S^{-1} \mQ_{CB} & - \mQ_{CB}^\T \mQ_S^{-1} \\ 
    - \mQ_S^{-1}\mQ_{CB} & \mQ_S^{-1} \end{bmatrix},
\end{align}
where
\begin{align}
    \mQ_{CB} &= \mC^\T \mB^{-1} \in \R^{N \times MN} \\
    \mQ_S &= \mT - \mC^\T \mB^{-1} \mC \notag \\
    &= \mT - \mQ_{CB} \mC \in \R^{N \times N}
\end{align}
and $\mQ_S = \mT - \mC^\T \mB^{-1} \mC$ is the Schur complement of $\mB$. Furthermore because $\mB$ is a block diagonal matrix, $\mB^{-1}$ is also block diagonal with each block being the inverse of its counterpart block in $\mB$. Each block in $\mB$ has the same structure and due to this structure the Sherman-Morrison formula can be used to invert each block,
\begin{align}
    \left(\frac{\lambda_{a}}{2\lambda_{y}}\mI_N + h_m \vx \vx^\T \right)^{-1} &= \frac{2\lambda_{y}}{\lambda_{a}}\mI_N - \frac{\frac{4\lambda_{y}^2}{\lambda_{a}^2} h_m \vx \vx^\T}{1 + \frac{2\lambda_{y}}{\lambda_{a}} h_m \norm{\vx}_2^2}
\end{align}
and $\left(\mB^{-1}\right)^\T = \mB^{-1}$.

As we wish to understand the sensitivity of $\vx^\dag_{\mathrm{TLS}}$ \eqref{eq:last_rows_for_x} we only require the final $N$ rows of the inverse of \eqref{eq:tls_hessian}. More precisely we will only require the submatrix
\begin{align}
    &(\nabla^2_{\vq\vq} f(\vq,\vt))^{-1}_{-N} \notag \\
    &=
    \frac{1}{4 \lambda_{y}} \mQ_S^{-1} \begin{bmatrix} 
    - \mQ_{CB} & \mI_N 
    \end{bmatrix} \in \R^{N \times (MN + N)} . \label{eq:rows_of_hessian}
\end{align}


To calculate \eqref{eq:rows_of_hessian} we require $\mQ_{CB}$ which is a block matrix with $M$ matrices horizontally stacked. The $m$th block is
\begin{align}
    & (d_m\mI_N + h_m \wh{\va}_m \vx^\T ) 
    \left( \frac{2\lambda_{y}}{\lambda_{a}}\mI_N - \frac{4\lambda_{y}^2}{\lambda_{a}^2} \phi_m \vx \vx^\T \right) \notag \\
    &=
    p_m \mI_N 
    - \frac{2\lambda_{y}}{\lambda_{a}} \phi_m p_m \vx \vx^\T \notag \\
    & \:\:\:\:\: + \frac{2\lambda_{y}}{\lambda_{a}} \phi_m \left(\frac{h_m}{\phi_m} -  \frac{2\lambda_{y}}{\lambda_{a}} h_m \norm{\vx}_2^2 \right) \wh{\va}_m \vx^\T \notag \\
    &= p_m \mI_N + \frac{2\lambda_{y}}{\lambda_{a}} \phi_m (\wh{\va}_m - p_m \vx )\vx^\T . \label{eq:C^TBinverse}
\end{align}

To obtain $\mQ_S$ in \eqref{eq:rows_of_hessian} we can use \eqref{eq:C^TBinverse} and the block matrix structure of $\mC$ to calculate $\mC^\T \mB^{-1} \mC = \mQ_{CB} \mC$, 
\begin{align}
    &\mC^\T \mB^{-1} \mC = \mQ_{CB} \mC \notag \\
    &= \sum_{m=1}^M \left( p_m \mI_N + \frac{2\lambda_{y}}{\lambda_{a}} \phi_m (\wh{\va}_m - p_m \vx )\vx^\T \right) \notag \\
    & \qquad \:\:\:\:\: \left( d_m\mI_N + h_m \vx \wh{\va}_m^\T \right) \notag \\
    &= \sum_{m=1}^M \frac{\lambda_{a}}{2\lambda_{y}} p_m^2 \mI_N + p_m \phi_m \wh{\va}_m \vx^\T - p_m^2 \phi_m  \vx \vx^\T \notag \\
    & \qquad \:\:\:\:\:  + p_m \phi_m \left(\frac{h_m}{\phi_m} - \frac{2\lambda_{y}}{\lambda_{a}} h_m \norm{\vx}_2^2 \right) \vx \wh{\va}_m^\T \notag \\
    & \qquad \:\:\:\:\: + \frac{2\lambda_{y}}{\lambda_{a}} h_m \phi_m \norm{\vx}_2^2 \wh{\va}_m \wh{\va}_m^\T \notag \\
    &= \frac{\lambda_{a}}{2\lambda_{y}} \norm{\vp}^2 \mI_N + \wh{\mA} \mPhi \vp \vx^\T - \vx \vp^\T \mPhi \vp \vx^\T + \vx \vp^\T \mPhi \wh{\mA}^\T \notag \\
    & \:\:\:\:\: + \sum_{m=1}^M \frac{2\lambda_{y}}{\lambda_{a}} h_m \phi_m \norm{\vx}_2^2 \wh{\va}_m \wh{\va}_m^\T .
\end{align}

Then the Schur complement of $\mB$ is
\begin{align}
    & \mQ_S = \mT - \mC^\T \mB^{-1} \mC \notag \\
    &= - \frac{\lambda_{a}}{2\lambda_{y}} \norm{\vp}^2 \mI_N - \wh{\mA} \mPhi \vp \vx^\T + \vx \vp^\T \mPhi \vp \vx^\T - \vx \vp^\T \mPhi \wh{\mA}^\T \notag \\
    & \:\:\:\:\: + \sum_{m=1}^M \phi_m \left(\frac{h_m}{\phi_m} -  \frac{2\lambda_{y}}{\lambda_{a}} h_m \norm{\vx}_2^2 \right) \wh{\va}_m \wh{\va}_m^\T \notag \\
    &= - \frac{\lambda_{a}}{2\lambda_{y}} \norm{\vp}^2 \mI_N - \wh{\mA} \mPhi \vp \vx^\T + \vx \vp^\T \mPhi \vp \vx^\T - \vx \vp^\T \mPhi \wh{\mA}^\T \notag \\
    &+ \wh{\mA} \mPhi \wh{\mA}^\T \notag \\
    &= - \frac{\lambda_{a}}{2\lambda_{y}} \norm{\vp}_2^2 \mI_N
    + (\wh{\mA} - \vx \vp^\T) \mPhi (\wh{\mA} - \vx \vp^\T)^\T .
    \label{eq:schur_complement}
\end{align}

Using \eqref{eq:rows_of_hessian} with \eqref{eq:schur_complement} and \eqref{eq:C^TBinverse} we can compute the last $N$ rows of \eqref{eq:argmin_wrt_y}, $\frac{d}{d y_k}\vx^\dag_{\mathrm{TLS}}(\vt) \in \R^N$. First,
\begin{align}
    &-\frac{1}{4 \lambda_{y}}
    \begin{bmatrix} - \mQ_{CB} & \mI_N \end{bmatrix}
    \frac{d}{d y_k} \nabla_{\vq} f(\vq, \vt) \notag \\
    &= \begin{bmatrix} -p_k \mI_N - \frac{2\lambda_{y}}{\lambda_{a}} \phi_k (\wh{\va}_k - p_k \vx )\vx^\T & \mI \end{bmatrix} 
    \begin{bmatrix} \vx \wh{\va}_k^\T \vx \\
    \wh{\va}_k \wh{\va}_k^\T \vx \end{bmatrix} \notag \\
    &= \wh{\va}_k^\T \vx \begin{bmatrix} -p_k \mI_N - \frac{2\lambda_{y}}{\lambda_{a}} \phi_k (\wh{\va}_k - p_k \vx )\vx^\T & \mI \end{bmatrix} 
    \begin{bmatrix} \vx \\
    \wh{\va}_k \end{bmatrix} \notag \\
    &= \wh{\va}_k^\T \vx  \left(-p_k\vx - \frac{2\lambda_{y}}{\lambda_{a}} \phi_k (\wh{\va}_k - p_k \vx )\norm{\vx}_2^2 + \wh{\va}_k \right) \notag \\
    &= \wh{\va}_k^\T \vx (\wh{\va}_k - p_k\vx) \frac{\phi_k}{h_k},
\end{align}
and therefore,
\begin{align}
    &\frac{d}{d y_k}\vx^\dag_{\mathrm{TLS}}(\vt) \notag \\
    &=
     \left(  - \frac{\lambda_{a}}{2\lambda_{y}} \norm{\vp}_2^2 \mI_N
    + (\wh{\mA} - \vx \vp^\T) \mPhi (\wh{\mA} - \vx \vp^\T)^\T \right)^{-1} \notag \\
    & \:\:\:\:\: \wh{\va}_k^\T \vx (\wh{\va}_k - p_k\vx) \frac{\phi_k}{h_k} .
    \label{eq:tls_measurement_sensitivity}
\end{align}

Similarly using \eqref{eq:rows_of_hessian} with \eqref{eq:schur_complement} and \eqref{eq:C^TBinverse} we can compute the last $N$ rows of \eqref{eq:argmin_wrt_a}, $\nabla_{\va_k} \vx^\dag_{\mathrm{TLS}}(\vt) \in \R^{N \times N}$. Again first,
\begin{align}
    &-\frac{1}{4 \lambda_{y}}
    \begin{bmatrix} - \mQ_{CB} & \mI_N \end{bmatrix}
    \nabla^2_{\va_k\vq} f(\vq, \vt) \notag \\
    &=
    - \begin{bmatrix} -p_k \mI_N - \frac{2\lambda_{y}}{\lambda_{a}} \phi_k (\wh{\va}_k - p_k \vx )\vx^\T & \mI \end{bmatrix} \notag \\
    & \:\:\:\:\:\:\:\:\:\: \begin{bmatrix} h_k \vx\vx^\T \\
    h_k \wh{\va}_k \vx^\T 
    + \frac{\lambda_{a}}{2 \lambda_{y}}p_k \mI_N
    \end{bmatrix} \notag \\
    &=
    - \left( -p_k h_k \vx\vx^\T - \frac{2\lambda_{y}}{\lambda_{a}} h_k \phi_k \norm{\vx}_2^2 (\wh{\va}_k - p_k \vx )\vx^\T \right. \notag \\
    & \:\:\:\:\:\:\:\:\:\: \left. +
    h_k \wh{\va}_k \vx^\T 
    + \frac{\lambda_{a}}{2 \lambda_{y}}p_k \mI_N
    \right) \notag \\
    &=
    - \left( \frac{\lambda_{a}}{2 \lambda_{y}}p_k \mI_N + h_k( \wh{\va}_k -p_k \vx )\vx^\T \right. \notag \\
    & \:\:\:\:\:\:\:\:\:\: \left.- \frac{2\lambda_{y}}{\lambda_{a}} h_k \phi_k \norm{\vx}_2^2 (\wh{\va}_k - p_k \vx )\vx^\T
    \right) \notag \\
    &=
    - \left( \frac{\lambda_{a}}{2 \lambda_{y}}p_k \mI_N \right. \notag \\
    & \:\:\:\:\:\:\:\:\:\: \left. + \phi_k \left( \frac{h_k}{\phi_k} - \frac{2\lambda_{y}}{\lambda_{a}} h_k \norm{\vx}_2^2 \right) (\wh{\va}_k - p_k \vx )\vx^\T
    \right) \notag \\
    &=
    - \left( \frac{\lambda_{a}}{2 \lambda_{y}}p_k \mI_N + \phi_k (\wh{\va}_k - p_k \vx )\vx^\T
    \right) ,
\end{align}
and therefore,
\begin{align}
    & \nabla_{\va_k} \vx^\dag_{\mathrm{TLS}}(\vt) \notag \\
    &=
    - \left(  - \frac{\lambda_{a}}{2\lambda_{y}} \norm{\vp}_2^2 \mI_N
    + (\wh{\mA} - \vx \vp^\T) \mPhi (\wh{\mA} - \vx \vp^\T)^\T \right)^{-1} \notag \\
    & \:\:\:\:\:\:\:\:\:\: \left( \frac{\lambda_{a}}{2 \lambda_{y}}p_k \mI_N + \phi_k (\wh{\va}_k - p_k \vx )\vx^\T
    \right) .
    \label{eq:tls_vector_sensitivity}
\end{align}

\subsection{Gradients for LS problem}  \label{sec:ls_gradients}

We restate the optimization problem \eqref{eq:error_analysis_ls_objective}
\begin{align}
    \vx^\dag_{\mathrm{LS}}(\vt) &= \argmin_{\vx} \underbrace{\sum_{m = 1}^M \left(y_m -  \abs{\inprod{\va_m, \vx}}^2 \right)^2}_{s(\vx, \vt)} .
\end{align}

We denote similar quantities to those in the TLS derivation. The main differences are that there are no $\ve_m$, $\lambda_{y}$ and $\lambda_{a}$ in the LS approach,
\begin{align}
    d_m^- &= (\abs{\inprod{\va_m, \vx}}^2 - y_m)\inprod{\va_m, \vx} \in \R \\
    h_m^- &= 3\abs{\inprod{\va_m, \vx}}^2 - y_m \in \R \\
    \mA^- &= \mA^\T \in \R^{N \times M} \\
    \vp^- &= 2 \begin{bmatrix} d_1^-, \ldots, d_M^- \end{bmatrix}^\T \in \R^M \\
    \mPhi^- &= \diag\left(h_1^-, \ldots, h_M^- \right) \in \R^{M \times M}.
\end{align}

To derive $\frac{d}{d y_k}\vx^\dag_{\mathrm{LS}}(\vt) \in \R^N$ and $\nabla_{\va_k} \vx^\dag_{\mathrm{LS}}(\vt) \in \R^{N \times N}$ for LS, the expressions that were derived for TLS can be used. Set $\{\ve_m\}_{m=1}^M =0$, $\lambda_{y} = \lambda_{a} = 1$ in the TLS expressions \eqref{eq:tls_hessian}, \eqref{eq:second_gradient_wrt_y} and \eqref{eq:second_gradient_wrt_a}. Then take the bottom right $N \times N$ block of \eqref{eq:tls_hessian} and the bottom $N$ rows of \eqref{eq:second_gradient_wrt_y} and \eqref{eq:second_gradient_wrt_a} to get for LS
\begin{align}
    \nabla^2_{\vx\vx} s(\vx, \vt) &= 4 \sum_{i=1}^M h^-_m \va_m\va_m^\T \notag \\
    &= 4 \mA^- \mPhi^- (\mA^-)^\T \in \R^{N \times N} \\
    \frac{d}{d y_k} \nabla_{\vx} s(\vx, \vt) &= -4 
    \va_k \va_k^\T \vx = -4\va^-_k (\va^-_k)^\T \vx \in \R^N \\
    \nabla^2_{\va_k\vx} s(\vx, \vt) &= 
    4 (h^-_k \va_k\vx^\T + d^-_k \mI_N) \notag \\
    &= 4\left(\frac{1}{2}p^-_k \mI_N + \phi^-_k \va^-_k \vx^\T \right) \in \R^{N \times N}.
\end{align}

Therefore
\begin{align}
    \frac{d}{d y_k}\vx^\dag_{\mathrm{LS}}(\vt) =& \left(\mA^- \mPhi^- (\mA^-)^\T\right)^{-1} (\va^-_k)^\T \vx \va^-_k \label{eq:ls_measurement_sensitivity} \\
    \nabla_{\va_k} \vx^\dag_{\mathrm{LS}}(\vt) =& - \left(\mA^- \mPhi^- (\mA^-)^\T\right)^{-1} \notag \\
    & \:\:\:\: \left(\frac{1}{2}p^-_k \mI_N + \phi^-_k \va^-_k \vx^\T \right) . \label{eq:ls_vector_sensitivity}
\end{align}

\subsection{Derivation of TLS solution Taylor series expansion}
\label{sec:tls_taylor_derivation}

To calculate $\nabla_{\vt}\vx^\dag_{\mathrm{TLS}}(\vt)\big|_{\vt=\wt{\vt}}$ we need the last $N$ rows of $\nabla_{\va_k} g(\vt)\big|_{\vt=\wt{\vt}}$ and $\frac{d}{d y_k}g(\vt)\big|_{\vt=\wt{\vt}}$ for $1 \leq k \leq M$ as in \eqref{eq:stacked_derivatives}. With $\vt = \wt{\vt}$,  $\{\ve_m\}_{m=1}^M = 0$ and $\vx = \vx^\#$, the quantities defined when deriving the gradients in Appendix \ref{sec:tls_gradients} become
\begin{align}
    \wh{\va}_m &= \va_m + \ve_m  = \wt{\va}_m \in \R^N \\
    d_m &= (\abs{\inprod{\wh{\va}_m, \vx}}^2 - y_m)\inprod{\wh{\va}_m, \vx} = 0 \in \R \\
    l_m &= \abs{\inprod{\wh{\va}_m, \vx}}^2 - y_m = 0 \in \R \\
    m_m &= 2 \abs{\inprod{\wh{\va}_m, \vx}}^2 = 2 \wt{y}_m \in \R \\
    h_m &= l_m + m_m = 3\abs{\inprod{\wh{\va}_m, \vx}}^2 - y_m = 2 \wt{y}_m \in \R \\
    p_m &= \frac{2\lambda_{y}}{\lambda_{a}} d_m = 0 \in \R \\
    \phi_m &= \frac{2 \wt{y}_m}{1 + \frac{4\lambda_{y}}{\lambda_{a}} \wt{y}_m \norm{\vx^\#}_2^2} \in \R \\
    \wh{\mA} &= (\mA + \mE)^\T = \wt{\mA}^\T \in \R^{N \times M} \\
    \vp &= \begin{bmatrix} p_1 & \cdots p_M \end{bmatrix}^\T = \begin{bmatrix} \vzero & \cdots \vzero \end{bmatrix}^\T \in \R^M \\
    \mPhi &= \diag\left(\phi_1, \ldots, \phi_M \right) 
    = 2 \wt{\mY} \mD \in \R^{M \times M} .
\end{align}

Using these quantities with \eqref{eq:tls_measurement_sensitivity} and \eqref{eq:tls_vector_sensitivity} in Appendix \ref{sec:tls_gradients},
\begin{align}
    \nabla_{\va_k} \vx^\dag_{\mathrm{TLS}}(\vt)\big|_{\vt=\wt{\vt}} 
    &= 
    - \left( \wt{\mA}^\T \mPhi \wt{\mA} \right)^{-1} \left( \phi_k \wt{\va}_k (\vx^\#)^\T
    \right)
    \\ 
    \frac{d}{d y_k} \vx^\dag_{\mathrm{TLS}}(\vt)\big|_{\vt=\wt{\vt}}
    &= 
    \left( \wt{\mA}^\T \mPhi \wt{\mA} \right)^{-1} \wt{\va}_k^\T \vx^\# \wt{\va}_k  \frac{\phi_k}{h_k},
\end{align}

Therefore 
\begin{align}
    &\nabla_{\vt} \vx^\dag_{\mathrm{TLS}}(\vt)\big|_{\vt=\wt{\vt}} \, \vgamma \notag \\
    &= \left( \wt{\mA}^\T \mPhi \wt{\mA} \right)^{-1} \notag \\
    &\hphantom{=} \left( 
    \sum_{m=1}^M - \phi_m \wt{\va}_m (\vx^\#)^\T \vdelta_m + \wt{\va}_m^\T \vx^\# \wt{\va}_m  \frac{\phi_m}{h_m} \eta_m
    \right) \notag \\
    &= \left( \wt{\mA}^\T \mPhi \wt{\mA} \right)^{-1} \left( 
    \sum_{m=1}^M \frac{\phi_m}{h_m} \eta_m \wt{\va}_m  \wt{\va}_m^\T - \phi_m \wt{\va}_m \vdelta_m^\T
    \right) \vx^\#  \notag \\
    &= \left( \wt{\mA}^\T \mPhi \wt{\mA} \right)^{-1} \left( 
    \wt{\mA}^\T \mPhi (2 \wt{\mY})^{-1} \mE_{\mY} \wt{\mA} - \wt{\mA}^\T \mPhi \mE_{\mA}
    \right) \vx^\#  \notag \\
    &= \left( \wt{\mA}^\T \mPhi \wt{\mA} \right)^{-1} \wt{\mA}^\T \mPhi \left( 
    (2 \wt{\mY})^{-1} \mE_{\mY} \wt{\mA} - \mE_{\mA}
    \right) \vx^\#  \notag \\
    &= \left( \wt{\mA}^\T 2 \wt{\mY} \mD \wt{\mA} \right)^{-1} \wt{\mA}^\T 2 \wt{\mY} \mD 
    \left( 
    (2 \wt{\mY})^{-1} \mE_{\mY} \wt{\mA} - \mE_{\mA}
    \right) \vx^\#  \notag \\
    &= \left( \wt{\mA}^\T \wt{\mY} \mD \wt{\mA} \right)^{-1} \wt{\mA}^\T \wt{\mY} \mD \vw .
    %
    %
\end{align}

\subsection{Derivation of LS solution Taylor series expansion}
\label{sec:ls_taylor_derivation}

Following the same procedure as in Appendix \ref{sec:tls_taylor_derivation} and using \eqref{eq:ls_measurement_sensitivity} and \eqref{eq:ls_vector_sensitivity} in Appendix \ref{sec:ls_gradients},
\begin{align}
    \nabla_{\va_k} \vx^\dag_{\mathrm{LS}}(\vt)\big|_{\vt=\wt{\vt}} 
    &= 
    - \left( \wt{\mA}^\T 2 \wt{\mY} \wt{\mA} \right)^{-1} \left( 2\wt{y}_k \wt{\va}_k (\vx^\#)^\T
    \right)
    \\ 
    \frac{d}{d y_k} \vx^\dag_{\mathrm{LS}}(\vt)\big|_{\vt=\wt{\vt}}
    &= 
    \left( \wt{\mA}^\T 2 \wt{\mY} \wt{\mA} \right)^{-1} \wt{\va}_k^\T \vx^\# \wt{\va}_k.
\end{align}

Therefore for LS
\begin{align}
    &\nabla_{\vt} \vx^\dag_{\mathrm{LS}}(\vt)\big|_{\vt=\wt{\vt}} \, \vgamma \notag \\
    &= \left( \wt{\mA}^\T 2 \wt{\mY} \wt{\mA} \right)^{-1} \left( 
    \sum_{m=1}^M \eta_m \wt{\va}_m  \wt{\va}_m^\T - 2\wt{y}_m \wt{\va}_m \vdelta_m^\T
    \right) \vx^\#  \notag \\
    &= \left( \wt{\mA}^\T 2 \wt{\mY} \wt{\mA} \right)^{-1} \left( 
    \wt{\mA}^\T 2 \wt{\mY} (2 \wt{\mY})^{-1} \mE_{\mY} \wt{\mA} - \wt{\mA}^\T 2 \wt{\mY} \mE_{\mA}
    \right) \vx^\#  \notag \\
    &= \left( \wt{\mA}^\T 2 \wt{\mY} \wt{\mA} \right)^{-1} \wt{\mA}^\T 2 \wt{\mY} \left( 
    (2 \wt{\mY})^{-1} \mE_{\mY} \wt{\mA} - \mE_{\mA}
    \right) \vx^\#  \notag \\
    &= \left( \wt{\mA}^\T \wt{\mY} \wt{\mA} \right)^{-1} \wt{\mA}^\T \wt{\mY} \vw .
    %
    %
\end{align}

\rev{
\section{Proof of Proposition \ref{proposition:expected_squared_error}}
\label{sec:expected_squared_error_derivation}

We begin by noting that $e_{\mathrm{TLS}}^2$ and $e_{\mathrm{LS}}^2$ can both be written in the form $e^2 = (\mR \vw)^\T (\mR \vw)$ where $\mR \in \R^{N \times M}$ is $\left( \wt{\mA}^\T \wt{\mY} \mD \wt{\mA} \right)^{-1} \wt{\mA}^\T \wt{\mY} \mD$ and $\left( \wt{\mA}^\T \wt{\mY} \wt{\mA} \right)^{-1} \wt{\mA}^\T \wt{\mY}$ for TLS and LS. The vector $\vw \in \R^M$ is as defined in Proposition \ref{proposition:linearized_error} and contains all the random quantities. Let $r_{ij}$ be the $i,j$th entry of $\mR$ and let $w_i$ be the $i$th entry of $\vw$. Then
\begin{align}
    e^2 &=
    \begin{bmatrix}
    \sum_{i=1}^M r_{1, i} w_i
    & \cdots &
    \sum_{i=1}^M r_{N, i} w_i
    \end{bmatrix}
    \begin{bmatrix}
    \sum_{j=1}^M r_{1, j} w_j \\
    \vdots \\
    \sum_{j=1}^M r_{N, j} w_j
    \end{bmatrix} \\
    &=
    \sum_{i=1}^M \sum_{j=1}^M r_{1, i} r_{1, j} w_i w_j
    + \ldots +
    \sum_{i=1}^M \sum_{j=1}^M r_{N, i} r_{N, j} w_i w_j .
\end{align}

Further, $w_i = \dfrac{\eta_i}{2\wt{y}_i} \inprod{\wt{\va}_i, \vx^\#} - \inprod{\vdelta_i, \vx^\#}$ and so all the entries of $\vw$ are independent of each other. As a result, $\mathbb{E}[w_i w_j] = 0$ if $i \neq j$ and
\begin{align}
    \mathbb{W}_i := \mathbb{E}[w_i^2]
    &=
    \mathbb{E}[\eta_i^2]\frac{\inprod{\wt{\va}_i, \vx^\#}^2}{4 \wt{y}_i^2} + (\vx^\#)^\T \mathbb{E}[\vdelta_i \vdelta_i^\T] \vx^\# \\
    &=
    \frac{\sigma^2_\eta}{4 \wt{y}_i} + \sigma^2_{\vdelta} \norm{\vx^\#}_2^2 .
\end{align}

Denoting $\vr_m$ as the $m$th column of $\mR$ and using $\mathbb{W}_i$, 
\begin{align}
    \mathbb{E}[e^2] &= 
    \sum_{i=1}^M r_{1, i}^2 \mathbb{W}_i
    + \ldots +
    \sum_{i=1}^M r_{N, i}^2 \mathbb{W}_i \\
    &=
    \sum_{i=1}^M \mathbb{W}_i (r_{1, i}^2 
    + \ldots +
    r_{N, i}^2)\\
    &=
    \sum_{i=1}^M \mathbb{W}_i \norm{\vr_i}_2^2 \\
    &=
    \sigma_{\vdelta}^2 \norm{\vx^\#}_2^2 \sum_{i=1}^M \norm{\vr_i}_2^2
    +
    \frac{\sigma_\eta^2}{4} \sum_{i=1}^M \frac{1}{\wt{y}_i} \norm{\vr_i}_2^2 \\
    &=
    \sigma_{\vdelta}^2 \norm{\vx^\#}_2^2 \norm{\mR}_F^2
    +
    \frac{\sigma_\eta^2}{4} \norm{\mR \wt{\mY}^{-\frac{1}{2}}}_F^2 .
\end{align}

The result in Proposition \ref{proposition:expected_squared_error} then follows by substituting the TLS and LS values for $\mR$. We also use the fact that the matrix multiplication of $\wt{\mY}$ and $\mD$ in the TLS expression is commutative because both matrices are diagonal.
}

\section{Experimental verification of Proposition \ref{proposition:linearized_error}}
\label{sec:taylor_verification}

We verify the derived reconstruction errors in Proposition \ref{proposition:linearized_error}. As the results are for a first-order approximation, we expect their accuracy to reduce as $\norm{\vgamma}$ increases. To vary $\norm{\vgamma}$ we vary the sensing vector SNR and pin the measurement SNR to be twice the sensing vector SNR. For each SNR combination we perform 100 trials. In each trial we generate new real-valued ground truth signals, Gaussian sensing vectors and Gaussian errors for sensing vectors and measurements. The ground truth signals are iid standard real Gaussian with $N=100$ and $\frac{M}{N} = 8$.

\begin{figure}[t]
    \centering
    \includegraphics[width=0.75\linewidth]{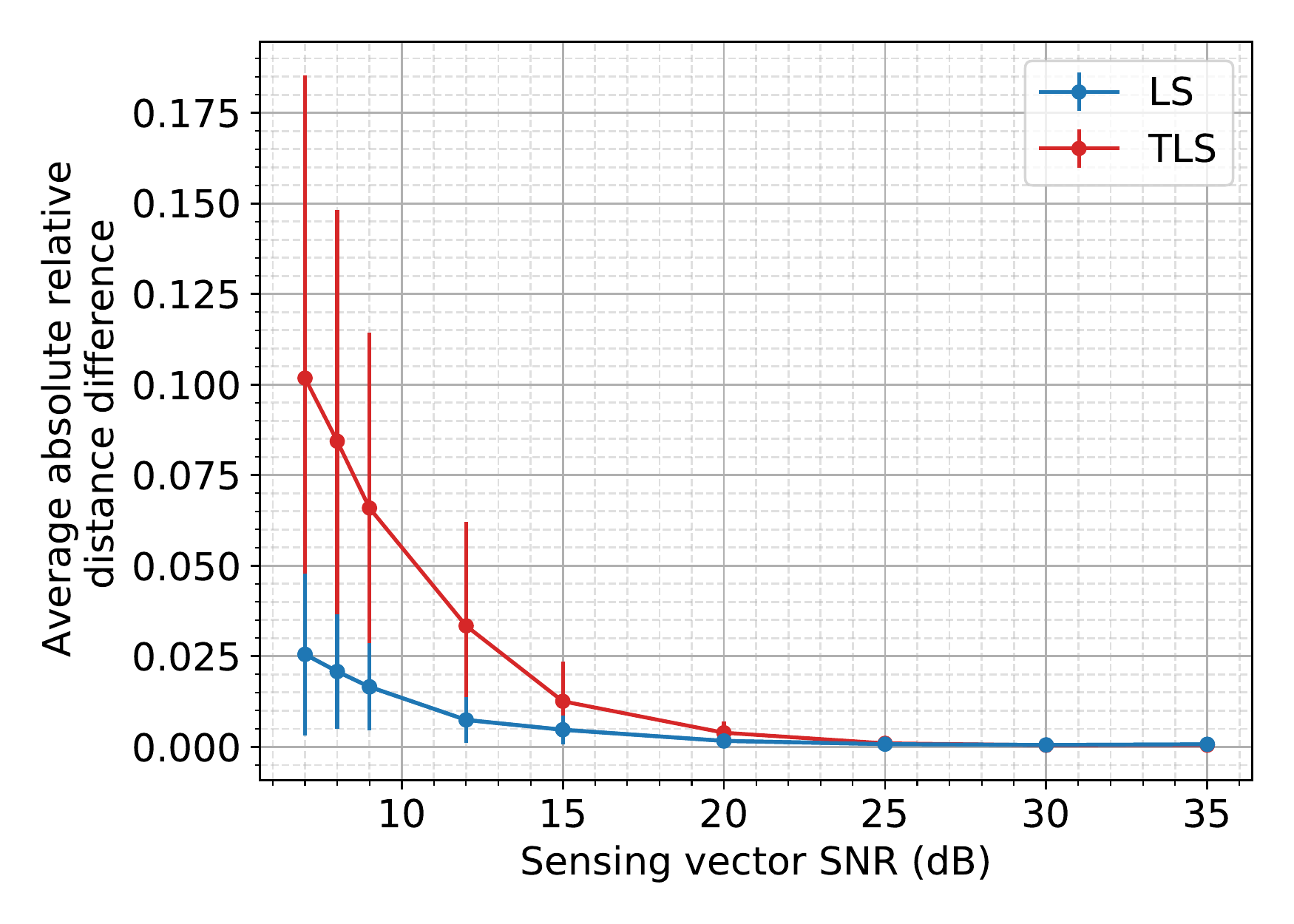}
    \caption{The average of the absolute value difference between the actual relative distance and the relative error reconstruction from Proposition \ref{proposition:linearized_error} when $\frac{M}{N} = 8$. Measurement SNR is twice the sensing vector SNR.}
    \label{fig:linearization_accuracy}
\end{figure}

We plot the average of the absolute difference between the relative distance from the solution of Algorithm \ref{algo:tls_pr_algorithm} and the relative reconstruction error, $\abs{\mathrm{rel.dist}(\vx^\#, \vx^\dag_{\mathrm{LS}}) - \text{rel.}e_{\mathrm{LS}}}$ for LS and $\abs{\mathrm{rel.dist}(\vx^\#, \vx^\dag_{\mathrm{TLS}}) - \text{rel.}e_{\mathrm{TLS}}}$ for TLS in Fig. \ref{fig:linearization_accuracy}. The step sizes are $\frac{1.0}{\lambda_{a}}$ for TLS and $0.05$ for LS. We set $\lambda_{a} = \frac{1}{N}$ and $\lambda_{y} = \frac{1}{\norm{\vx^{(0)}}_2^4}$. As expected the first-order approximations are accurate for high SNR and decrease in accuracy with decreasing SNR. The high accuracy for the moderate to high SNR values also confirms that Algorithm \ref{algo:tls_pr_algorithm} can optimize \eqref{eq:quadratic_TLS_geometric_2}.

\rev{

\section{Handcrafted errors}
\label{sec:handcrafted_errors}

Using the results and notation of Section \ref{sec:linearization}, we show that there exist error models which can significantly change the relative performance of TLS and LS. With scalars $k_a, \, k_y \in \R^+$ to control the SNR of $\mE_{\mA}$ and $\mE_{\mY}$, we create errors $\mE_{\mA} = k_a\mD^{-1}\mE_{\mA}'$ for some $\mE_{\mA}'$ and $\mE_{\mY} = k_y\mD^{-1}\mE_{\mY}'$ for some diagonal $\mE_{\mY}'$. With these errors, the expressions from Proposition \ref{proposition:linearized_error} become
\begin{align}
    e_{\mathrm{TLS}} &= \norm{\left( \wt{\mA}^\T \wt{\mY} \mD \wt{\mA} \right)^{-1} \wt{\mA}^\T \wt{\mY} \vw'}_2 \label{eq:tls_linearized_error_handcrafted} \\
    e_{\mathrm{LS}} &= \norm{\left( \wt{\mA}^\T \wt{\mY} \wt{\mA} \right)^{-1} \wt{\mA}^\T \wt{\mY} \mD^{-1} \vw'}_2  \label{eq:ls_linearized_error_handcrafted}
\end{align}
where $\vw' = ((2\wt{\mY})^{-1} k_y \mE_{\mY}' \wt{\mA} - k_a \mE_{\mA}') \vx^\# $. Compared to \eqref{eq:tls_linearized_error}, $e_{\mathrm{TLS}}$ in \eqref{eq:tls_linearized_error_handcrafted} does not multiply $\vw'$ by $\mD$. Additionally, compared to \eqref{eq:ls_linearized_error}, $e_{\mathrm{LS}}$ in \eqref{eq:ls_linearized_error_handcrafted} multiplies $\vw'$ by $\mD^{-1}$. The elements of diagonal matrix $\mD^{-1}$ are greater than one, $d^{-1}_{mm} =  \left( 1 + \frac{4 \lambda_{y}}{\lambda_{a}}\norm{\vx^\#}_2^2 \wt{y}_m \right) > 1$, and we investigate how this alters performance when the sensing vectors follow the  iid standard real Gaussian measurement model. Appendix \ref{sec:cdp_experiments_appendix} contains experiments using the coded diffraction pattern model.



\subsection{First-order reconstruction error numerical experiments}

In the next set of experiments, $\mE_{\mA} = k_a\mD_1^{-1}\mE_{\mA}'$ and $\mE_{\mY} = k_y\mD_1^{-1}\mE_{\mY}'$, where $\mD_1 = ( \mI_M + 4 \norm{\vx^\#}_2^2 \wt{\mY} )^{-1}$ is free of the regularization parameters. Matrix $\mE_{\mA}'$ and diagonal matrix $\mE_{\mY}'$ are iid zero-mean Gaussian.
We repeat the experiment of Fig. \ref{fig:linearization_varying_m} using these created errors in Fig. \ref{fig:linearization_varying_m_handcrafted}. All other experimental details are unchanged. We see that now TLS outperforms LS with this error model.

\begin{figure}[t]
    \centering
    \includegraphics[width=0.75\linewidth]{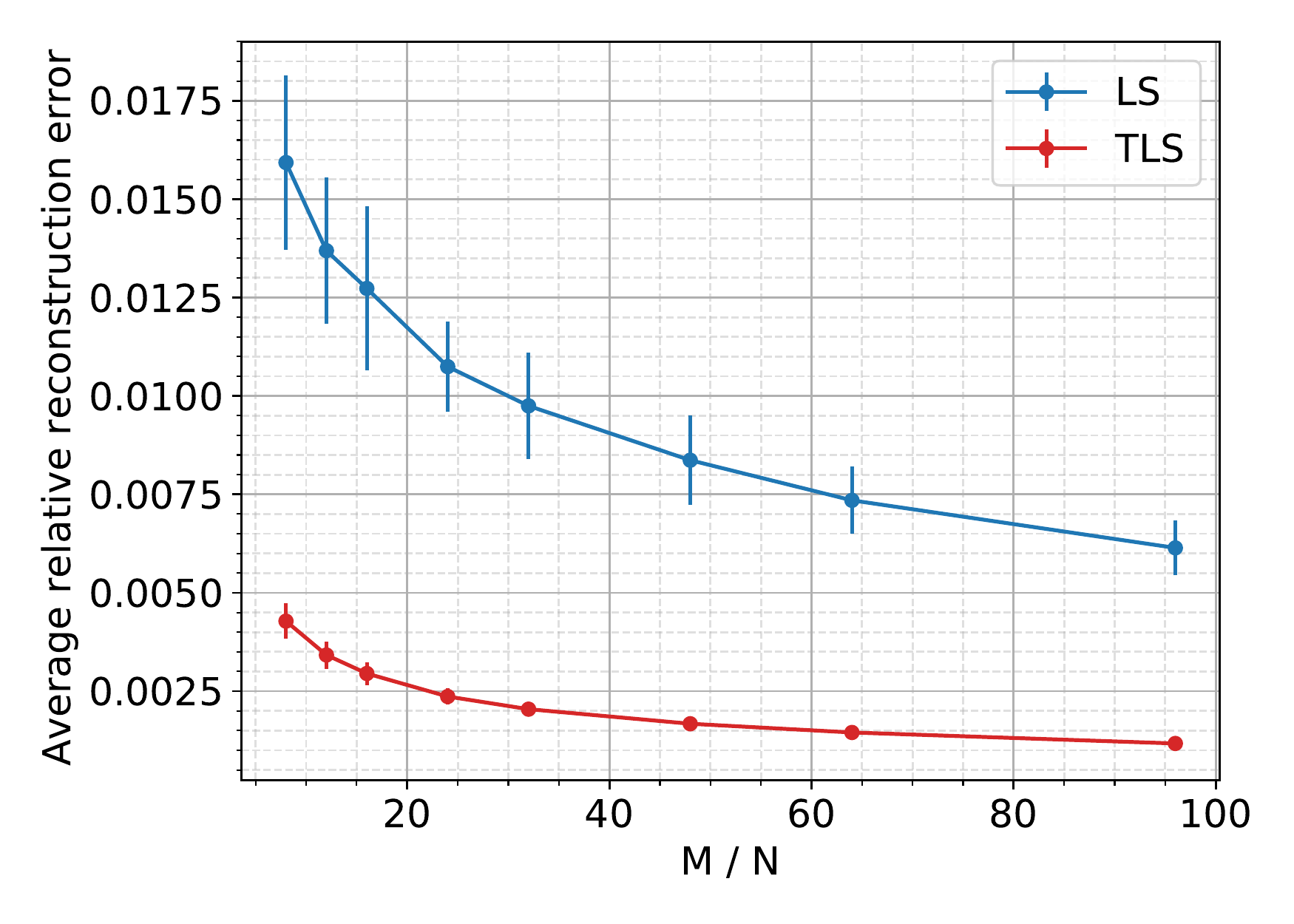}
    \caption{Relative reconstruction errors, \eqref{eq:tls_linearized_error} and \eqref{eq:ls_linearized_error} for different values of $\frac{M}{N}$ when handcrafted errors are used. Measurement and sensing vector SNRs are 40 dB.}
    \label{fig:linearization_varying_m_handcrafted}
\end{figure}

Next we fix $\frac{M}{N} = 8$ and the sensing vector SNR to 100 dB so there is virtually no sensing vector error. We vary the measurement SNR over 100 trials and use the handcrafted errors. The sensing vectors and ground truth signals are generated in the same way as in the numerical experiments of Section \ref{sec:linearization}. In Fig. \ref{fig:linearization_target_y} we plot the average first-order relative reconstruction error and see that with this setting TLS outperforms LS. This occurs despite there only being measurement error, a setting where we may expect LS to outperform TLS.

The results in Figs. \ref{fig:linearization_varying_m_handcrafted} and \ref{fig:linearization_target_y} show that the type of measurement and sensing vector error can impact performance.

\begin{figure}[t]
    \centering
    \includegraphics[width=0.75\linewidth]{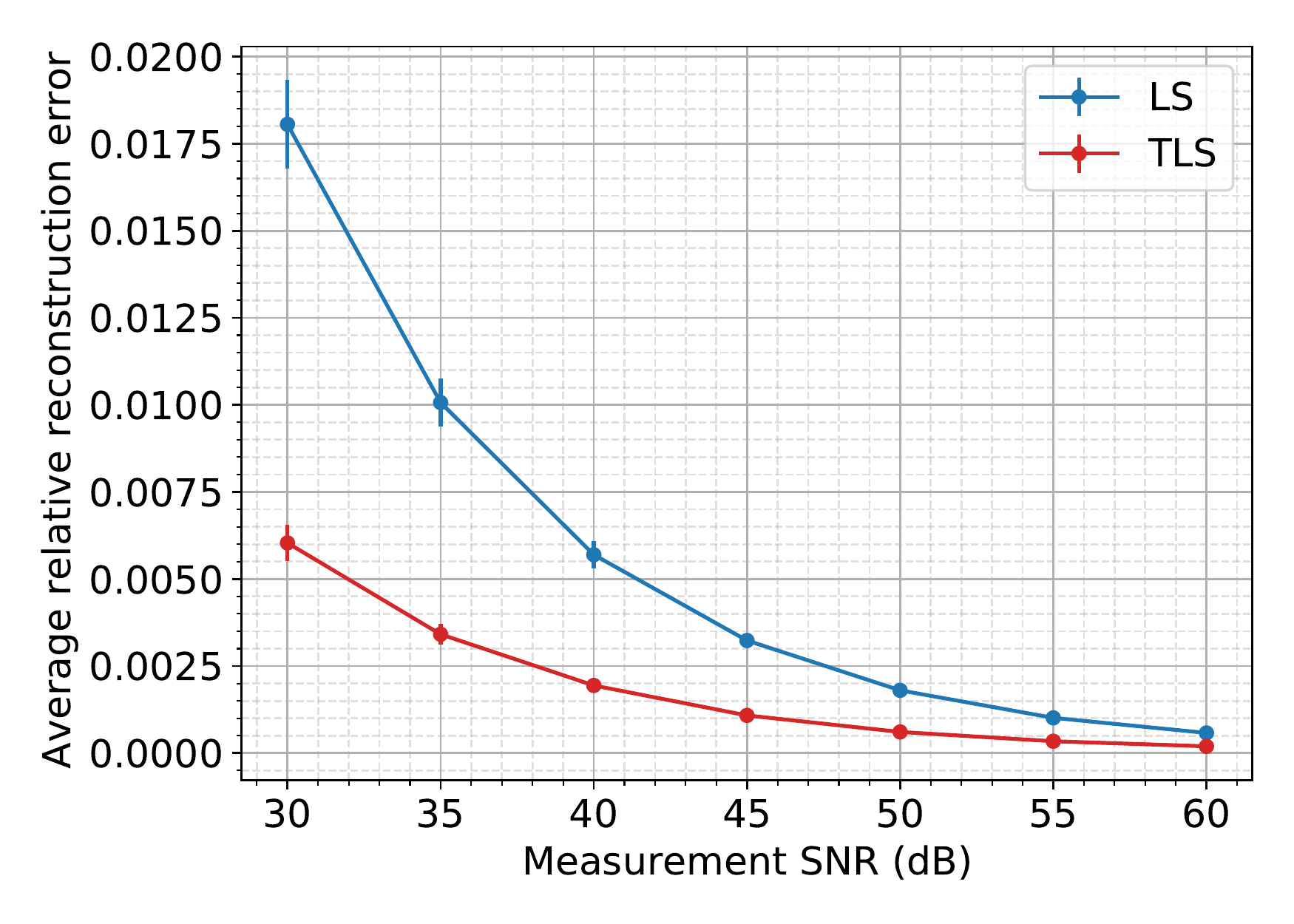}
    \caption{Relative reconstruction errors with handcrafted errors when there is only error in measurements. Here $\frac{M}{N} = 8$.}
    \label{fig:linearization_target_y}
\end{figure}

\subsection{Simulations with actual reconstruction error}

To investigate the impact of this error model on the actual reconstruction error, we design handcrafted errors in the same manner as above and calculate the actual reconstruction error. Despite the earlier analysis using real-valued errors, we show that the ideas carry through when we use the complex Gaussian measurement model with $\mE_{\mA}$ being complex Gaussian.

In the first simulation we use a step size of $\frac{0.5}{\lambda_{a}}$ for TLS and 0.02 for LS and repeat the experiment of Fig. \ref{fig:gaussian_varymn_largerA_SNR} with handcrafted errors instead. Fig. \ref{fig:gaussian_varymn_handcrafted} shows that in this setting TLS outperforms LS, the opposite of what is seen with Gaussian errors in Fig. \ref{fig:gaussian_varymn_largerA_SNR}. This is consistent with the first-order reconstruction error numerical experiment of Fig. \ref{fig:linearization_varying_m_handcrafted}.

\begin{figure}[t]
    \centering
    \includegraphics[width=0.75\linewidth]{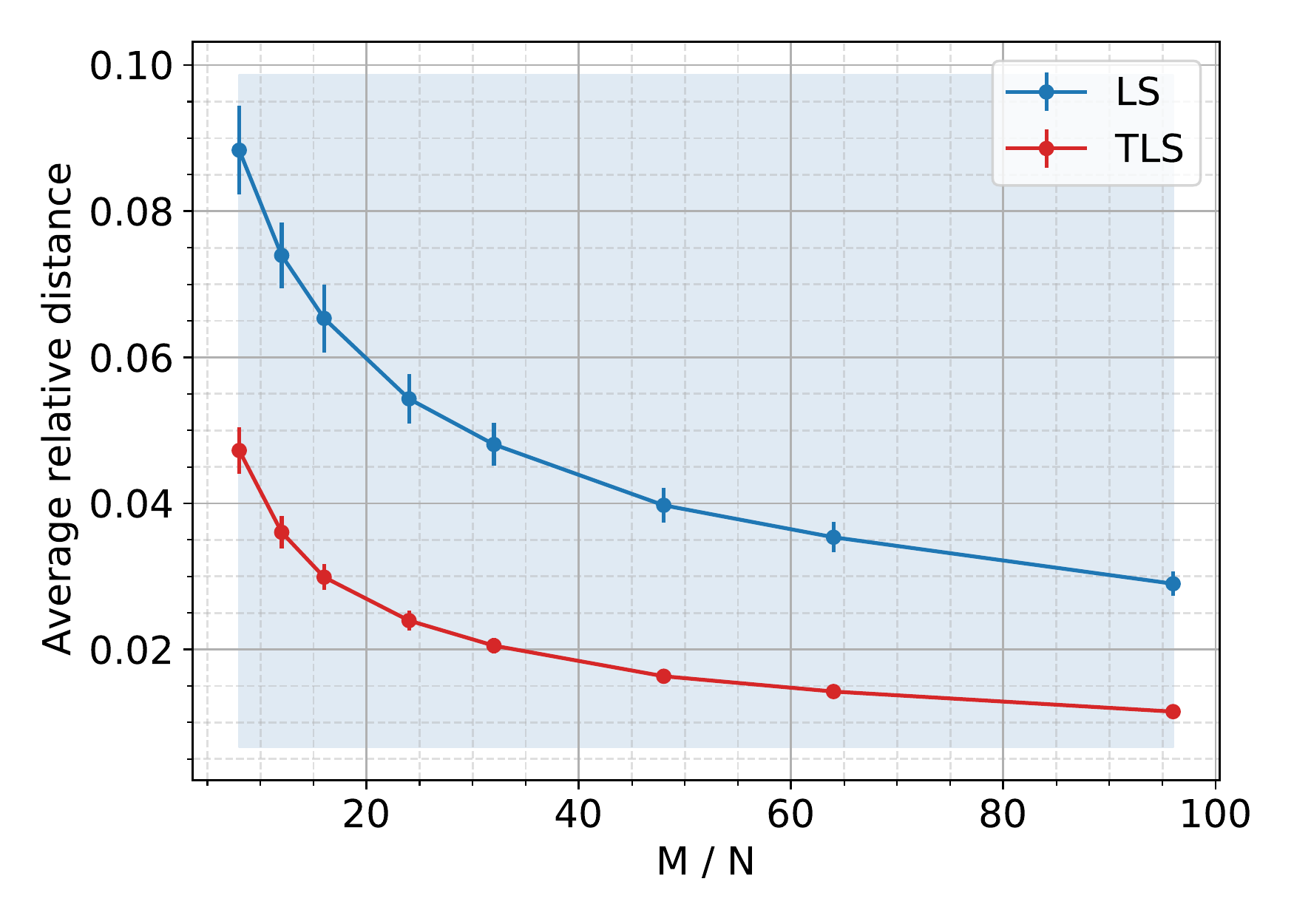}
    \caption{Relative distance of reconstructions using TLS and LS for the Gaussian measurement model for different values of $\frac{M}{N}$ when handcrafted errors are used. Measurement SNR is 20 dB and sensing vector SNR is 30 dB.}
    \label{fig:gaussian_varymn_handcrafted}
\end{figure}

Next we use $\frac{M}{N} = 8$ and a step size of $\frac{0.2}{\lambda_{a}}$ for TLS and 0.02 for LS. Following the experiment of Fig. \ref{fig:linearization_target_y}, in Fig. \ref{fig:y_targeted} the sensing vector SNR is 100 dB and there is virtually no sensing vector error. The measurement SNR is varied and the performance of TLS and LS with handcrafted errors is compared to TLS and LS with iid Gaussian errors. We do 100 trials at each measurement SNR. Even though there is significant error only in the measurements, TLS with handcrafted errors outperforms LS as was suggested by Fig. \ref{fig:linearization_target_y}. With Gaussian errors, LS outperforms TLS when there are only measurement errors. 

\begin{figure}[t]
    \centering
    \includegraphics[width=0.75\linewidth]{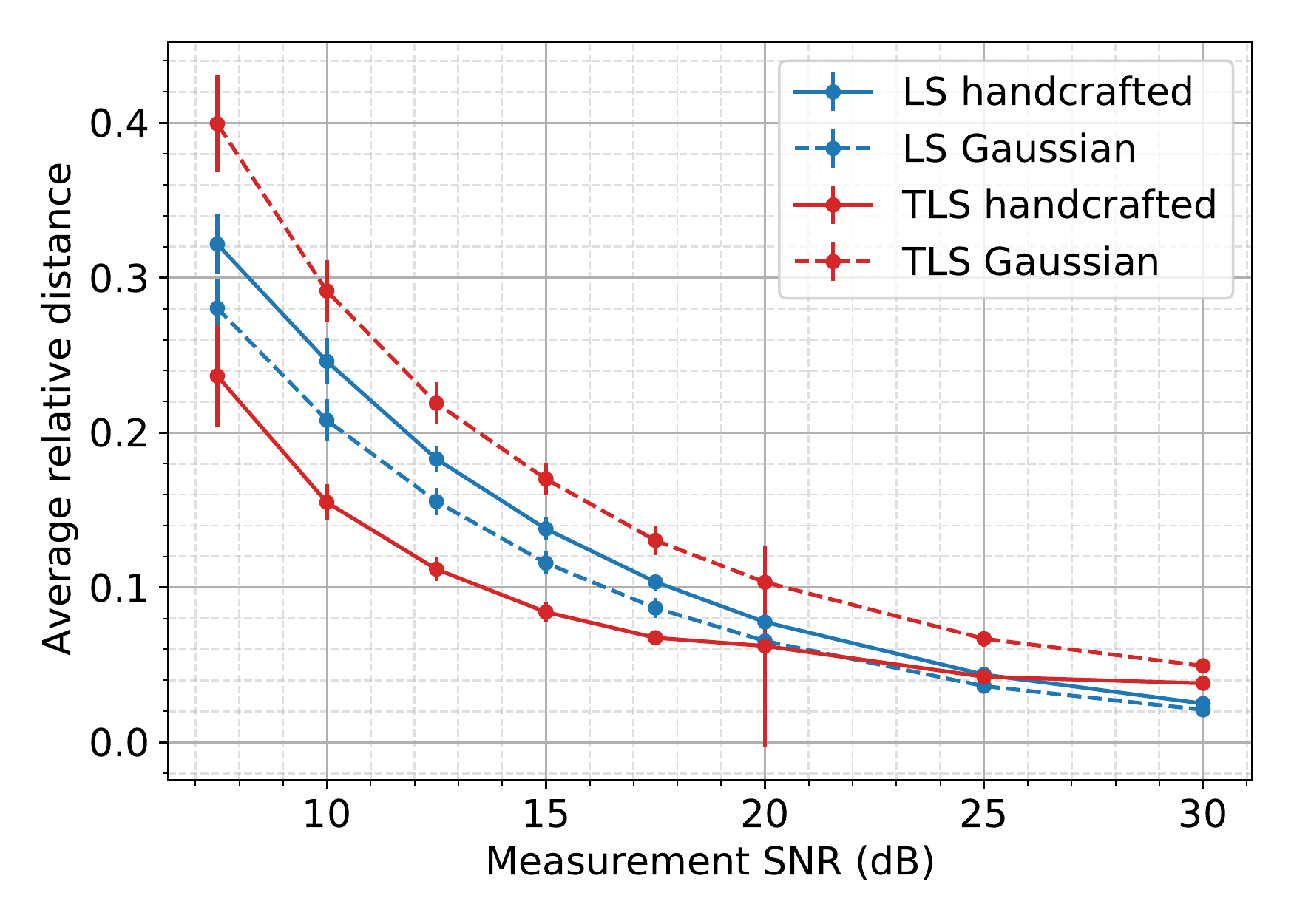}
    \caption{Performance of TLS and LS with handcrafted errors when there is only error in measurements for the Gaussian measurement model. Here $\frac{M}{N} = 8$.}
    \label{fig:y_targeted}
\end{figure}

With $\frac{M}{N} = 8$ and handcrafted errors, Fig. \ref{fig:targeted_mn8} shows an identical experiment to that of Fig. \ref{fig:gaussian_model_random_perturbations}. We see that the relative performance of TLS over LS improves with handcrafted errors compared to Fig. \ref{fig:gaussian_model_random_perturbations} where random Gaussian errors were used. 

\begin{figure}[t]
    \centering
    \includegraphics[width=0.75\linewidth]{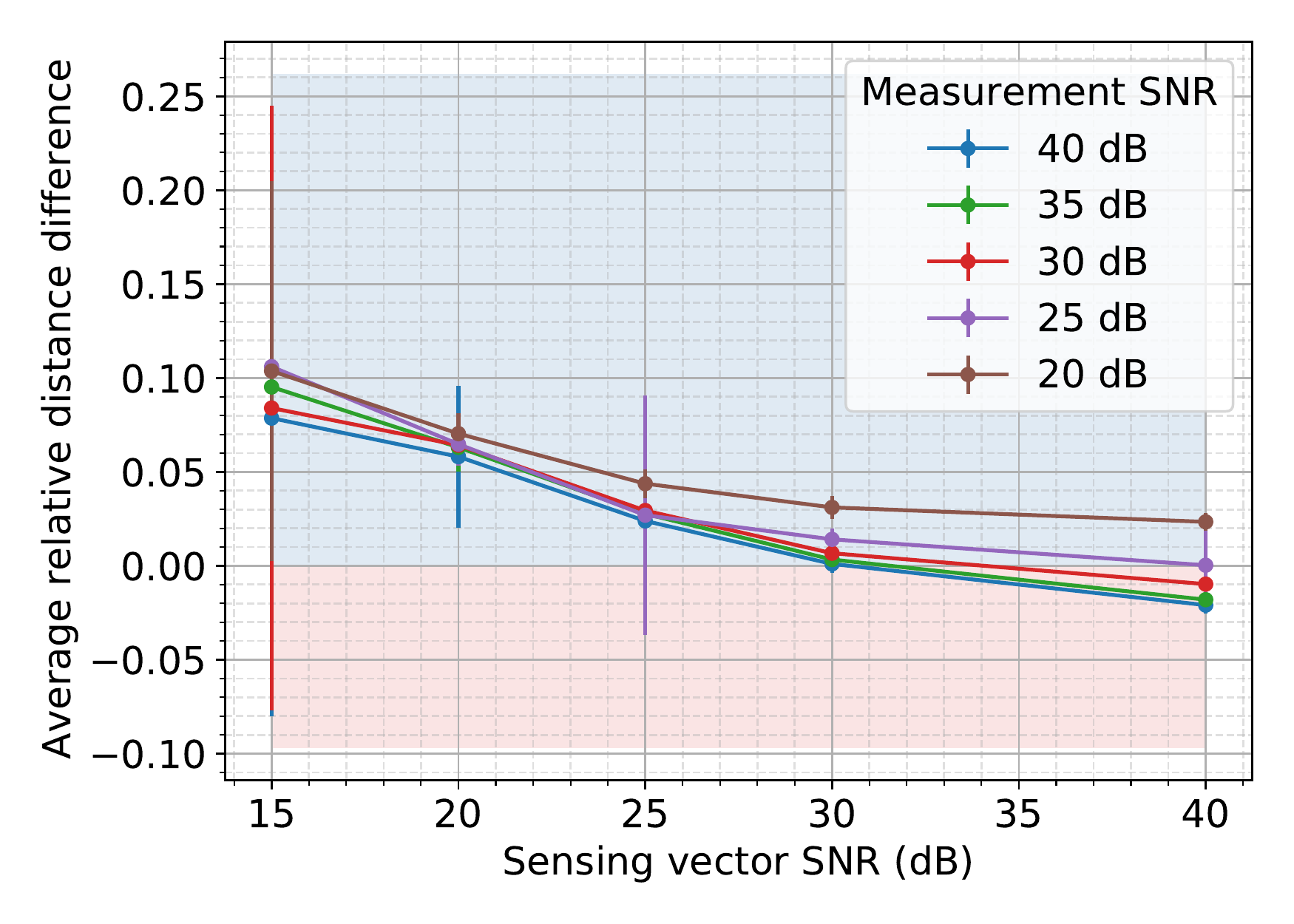}
    \caption{Average difference in relative distance between TLS and LS solutions for the Gaussian measurement model when $\frac{M}{N} = 8$ for different measurement and sensing vector SNR combinations when the errors are handcrafted. This can be compared to Fig. \ref{fig:gaussian_model_random_perturbations} when $\frac{M}{N} = 8$.}
    \label{fig:targeted_mn8}
\end{figure}




}

\section{Coded diffraction pattern (CDP) measurement model} \label{sec:cdp_appendix}

Denoting row $n$ of the $N$-point DFT matrix as $\vf_n^*$ and $L$ modulation patterns $\{\vp_l\}_{l=1}^L \in \C^N$, the $M = LN$ quadratic coded diffraction pattern measurements are then
\begin{align}
    y_m \approx |\underbrace{\vf_n^* \diag(\vp_l)^* }_{\va_m^*} \vx |^2, \qquad m = (n, l)
\end{align}
where $1 \leq n \leq N$, $1 \leq l \leq L$ \cite{candes2015phasecdp}. The modulation patterns $\vp_l \in \C^N$ follow the octanary pattern which means its entries are iid and follow the distribution of $p$ where $p = q_1 q_2$. The random variable $q_1$ is one of $\{-1, 1, -j, j\}$ with equal probability and $q_2 = \frac{\sqrt{2}}{2}$ with probability 0.8 or $q_2 = \sqrt{3}$ with probability 0.2. Note that $M$ can only be an integer multiple of $N$ and depends on the number of patterns used.



\subsection{Experiments} \label{sec:cdp_experiments_appendix}

In this section we repeat the experiments that were done for the Gaussian measurement model in Section \ref{sec:simulations} for the CDP measurement model. The experimental setup such as the number of trials, type of ground truth signal, step sizes and iteration stopping criteria are the same as those used for the equivalent simulation with the Gaussian model.

\paragraph{Random errors}

\begin{figure}[!t]
    \centering
    \subfloat[$L = \frac{M}{N} = 16$]{\includegraphics[width=0.75\linewidth]
        {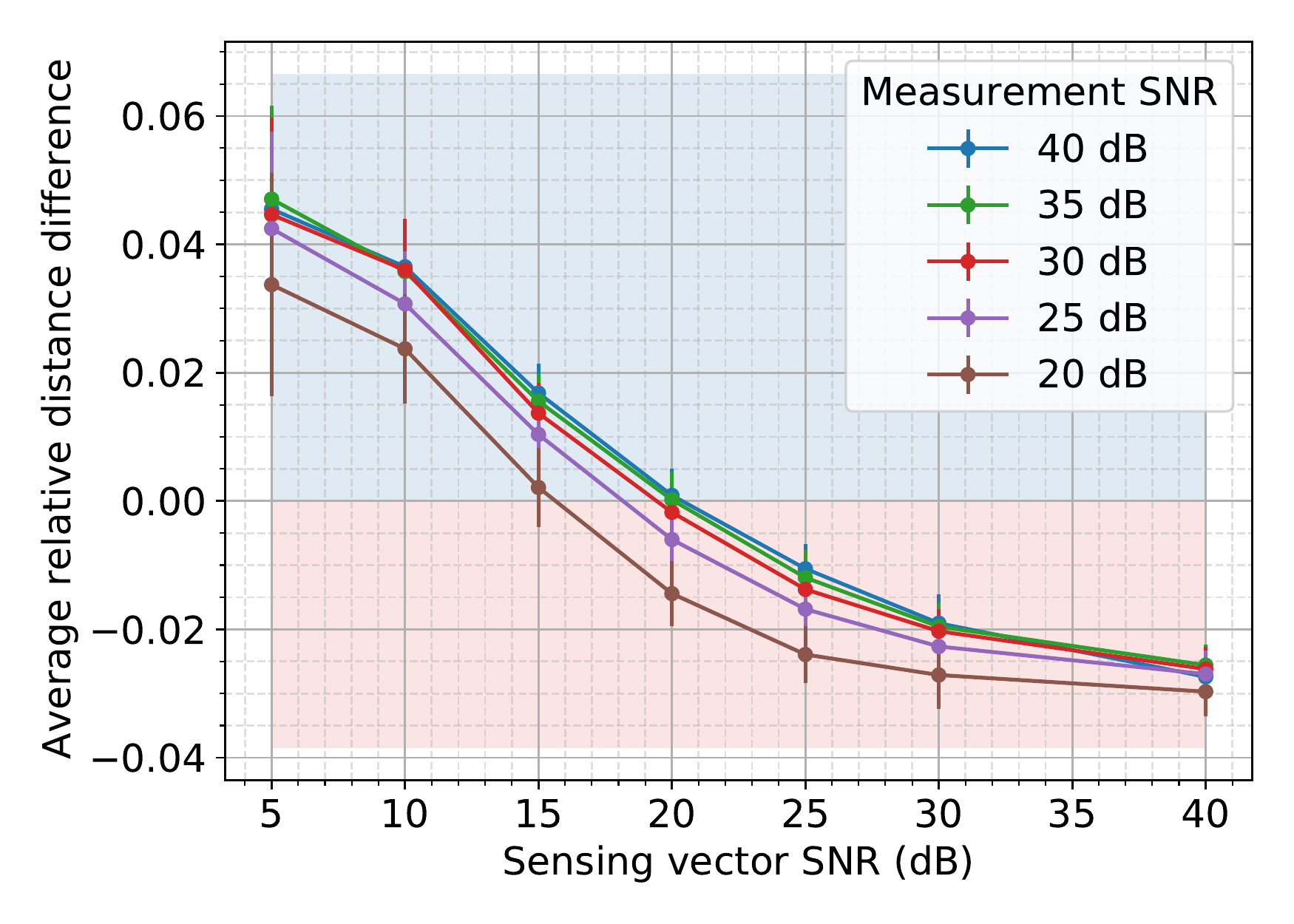}}

    \subfloat[$L = \frac{M}{N} = 32$]{\includegraphics[width=0.75\linewidth]
        {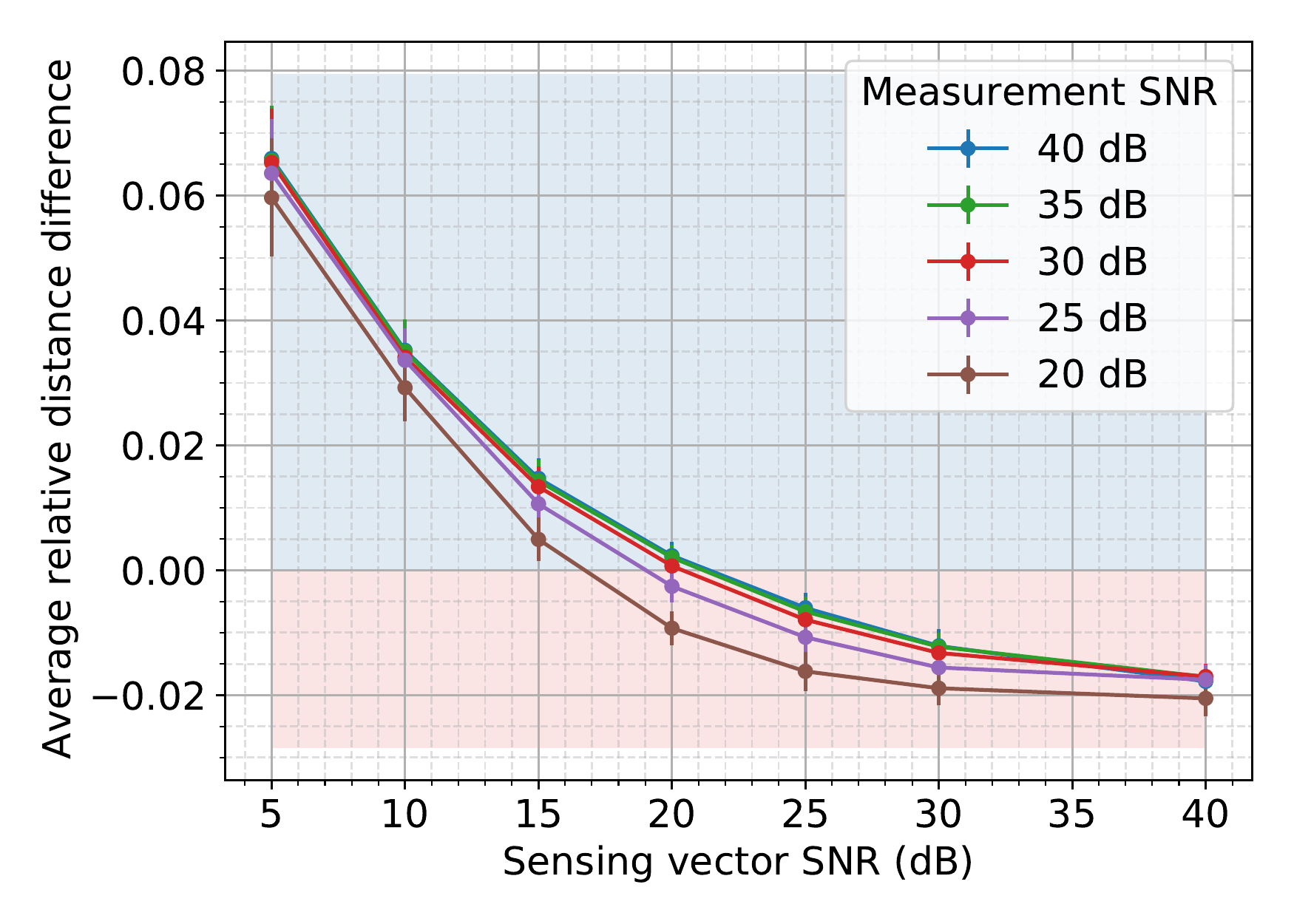}}
    
    \caption{Average difference in relative distance of TLS and LS solutions, $\mathrm{rel.dist}(\vx^\#, \vx^\dag_{\mathrm{LS}}) - \mathrm{rel.dist}(\vx^\#, \vx^\dag_{\mathrm{TLS}})$, for the octanary CDP measurement model for different measurement and sensing vector SNR combinations when the number of patterns is $L = \frac{M}{N} \in \{16, 32\}$.}
    \label{fig:cdp_model_random_perturbations}
\end{figure}

\begin{figure}[!t]
    \centering
    \subfloat[Sensing vector SNR is 10 dB. Gaussian errors. \label{fig:cdp_varymn}]
    {\includegraphics[width=0.75\linewidth]
        {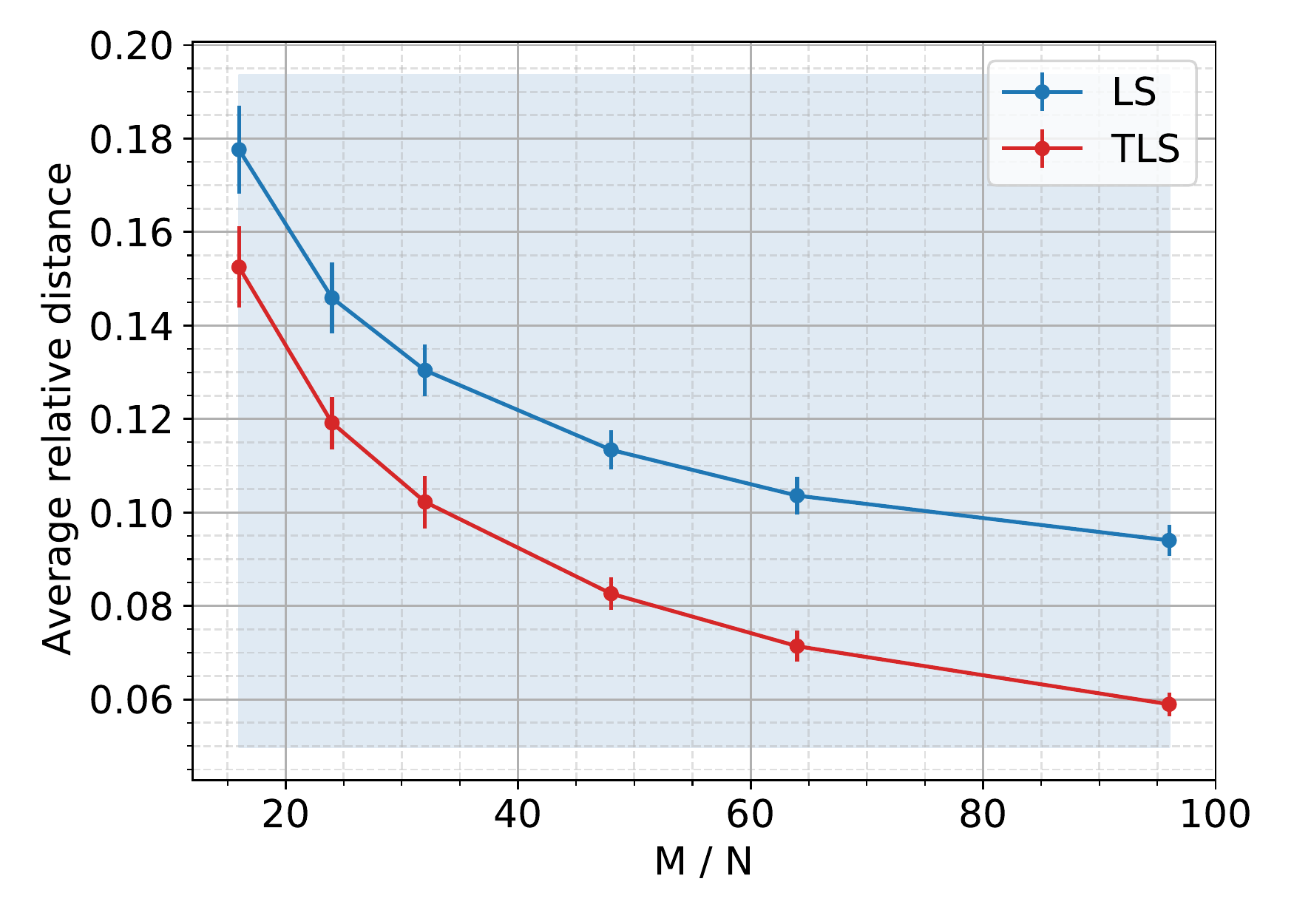}}
    
    \subfloat[Sensing vector SNR is 30 dB. Gaussian errors. \label{fig:cdp_varymn_higher_A_snr}]
    {\includegraphics[width=0.75\linewidth]
        {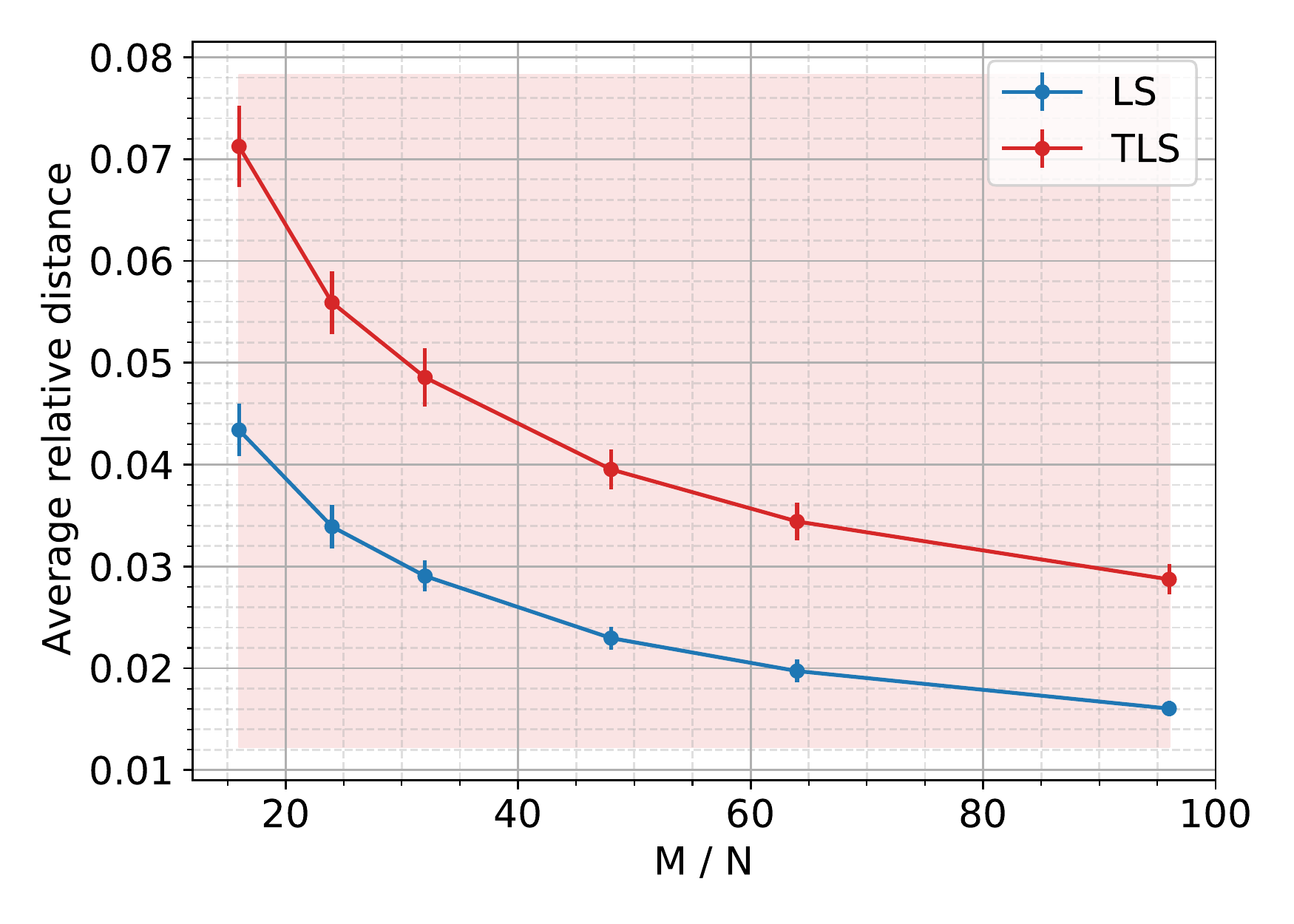}}
    
    \subfloat[Sensing vector SNR is 30 dB. Handcrafted errors. \label{fig:cdp_varymn_handcrafted}]
    {\includegraphics[width=0.75\linewidth]
        {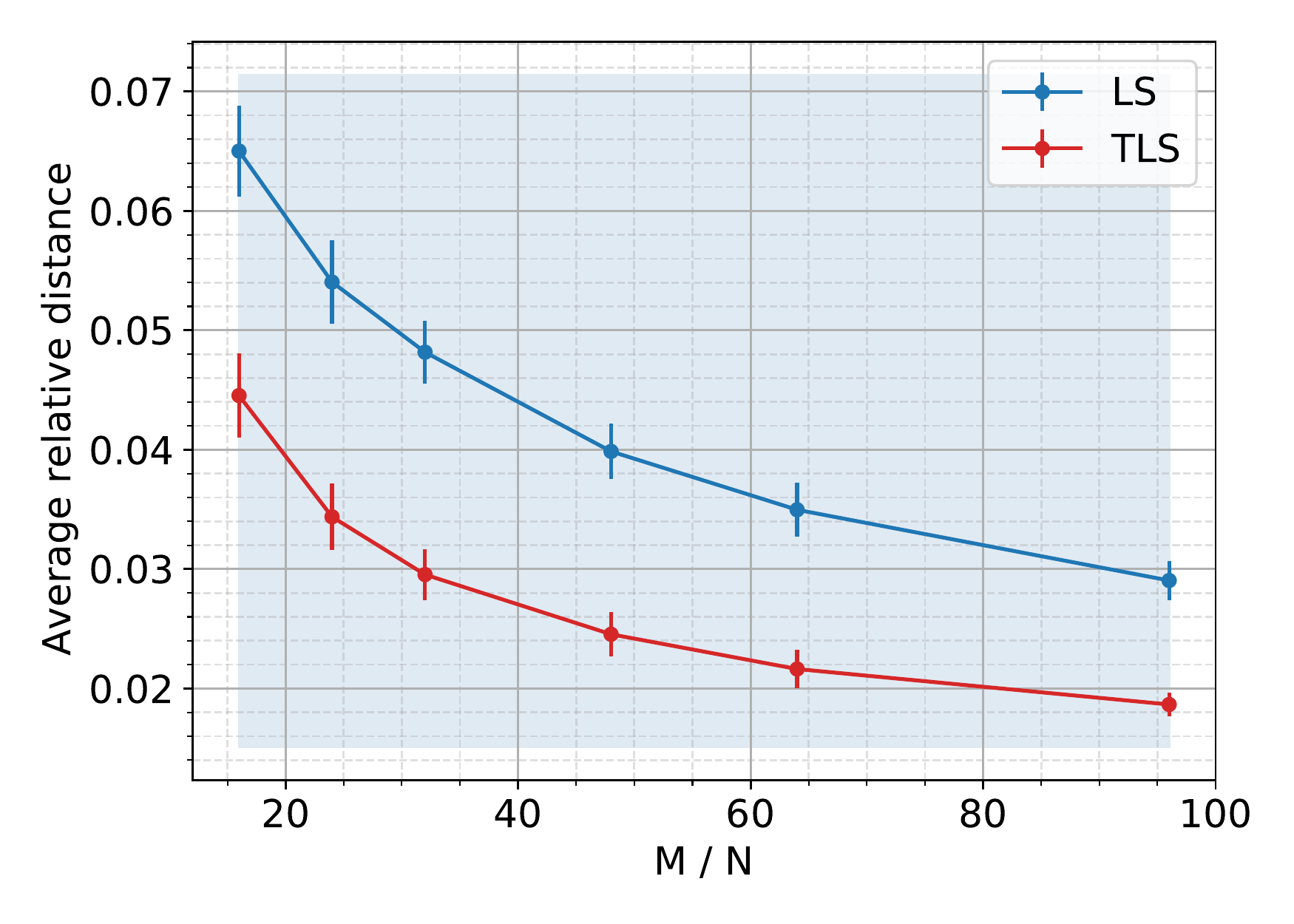}}
    
    \caption{Relative distance of reconstructions using TLS and LS for the CDP measurement model for different \rev{number of patterns $L = \frac{M}{N}$} when measurement SNR is 20 dB.}
\end{figure}

The experiments in Figs. \ref{fig:cdp_model_random_perturbations}, \ref{fig:cdp_varymn}, \ref{fig:cdp_varymn_higher_A_snr} are for the CDP measurement model and are the same as Figs. \ref{fig:gaussian_model_random_perturbations}, \ref{fig:gaussian_varymn} and \ref{fig:gaussian_varymn_largerA_SNR} for the Gaussian measurement model.

\paragraph{Handcrafted errors}

\begin{figure}[t]
    \centering
    \includegraphics[width=0.75\linewidth]{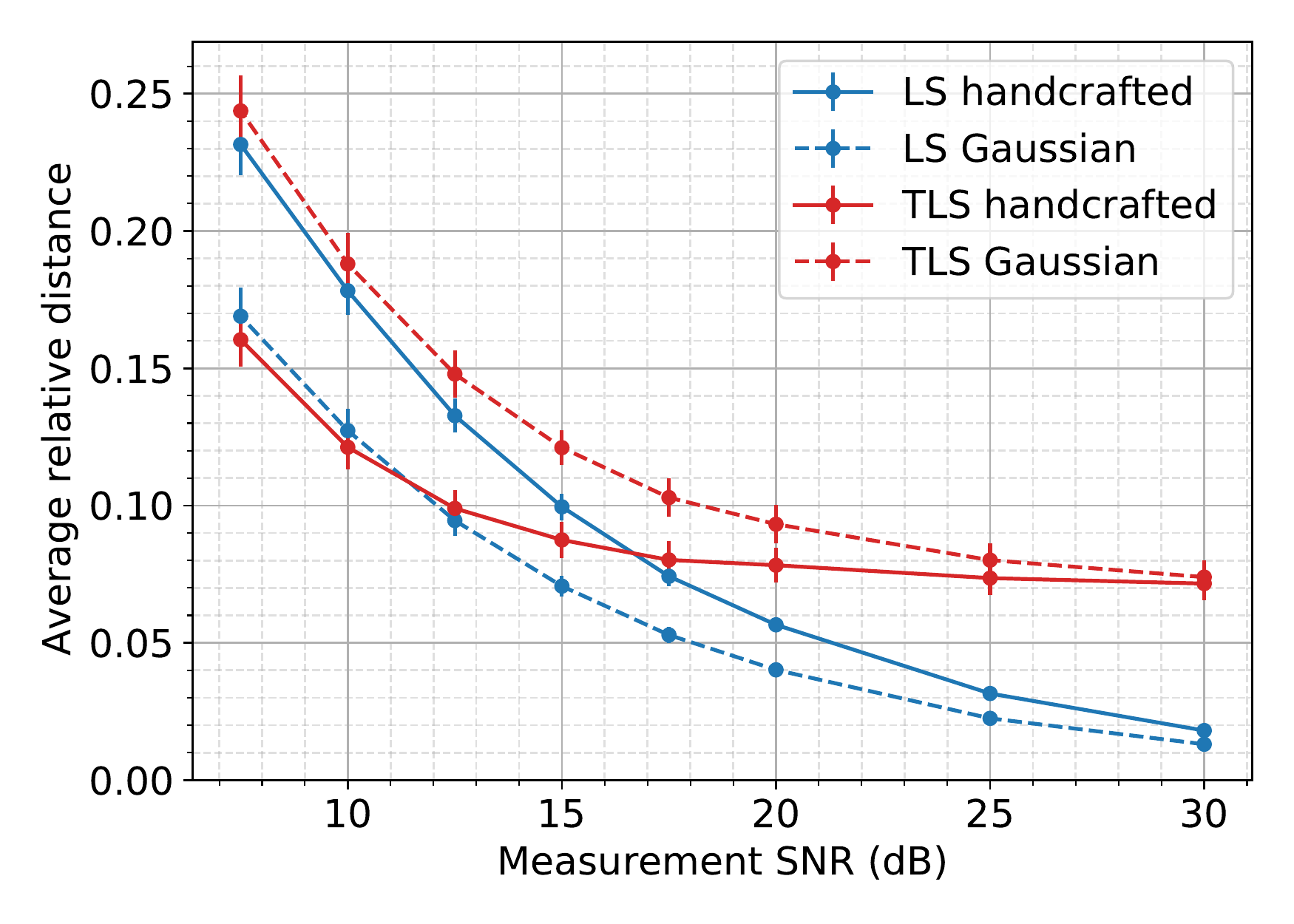}
    \caption{Performance of TLS and LS with handcrafted errors when there is only error in measurements for the CDP measurement model. Here $L = \frac{M}{N} = 16$ octanary patterns are used.}
    \label{fig:cdp_y_targeted}
\end{figure}

\begin{figure}[t]
    \centering
    \includegraphics[width=0.75\linewidth]{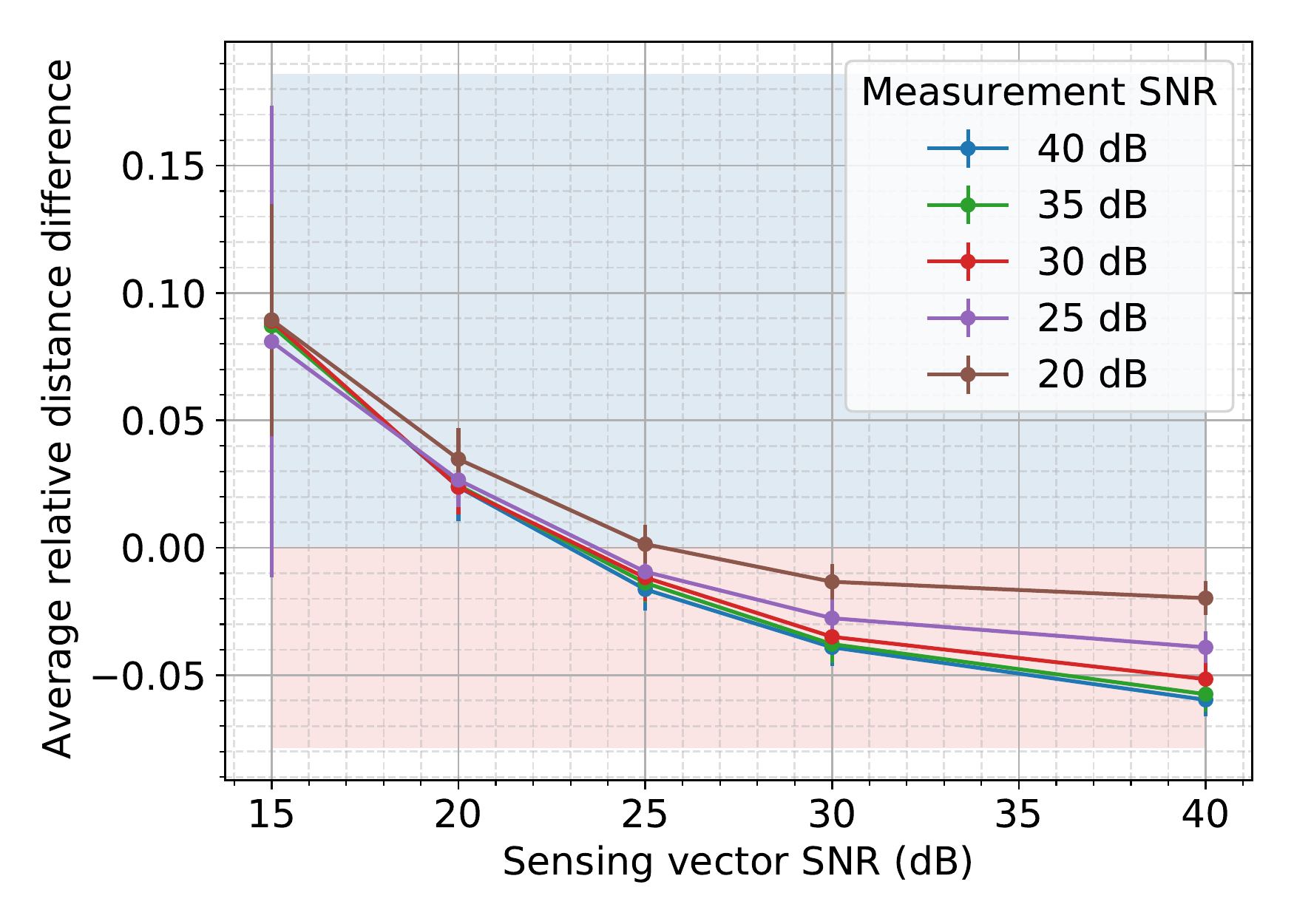}
    \caption{Average difference in relative distance between TLS and LS solutions for the CDP measurement model when $L = \frac{M}{N} = 16$ octanary patterns are used. Different measurement and sensing vector SNR combinations are used and the errors are handcrafted. This should be compared to Fig. \ref{fig:cdp_model_random_perturbations} when $L = \frac{M}{N} = 16$.}
    \label{fig:cdp_targeted_mn16}
\end{figure}

Figs. \ref{fig:cdp_varymn_handcrafted}, \ref{fig:cdp_y_targeted} and \ref{fig:cdp_targeted_mn16} show the performance with handcrafted errors for the CDP measurement model. With only handcrafted measurement error and 100 dB SNR sensing vector error, the performance of TLS is better than LS for low measurement SNR in Fig. \ref{fig:cdp_y_targeted} compared to when there are random Gaussian errors. The performance with different error combinations for handcrafted errors in Fig. \ref{fig:cdp_targeted_mn16} should be compared against Fig. \ref{fig:cdp_model_random_perturbations}.


\rev{

\paragraph{Sensing vector correction verification}

The experiments done to verify the sensing vector corrections for the Gaussian measurement model in Figs. \ref{fig:gaussian_model_random_perturbations_A_correction} and \ref{fig:gaussian_varymn_A_error} are done for the CDP measurement model in Figs. \ref{fig:cdp_model_random_perturbations_A_correction} and \ref{fig:cdp_varymn_A_error}.

\begin{figure}[!t]
    \centering

    \subfloat[$L = \frac{M}{N}=16$]{\includegraphics[width=0.75\linewidth]
        {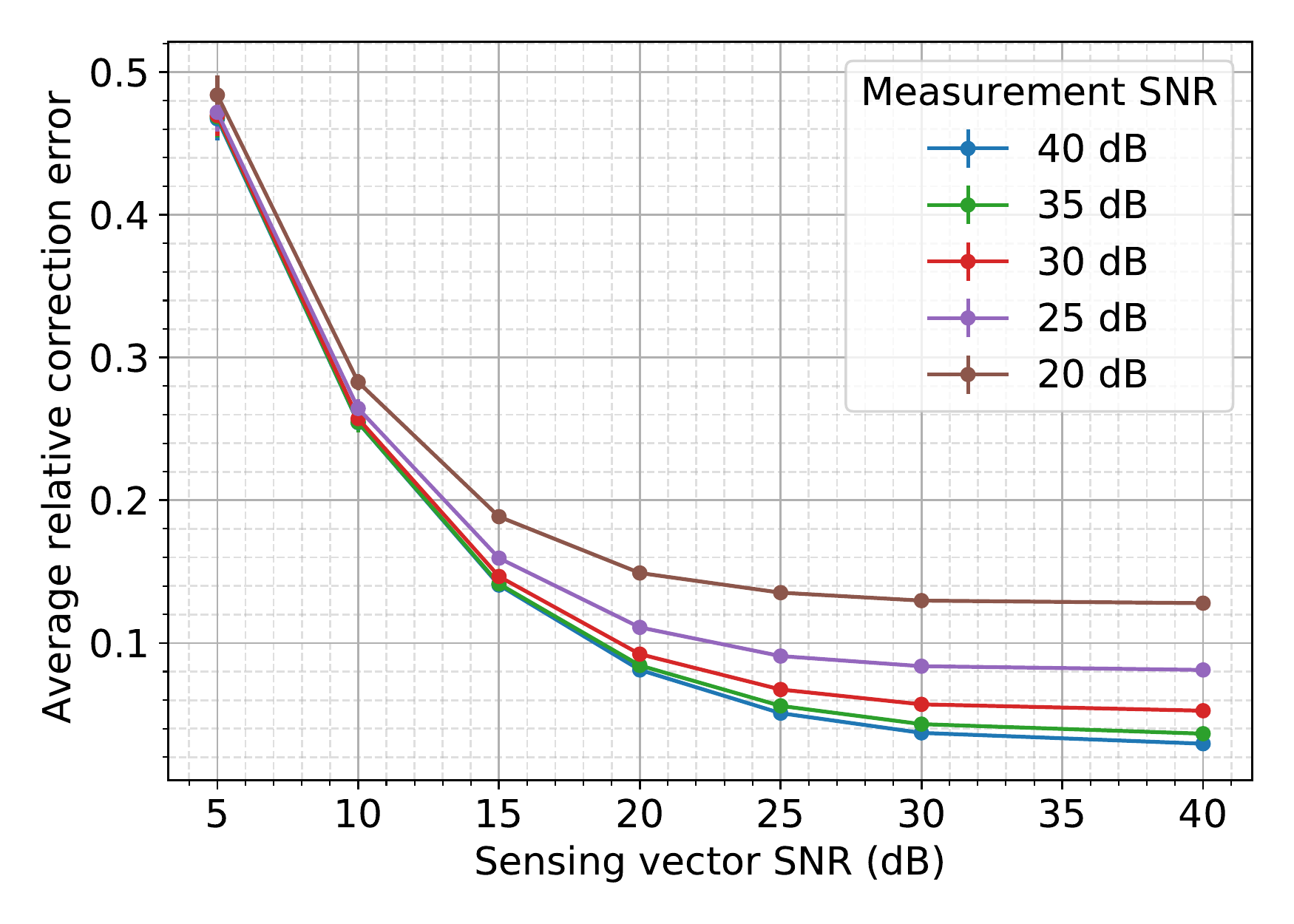}}
    
    \subfloat[$L = \frac{M}{N}=32$]{\includegraphics[width=0.75\linewidth]
        {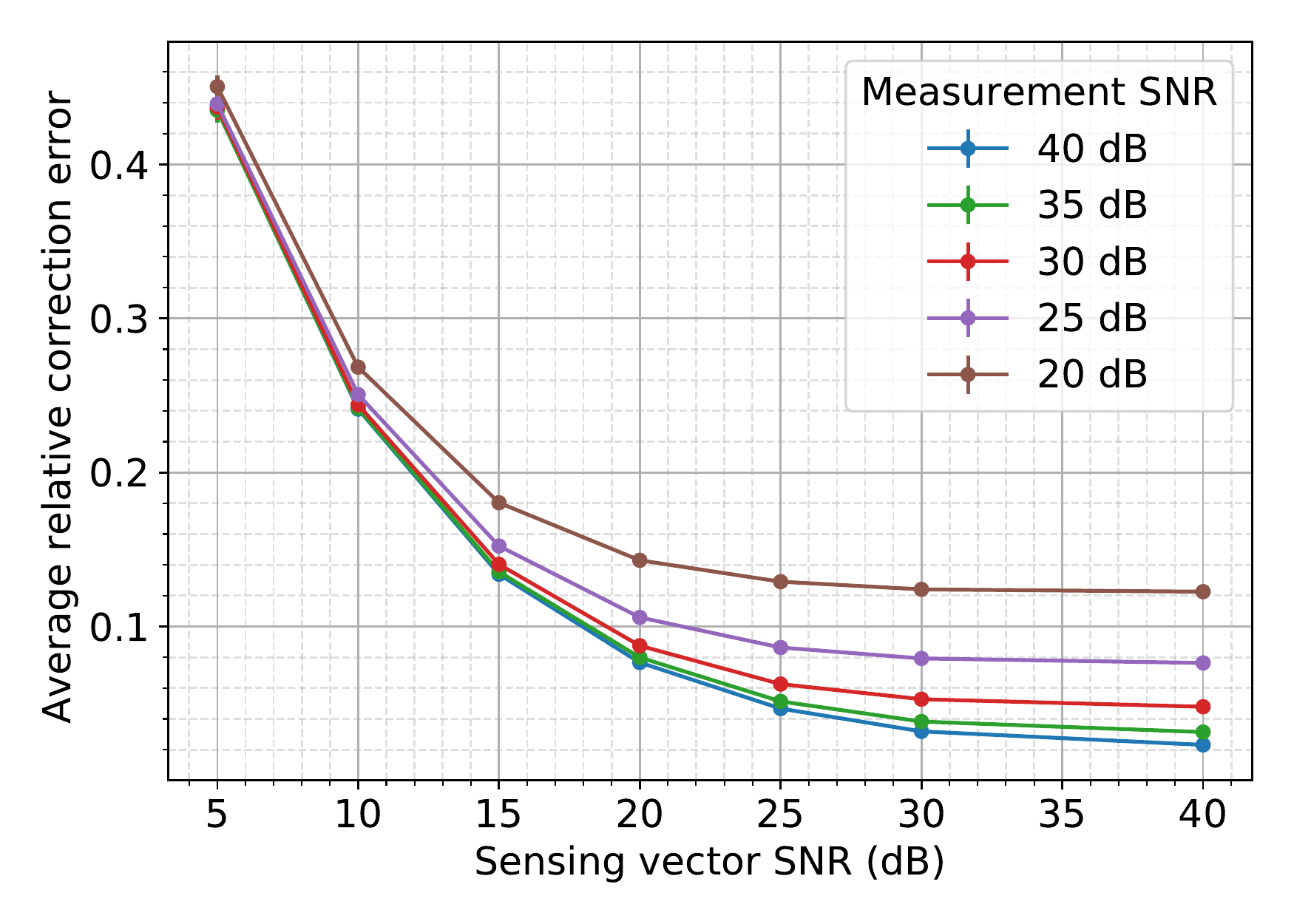}}
    
    \caption{\rev{Average relative sensing vector correction error when using TLS, $\mathrm{rel.corr}(\{\wt{\va},\, \vx^\#\}, \{\wh{\va}^\dag, \,\e^{j \varphi} \vx^\dag_{\mathrm{TLS}})$, for the CDP measurement model for different measurement and sensing vector SNR combinations when the number of patterns is $L = \frac{M}{N} \in \{16, 32\}$.}}
    \label{fig:cdp_model_random_perturbations_A_correction}
\end{figure}

\begin{figure}[t]
    \centering
    \includegraphics[width=0.75\linewidth]{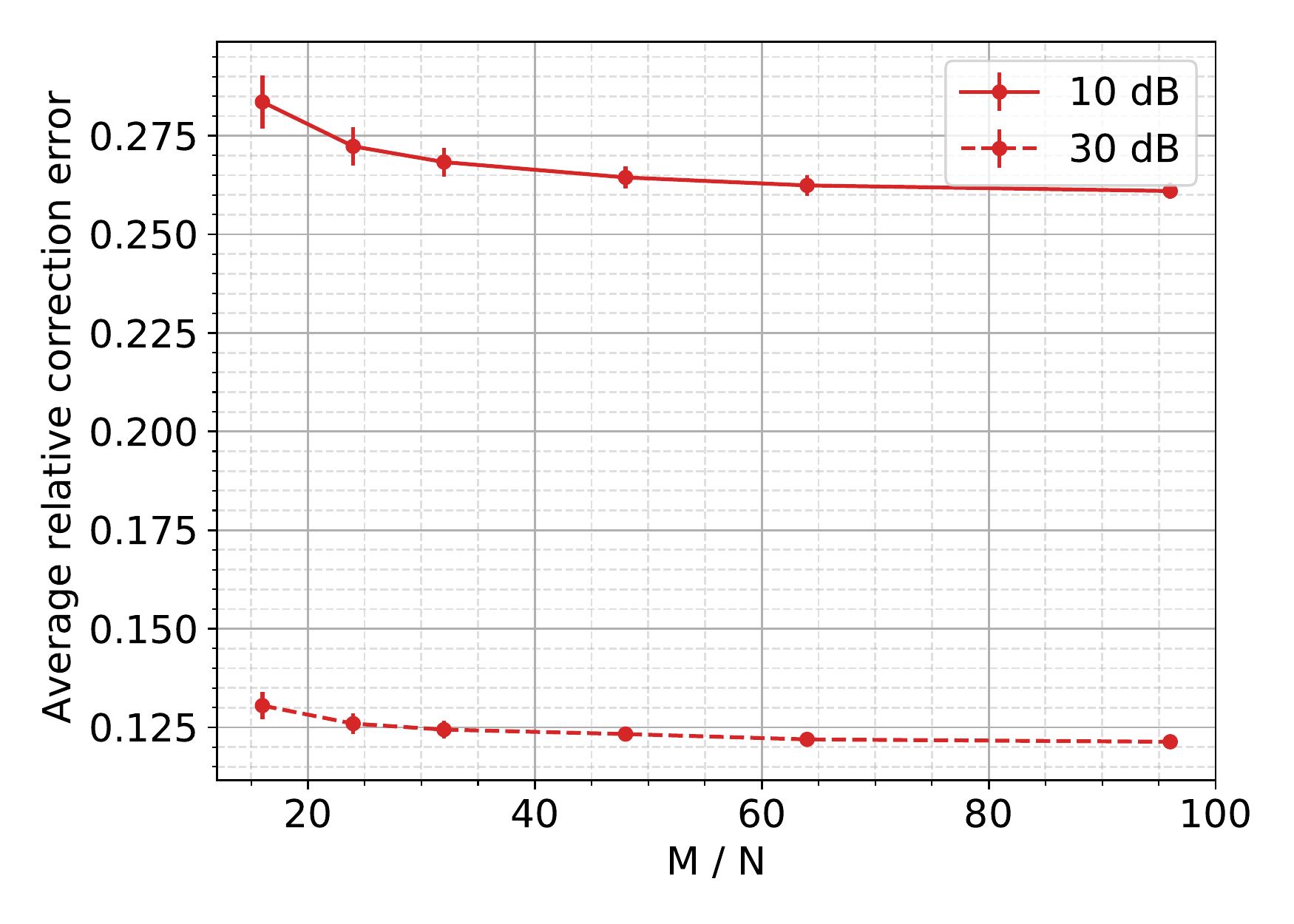}
    \caption{\rev{Relative sensing vector correction error when using TLS for the CDP measurement model for different number of patterns $L = \frac{M}{N}$ when measurement SNR is 20 dB. The sensing vector SNR is 10 dB or 30 dB.}}
    \label{fig:cdp_varymn_A_error}
\end{figure}

}

\section{Additional OPU experiment information} \label{sec:opu_appendix}

\subsection{Sensing vector calibration}

Sensing vector calibration is typically time consuming due to the quadratic nature of \eqref{eq:quadratic_inverse_problem}. We use a rapid numerical interferometry calibration procedure that first inputs $K$ calibration signals, $\mXi = [\vxi_1, \ldots, \vxi_K] \in \R^{N \times K}$ into the OPU and obtains the phase of the corresponding optical \emph{measurements}, $\mS$ in $\mQ := \abs{\mS}^2 \approx \abs{\mA \mXi}^2 \in \R^{M \times K}$. Transmission matrix $\mA \in \C^{M \times N}$ is then recovered by solving the linear system $\mS = \mA \mXi$ \cite{gupta2020fast}. If $\mS$ has errors, the calibrated $\mA$ may have errors. We implemented this method with $1.5N$ calibration signals and 20 anchor signals. We use the same procedure as Gupta et al. to design calibration signals \cite{gupta2020fast}.

\subsection{Experiment details}

When doing the experiments with random images on the OPU, we set the camera exposure time to 700 $\mu s$ to utilize and not saturate the full zero to 255 8-bit measurement range of the camera. The input display frametime is set to 1200 $\mu s$. For the experiments with the real images, the camera exposure is 400 $\mu s$ and the frametime is 500 $\mu s$.

To use a new set of sensing vectors in each trial we calibrate a complex-valued transmission matrix with $2^{17}$ rows. In each trial in Fig. \ref{fig:opu_trials} we then do phase retrieval by choosing a new set of $M$ rows. The optical measurements corresponding to the chosen $M$ rows are used. As previously, the TLS and LS iterations in Algorithm \ref{algo:tls_pr_algorithm} are stopped when the objective function value between successive iterates changes by less than $10^{-6}$ and the initialization is done using 50 iterations of the power method.

Because the output device exposure time controls the range of the measurements, the entries of the calibrated OPU transmission matrix are iid complex Gaussian and the calibrated Gaussian sensing vectors are scaled versions of the sensing vectors from the complex Gaussian measurement model. This does not impact the sensing vector updates in Section \ref{sec:measurement_vector_update} because the procedure does not assume a measurement model. However, the initialization scaling and signal gradient descent updates \eqref{eq:gradient_update} for both TLS and LS require minor changes. Instead of altering these steps we estimate the standard deviation and variance of the calibrated transmission matrix from its entries and divide the calibrated matrix by the estimated standard deviation. Correspondingly, we also divide the measurements by the estimated variance.


\end{document}